\documentclass[jounral]{IEEEtran}
\usepackage{amsmath}
\usepackage{amsfonts}
\usepackage{graphicx}  
\usepackage{caption}
\usepackage{subcaption}
\usepackage{multirow}
\usepackage{graphicx}
\usepackage[table,xcdraw]{xcolor}
\usepackage{amssymb}
\usepackage{color}
\usepackage{latexsym}
\usepackage{cite}
\usepackage{url}
\usepackage{tikz}
\def\checkmark{\tikz\fill[scale=0.4](0,.35) -- (.25,0) -- (1,.7) -- (.25,.15) -- cycle;} 
\def\crossmark{{\Large $\times$}}
\newcommand{\mcell}[1]{\begin{tabular}[c]{@{}l@{}}#1\end{tabular}}

\title{Jamming Attacks and Anti-Jamming Strategies in Wireless Networks: A Comprehensive Survey}
\author{
Hossein Pirayesh
and
Huacheng Zeng

Department of Computer Science and Engineering, Michigan State University, East Lansing, MI USA
}

\begin{document}
\maketitle

\begin{abstract}
Wireless networks are a key component of the telecommunications infrastructure in our society, and wireless services become increasingly important as the applications of wireless devices have penetrated every aspect of our lives. 
Although wireless technologies have significantly advanced in the past decades, most wireless networks are still vulnerable to radio jamming attacks due to the openness nature of wireless channels, and the progress in the design of jamming-resistant wireless networking systems remains limited. 
This stagnation can be attributed to the lack of practical physical-layer wireless technologies that can efficiently decode data packets in the presence of jamming attacks. 
This article surveys existing jamming attacks and anti-jamming strategies in wireless local area networks (WLANs), cellular networks, cognitive radio networks (CRNs), ZigBee networks, Bluetooth networks, vehicular networks, LoRa networks, RFID networks, and GPS system, with the objective of offering a comprehensive knowledge landscape of existing jamming/anti-jamming strategies and stimulating more research efforts to secure wireless networks against jamming attacks. 
Different from prior survey papers, this article conducts a comprehensive, in-depth review on jamming and anti-jamming strategies, casting insights on the design of jamming-resilient wireless networking systems. 
An outlook on promising anti-jamming techniques is offered at the end of this article to delineate important research directions. 


\end{abstract}

\section{Introduction}
\label{sec:introduction}

With the rapid proliferation of wireless devices and the explosion of Internet-based mobile applications under the driving forces of 5G and artificial intelligence, wireless services have penetrated every aspect of our lives and become increasingly important as an essential component of the telecommunications infrastructure in our society. 
In the past two decades, we have witnessed the significant advancement of wireless communication and networking technologies such as polar code \cite{condo2020practical,bioglio2020design}, massive multiple-input multiple-output (MIMO) \cite{albreem2019massive,bjornson2020scalable,liu2018massive}, millimeter-wave (mmwave) \cite{wang2018millimeter,shen2019miniaturized}, non-orthogonal multiple access (NOMA) \cite{sangdeh2020practical,makki2020survey,yuan20205g,chen2019full}, carrier aggregation \cite{goyal2020lte}, novel interference management \cite{naderializadeh2019cache}, learning-based resource allocation \cite{yu2020deep,chen2019joint}, software-defined radio \cite{wyglinski2018software}, and software-defined wireless networking \cite{xia2014survey}. These innovative wireless technologies have dramatically boosted the capacity of wireless networks and the quality of wireless services, leading to a steady evolution of cellular networks towards 5th generation (5G) and Wi-Fi networks towards 802.11ax. 
With the joint efforts from academia, federal governments, and private sectors, it is expected that high-speed wireless services will become ubiquitously available for massive devices to realize the vision of Internet of Everything (IoE) in the near future \cite{miao2019fair}.

\begin{table*}[]
\centering
\caption{This survey article versus prior survey papers.}
\label{tab:survey_papers}
\begin{tabular}{|cccccc|}
\hline
\textbf{Ref.}    & \textbf{Studied networks} & \textbf{Studied layers}    & \textbf{Attacks techniques} & \textbf{Anti-attack strategies}   & \textbf{Attack detection} \\ \hline
\cite{mpitziopoulos2009survey}      & WSNs  & PHY    & \checkmark & \checkmark     & \checkmark   \\ \hline
\cite{zhou2008securing}      & WSNs  & PHY/Network/Session& \checkmark & \checkmark     & \checkmark   \\ \hline
\cite{amin2016comprehensive} & ZigBee networks      & PHY/MAC& \checkmark & \crossmark   & \crossmark \\ \hline
\cite{raymond2008denial}     & WSNs  & \!\!\!\!\!\!\!\!\!\!PHY/MAC/Network/Transport/Application\!\!\!\!\!\!\!\!\!\!     & \checkmark & \checkmark     & \checkmark   \\ \hline
\cite{vadlamani2016jamming}  & WSNs/WLANs   & PHY    & \checkmark & \checkmark     & \checkmark   \\ \hline
\cite{xu2006jamming}  & WSNs  & PHY/MAC& \checkmark & \checkmark     & \checkmark   \\ \hline
\cite{zhang2015byzantine}    & CRNs  & MAC    & \checkmark & \checkmark     & \checkmark   \\ \hline
\cite{das2013primary} & CRNs  & MAC    & \checkmark & \checkmark     & \checkmark   \\ \hline
\cite{fragkiadakis2012survey}& CRNs  & MAC    & \checkmark & \checkmark     & \checkmark   \\ \hline
\cite{di2013jamming}  & CRNs  & PHY/MAC& \crossmark      & \checkmark     & \checkmark   \\ \hline
\cite{attar2012survey}& CRNs  & MAC    & \checkmark & \crossmark & \crossmark\\ \hline
\cite{jover2013security}     & Cellular networks   & PHY/Link/Network     & \checkmark & \crossmark & \crossmark \\ \hline
\cite{grover2014jamming}     & Ad-hoc networks     & PHY/MAC& \checkmark & \checkmark     & \checkmark   \\ \hline
\cite{shahriar2014phy}& OFDM networks & PHY    & \checkmark & \crossmark  & \crossmark\\ \hline
\cite{pelechrinis2010denial} & Ad-hoc networks     & PHY/MAC& \checkmark & \checkmark     & \checkmark   \\ \hline
\multicolumn{1}{|c}{\mcell{~This \\ article}} & \begin{tabular}[c]{@{}c@{}}WLANs/Cellular/CRNs/\\ ZigBee/Bluetooth/Vehicular/\\ GPS/RFID networks\end{tabular} & PHY/MAC/Implementation  & \checkmark & \checkmark     & \multicolumn{1}{c|}{\checkmark}   \\ \hline
\end{tabular}

\end{table*}

As we are increasingly reliant on wireless services, security threats have become a big concern about the confidentiality, integrity, and availability of wireless communications. 
Compared to other security threats such as eavesdropping and data fabrication, wireless networks are particularly vulnerable to radio jamming attacks for the following reasons. 
First, jamming attacks are easy to launch. 
With the advances in software-defined radio, one can easily program a small \$10 USB dongle device to a jammer that covers 20~MHz bandwidth below 6 GHz and up to 100~mW transmission power \cite{vanhoef2014advanced}. 
Such a USB dongle suffices to disrupt the Wi-Fi services in a home or office scenario.
Other off-the-shelf SDR devices such as USRP \cite{ettus2008universal} and WARP \cite{warpv3} are even more powerful and more flexible when using as a jamming emitter. 
The ease of launching jamming attacks makes it urgent to secure wireless networks against intentional and unintentional jamming threats.
Second, jamming threats can only be thwarted at the physical (PHY) layer but not at the MAC or network layer. 
When a wireless network suffers from jamming attacks, its legitimate wireless signals are typically overwhelmed by irregular or sophisticated radio jamming signals, making it hard for legitimate wireless devices to decode data packets.
Therefore, any strategies at the MAC layer or above are incapable of thwarting jamming threats, and innovative anti-jamming strategies are needed at the physical layer. 
Third, the effective anti-jamming strategies for real-world wireless networks remain limited. 
Despite the significant advancement of wireless technologies, most of current wireless networks (e.g., cellular and Wi-Fi networks) can be easily paralyzed by jamming attacks due to the lack of protection mechanism. 
The vulnerability of existing wireless networks can be attributed to the lack of effective anti-jamming mechanisms in practice.
The jamming vulnerability of existing wireless networks also underscores the critical need and fundamental challenges in designing practical anti-jamming schemes.

This article provides a comprehensive survey on jamming attacks and anti-jamming strategies in various wireless networks, with the objectives of providing readers with a holistic knowledge landscape of existing jamming/anti-jamming techniques and stimulating more research endeavors in the design of jamming-resistant wireless networking systems.
Specifically, our survey covers wireless local area networks (WLANs), cellular networks, cognitive radio networks (CRNs), vehicular networks, Bluetooth networks, ad hoc networks, etc. 
For each type of wireless network, we first offer an overview on the system design and then provide a primer on its PHY/MAC layers, followed by an in-depth review of the existing PHY-/MAC-layer jamming and defense strategies in the literature. 
Finally, we offer discussions on open issues and promising research directions.

Prior to this work, there are several survey papers on jamming and/or anti-jamming attacks in wireless networks \cite{mpitziopoulos2009survey,zhou2008securing,amin2016comprehensive,raymond2008denial,vadlamani2016jamming,xu2006jamming,zhang2015byzantine,das2013primary,fragkiadakis2012survey,di2013jamming,attar2012survey,jover2013security,grover2014jamming,shahriar2014phy,pelechrinis2010denial}.
In \cite{mpitziopoulos2009survey}, the authors surveyed the jamming attacks and defense mechanisms in WSNs.
In \cite{zhou2008securing}, Zhou et al.	surveyed the security challenges on WSNs' network protocols, including key establishment, authentication, integrity protection, and routing.
In \cite{amin2016comprehensive}, Amin et al. surveyed PHY and MAC layer attacks on IEEE 802.15.4 (ZigBee networks).
In \cite{raymond2008denial}, Raymond et al. focused on denial-of-service attacks and the countermeasures in higher WSNs' network protocols (e.g., transport and application layers).
\cite{vadlamani2016jamming}	classified the attacks and countermeasure techniques from both the attacker and the defender's perspective, the game-theoretical models, and the solutions used in WSNs and WLANs.
In \cite{xu2006jamming}, Xu et al. surveyed jamming attacks, jamming detection strategies, and defense techniques in WSNs.
In \cite{zhang2015byzantine}, Zhang et al. surveyed a MAC layer attack, known as Byzantine attack (a.k.a. false-report attack), and its possible countermeasures in cooperative spectrum sensing CRNs.
In \cite{das2013primary}, Das et al. surveyed a MAC layer security threat called primary user emulation attack, its detection mechanisms, and defense techniques in CRNs.
\cite{fragkiadakis2012survey, di2013jamming, attar2012survey} summarized the jamming attacks, MAC layer security challenges, and detection techniques in CRNs.
\cite{jover2013security} surveyed the denial of service attacks in LTE cellular networks.
\cite{grover2014jamming, pelechrinis2010denial} reviewed the generic PHY-layer jamming attacks, detection, and countermeasures in wireless ad hoc networks.
In \cite{shahriar2014phy}, Shahriar et al. offered a comprehensive overview of PHY layer security challenges in OFDM networks.
Table~\ref{tab:survey_papers} summarizes the existing survey works on jamming and/or anti-jamming attacks in wireless networks.

Unlike prior survey papers, this article conducts a comprehensive review on up-to-date jamming/anti-jamming strategies, and provides the necessary PHY/MAC-layer knowledge to understand the jamming/anti-jamming strategies in various wireless networks.
The contributions of this paper are summarized as follows.
\begin{itemize}
\item 
We conduct a comprehensive, in-depth review on existing jamming attacks in various wireless networks, including WLANs, cellular networks, CRNs, Bluetooth and ZigBee networks, LoRaWANs, VANETs, UAVs, RFID systems, and GPS systems. 
We offer the necessary PHY/MAC-layer knowledge to understand the destructiveness of jamming attacks in those networks.

\item 
We conduct an in-depth survey on existing anti-jamming strategies in different wireless networks, including power control, spectrum spreading, frequency hopping, MIMO-based jamming mitigation, and jamming-aware protocols. 
We quantify their jamming mitigation capability and discuss their applications.

\item 
In addition to the review of jamming and anti-jamming strategies, we discuss the open issues of jamming threats in wireless networks and point out promising research directions. 

\end{itemize}

\begin{figure}
	\centering
	\includegraphics[width=3.4in]{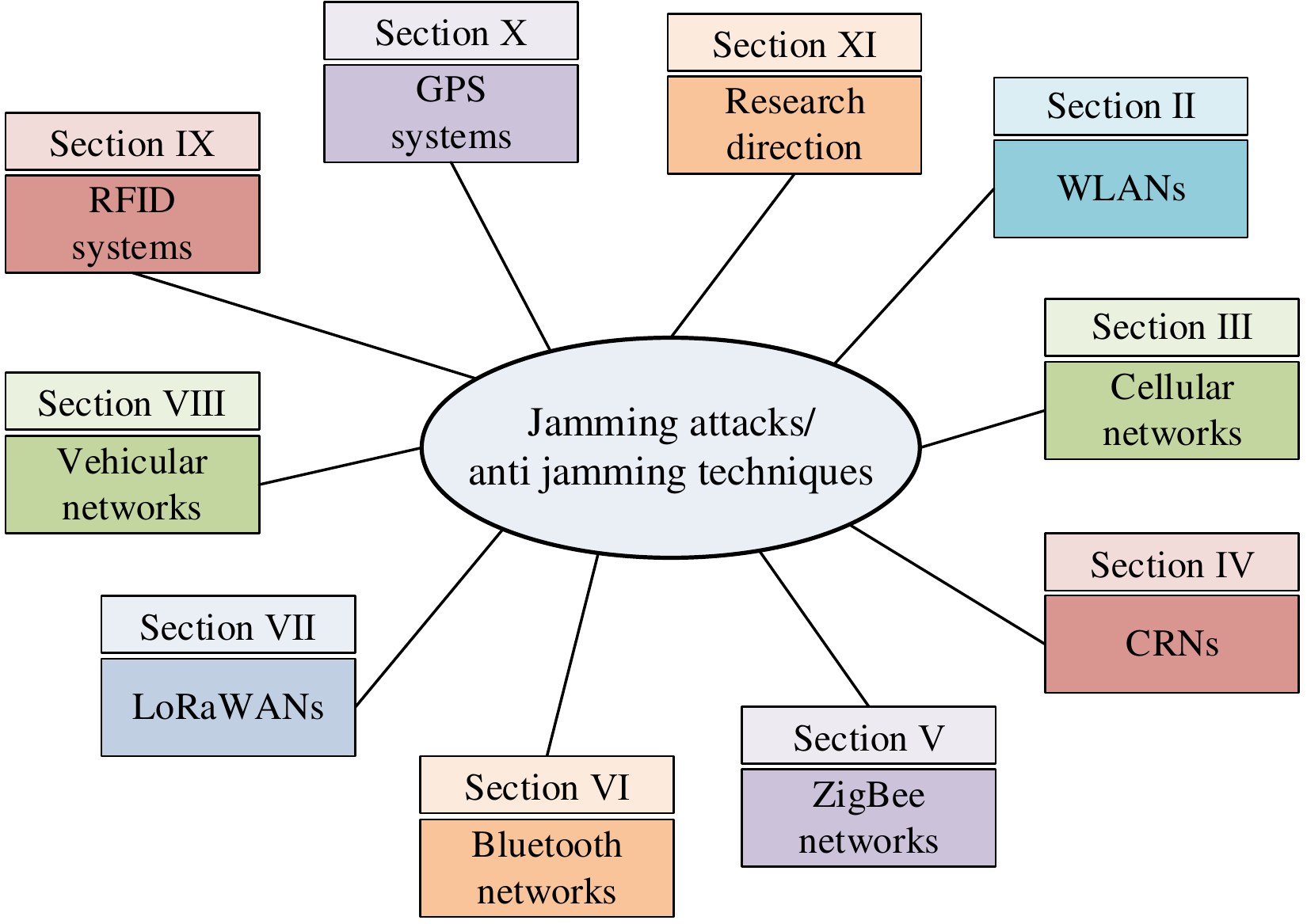}
	\caption{The structure of this article.}
	\label{fig:summary}
\end{figure}
\begin{table}[t]
\caption{List of abbreviations.}
\label{tab:acr}
\centering
\begin{tabular}{|l|l|}
\hline
\textbf{Abbreviation} & \textbf{Explanation}                                     \\ \hline
AP               & Access Point                                           \\ \hline
ARF              & Automatic Rate Fallback                                \\ \hline
ARQ              & Automatic Repeat Request                               \\ \hline
BLE              & Bluetooth Low Energy                                   \\ \hline
COTS             & Commercial Off-The-Shelf                               \\ \hline
CP               & Cyclic Prefix                                          \\ \hline
CRC              & Cyclic Redundancy Check                                \\ \hline
CRN              & Cognitive Radio Network                                \\ \hline
CSMA/CA          & Carrier-Sense Multiple Access/Collision Avoidance      \\ \hline
CSS              & Cooperative Spectrum Sensing                           \\ \hline
DCI/UCI          & Downlink/Uplink Control Information                    \\ \hline
DoS              & Denial of  Service                                     \\ \hline
DSSS             & Direct-Sequence Spread Spectrum                        \\ \hline
FC               & Fusion Center                                          \\ \hline
FHSS             & Frequency Hopping Spread Spectrum                      \\ \hline
GMSK             & Gaussian Minimum Shift Keying                          \\ \hline
GPS              & Global Positioning System                              \\ \hline
ICI              & Inter Channel Interference                             \\ \hline
LoRaWAN          & LoRa Wide Area Network                                 \\ \hline
LTE              & Long Term Evolution                                    \\ \hline
LTF              & Long Training Field                                    \\ \hline
MAC              & Medium Access Control                                  \\ \hline
MCS              & Modulation and Coding Scheme                           \\ \hline
MIB/SIB          & Master/System Information Block                        \\ \hline
MIMO             & Multiple Input Multiple Output                         \\ \hline
MMSE             & Minimum Mean Square Error                              \\ \hline
MU-MIMO          & Multi User Multiple Input Multiple Output              \\ \hline
NAV              & Net Allocation Vector                                  \\ \hline
NDP              & Null Data Packet                                       \\ \hline
NOMA             & Non-Orthogonal Multiple Access                         \\ \hline
OFDM             & Orthogonal Frequency Division Multiplexing             \\ \hline
PBCH             & Physical Broadcast Channel                             \\ \hline
PDCCH/PUCCH      & Physical Downlink/Uplink Control Channel               \\ \hline
PDSCH/PUSCH      & Physical Downlink/Uplink Shared Channel                \\ \hline
PHICH            & Physical Hybrid ARQ Indicator Channel                  \\ \hline
PRACH            & Physical Random Access Channel                         \\ \hline
PRB              & Physical Resource Block                                \\ \hline
PSS/SSS          & Primary/Secondary Synchronization Signal               \\ \hline
PU               & Primary User                                           \\ \hline
PUE              & Primary User Emulation                                 \\ \hline
RAA              & Rate Adaptation Algorithm                              \\ \hline
RAT              & Radio Access Technology                                \\ \hline
RFID             & Radio-Frequency Identification                         \\ \hline
RSI              & Road Side Infrastructure                               \\ \hline
RSSI             & Received Signal Strength Indicator                     \\ \hline
RTS/CTS          & Request To Send/Clear To Send                          \\ \hline
SC-FDMA          & Single Carrier Frequency Division Multiple Access      \\ \hline
SDR              & Software Defined Radio                                 \\ \hline
SNR              & Signal to Noise Ratio                                  \\ \hline
STF              & Short Training Field                                   \\ \hline
SU               & Secondary User                                         \\ \hline
UAV              & Unmanned Aerial Vehicle                                \\ \hline
USRP             & Universal Software Radio Peripheral                    \\ \hline
VANET            & Vehicular Network                                      \\ \hline
VHT              & Very High Throughput                                   \\ \hline
WCDMA            & Wideband Code Division Multiple Access                 \\ \hline
WLAN             & Wireless Local Area Network                            \\ \hline
WSN              & Wireless Sensor Network                                \\ \hline
ZF               & Zero Forcing                                           \\ \hline
\end{tabular}
\end{table}

The remainder of this article is organized following the structure as shown in Fig.~\ref{fig:summary}.
In Section~\ref{sec:wlan}, we survey jamming and anti-jamming attacks in WLANs.
In Section~\ref{sec:cellular}, we survey jamming and anti-jamming attacks in cellular networks.
In Section~\ref{sec:crn}, we survey jamming and anti-jamming attacks in cognitive radio networks.
Sections \ref{sec:zigbee} and \ref{sec:bluetooth} offer an in-depth review on jamming attacks and anti-jamming techniques for ZigBee and Bluetooth networks, respectively.
Section \ref{sec:lora} presents an overview of jamming attacks and anti-jamming techniques in LoRa communications.
Section~\ref{sec:vanet} studies existing jamming and anti-jamming techniques for vehicular networks, including on-ground vehicular transportation networks (VANETs) and in-air unmanned aerial vehicular (UAV) networks.
Section~\ref{sec:rfid} reviews jamming and anti-jamming techniques for RFID systems, and
Second~\ref{sec:gps} reviews those techniques for GPS systems.
Section~\ref{sec:open_problems} discusses open problems and points out some promising research directions.
Section~\ref{sec:conclusion} concludes this article.
Table~\ref{tab:acr} lists the abbreviations used in this article.

\section{Jamming and Anti-Jamming Attacks in WLANs}
\label{sec:wlan}
\begin{figure}
	\centering
	\includegraphics[width=3.35in]{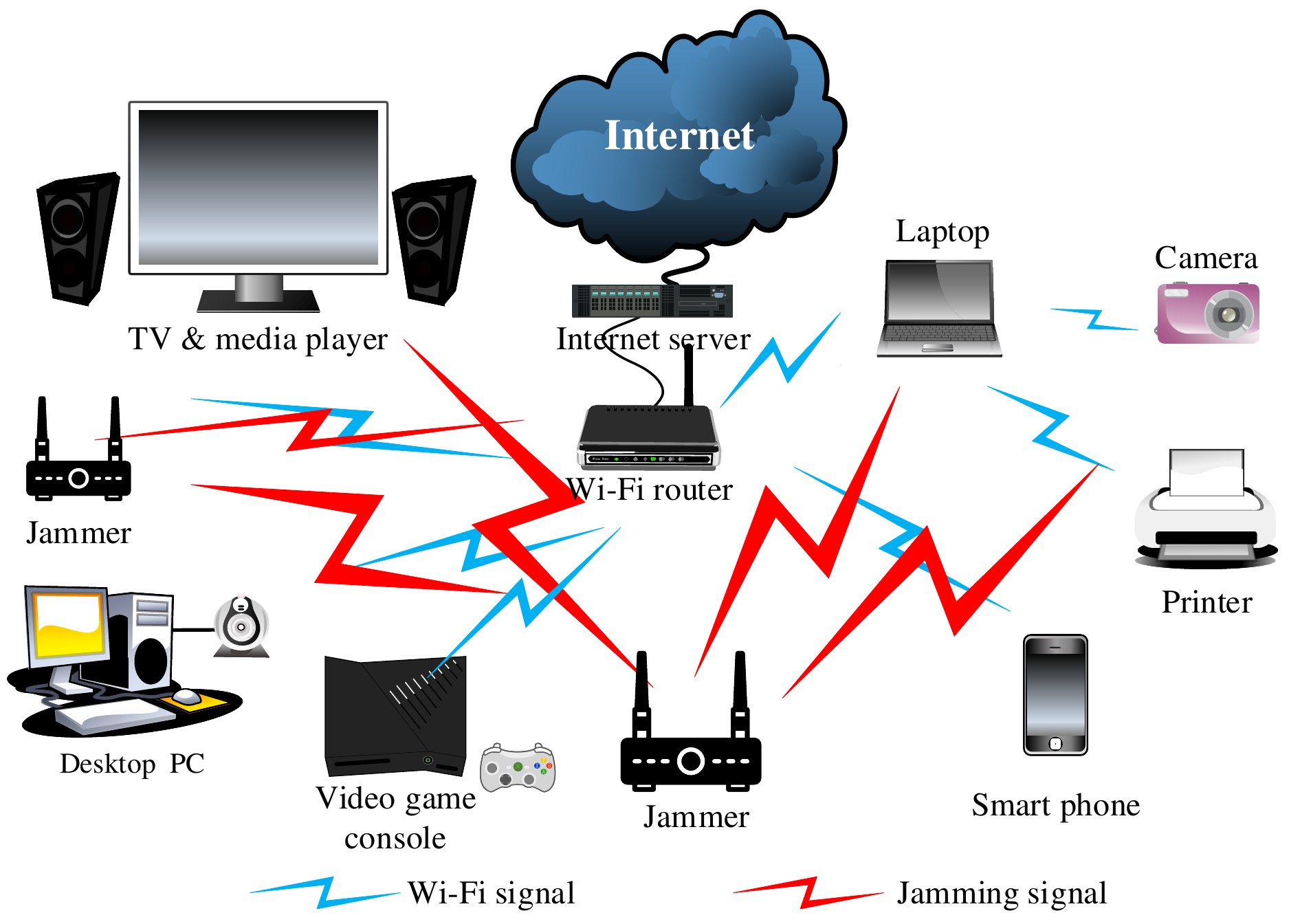}
	\caption{Illustration of jamming attacks in a Wi-Fi network.}
	\label{fig:wifi_jammer}
\end{figure}

WLANs become increasingly important as they carry even more data traffic than cellular networks. 
With the proliferation of wireless applications in smart homes, smart buildings, and smart hospital environments, securing WLANs against jamming attacks is of paramount importance. 
In this section, we study existing jamming attacks and anti-jamming techniques for a WLAN as shown in Fig.~\ref{fig:wifi_jammer}, where one or more malicious jamming devices attempt to disrupt wireless connections of Wi-Fi devices.
Prior to that, we first review the MAC and PHY layers of WLANs, which will lay the knowledge foundation for our review on existing jamming/anti-jamming strategies.

\subsection{A Primer of WLANs}

As shown in Fig.~\ref{fig:wifi_jammer}, WLANs are the most dominant wireless connectivity infrastructure for short-range and high-throughput Internet services and have been widely deployed in population-dense scenarios such as homes, offices, campuses, shopping malls, and airports.
Wi-Fi networks have been designed based on the IEEE 802.11 standards, and 802.11a/g/n/ac standards are widely used in various commercial Wi-Fi devices such as smartphones, laptops, printers, cameras, and smart televisions. 
Most of Wi-Fi networks operate in unlicensed industrial, scientific, and medical (ISM) frequency bands, which have 14 overlapping $20$~MHz channels on 2.4 GHz and 28 non-overlapping $20$~MHz channels bandwidth in 5 GHz \cite{7797535}.
Most Wi-Fi devices are limited to a maximum transmit power of $100$~mW, with a typical indoor coverage range of 35 m.
A Wi-Fi network can cover up to $1$~km range in outdoor environments in an extended coverage setting.  


\subsubsection{MAC-Layer Protocols}
Wi-Fi devices use CSMA/CA as their MAC protocols for channel access.
A Wi-Fi user requires to sense the channel before it sends its packets.
If the channel sensed busy, the user waits for a DIFS time window and backs off its transmissions for a random amount of time.
If the user cannot access the channel in one cycle, it cancels the random back-off counting and stands by for the channel to be idle for the DIFS duration.
In this case, the user can immediately access the channel as the longer waiting users have priority over the users recently joined the network.

\begin{figure}
	\centering
	\includegraphics[width=3.5in]{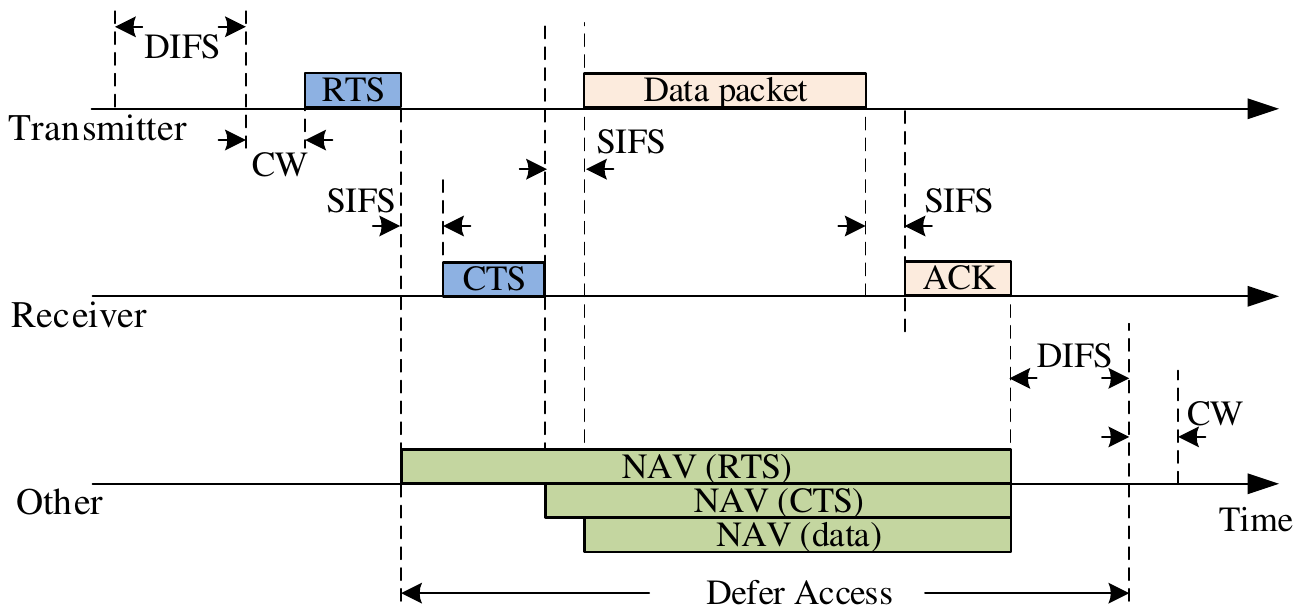}
	\caption{The RTS/CTS protocol in 802.11 Wi-Fi networks \cite{perahia2013next}.}\vspace{-0.15in}
	\label{fig:wifi_mac}
\end{figure}

The CSMA/CA MAC protocol, however, suffers from the hidden node problem.
The hidden node problem refers to the case where one access point (AP) can receive from two nodes, but those two nodes cannot receive from each other.
If both nodes sense the channel idle and send their data to the AP, then packet collision occurs at the AP. 
The RTS/CTS (Request-to-Send and Clear-to-Send) protocol was invented to mitigate the hidden node problem, and 
Fig.~\ref{fig:wifi_mac} shows the RTS/CTS protocol mechanism. 
The transmitter who intends to access the channel waits for the DIFS duration.
If the channel is sensed idle, the transmitter sends an RTS packet to identify the receiver and the required duration for data transmission.
Every node receiving the RTS sets its Net Allocation Vector (NAV) to defer its try for accessing the channel to the subsequent frame exchange.

While previous and current Wi-Fi networks (e.g., 802.11g, 802.11n and 802.11ac) use the distributed CSMA/CA protocol for medium access control, the next-generation 802.11ax Wi-Fi networks (marketed as Wi-Fi 6) come with a centralized architecture with features such as OFDMA, both uplink and downlink MU-MIMO, trigger-based random access, spatial frequency reuse, and target wake time (TWT) \cite{7422404,8468986}. 
Despite these new features, 802.11ax devices will be backward compatible with the predecessor Wi-Fi devices. 
Therefore, the jamming and anti-jamming attacks designed for 802.11n/ac Wi-Fi networks also apply to the upcoming 802.11ax Wi-Fi networks. 

\begin{figure}
	\centering
	\begin{subfigure}[b]{3.5in}
		\centering
		\includegraphics[width=3.5in]{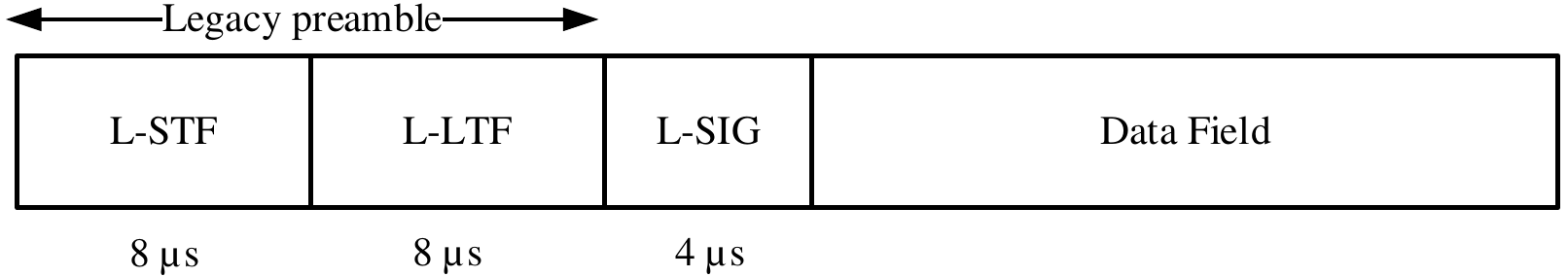}
		\caption{Legacy Wi-Fi frame structure.}\vspace{.1in}
	\end{subfigure}
	\begin{subfigure}[b]{3.5in}
		\centering
		\includegraphics[width=3.5in]{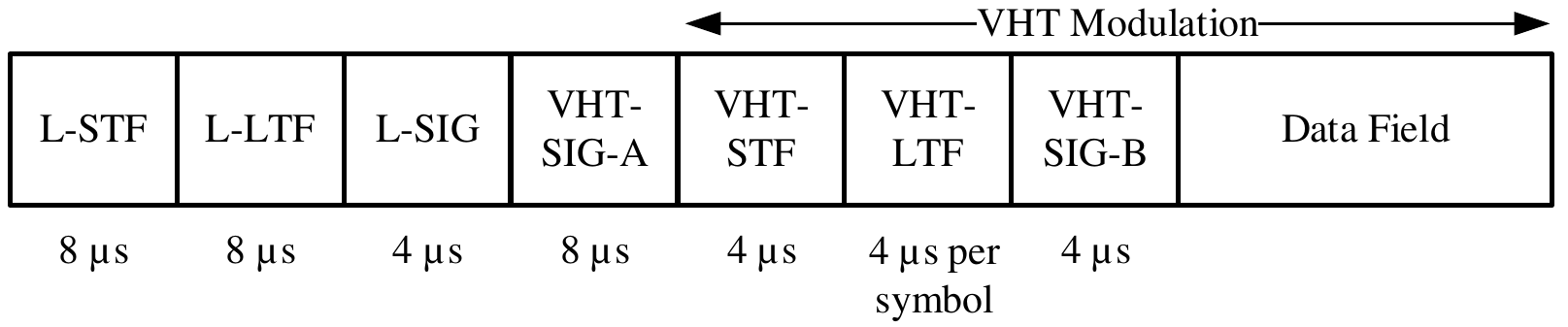}
		\caption{VHT Wi-Fi frame structure.}
	\end{subfigure}
	\caption{Two frame structures used in 802.11 Wi-Fi networks.}
	\label{fig:wifi_frame}

\end{figure}

\subsubsection{Frame Structures} 

Most Wi-Fi networks use OFDM modulation at the PHY layer for both uplink and downlink transmissions.
Fig.~\ref{fig:wifi_frame}(a) shows the legacy Wi-Fi (802.11a/g) frame, which consists of preamble, signal field, and data field.
The preamble comprises two STFs and two LTFs, mainly used for frame synchronizations and channel estimation purposes.
In particular, STF consists of ten identical symbols and is used for start-of-packet detection, coarse time and frequency synchronizations. 
LTF consists of two identical OFDM symbols and is used for fine packet and frequency synchronizations.
LTF is also used for channel estimation and equalization.
Following the preamble, the signal (SIG) field carries the necessary packet information such as the adopted modulation and coding scheme (MCS) and the data part's length.
SIG field is always transmitted using BPSK modulation for minimizing the error probability at the receiver side.
Data field carries user payloads and user-specific information.
Wi-Fi may use different MCS (e.g., OQPSK, 16-QAM, 64-QAM) for data bits modulation, depending on the link quality.
Four pilot signals are also embedded into four different tones (subcarriers) for further residual carrier and phase offset compensation in the data field.
 
\begin{figure}
	\centering
	\includegraphics[width=3.5in]{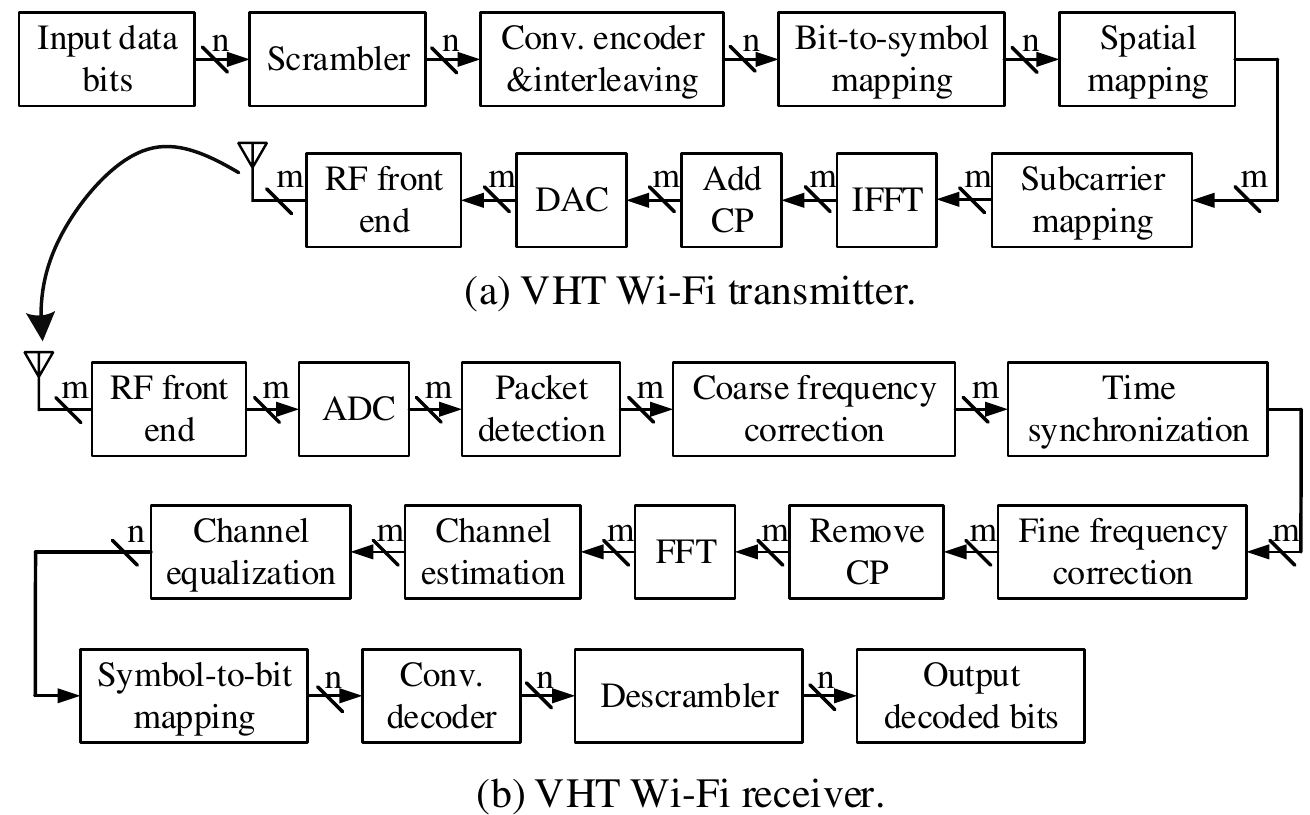}
	\caption{A schematic diagram of baseband signal processing for an 802.11 Wi-Fi transceiver \cite{perahia2013next} ($n \le 4$ and $m \le 8$).}\vspace{-0.15in}
	\label{fig:wifi_transceiver}
\end{figure} 

Fig.~\ref{fig:wifi_frame}(b) shows the VHT format structure used by 802.11ac.
As shown in the figure, it consists of L-STF, L-LTF, L-SIG, VHT-SIG-A, VHT-STF, VHT-LTF, VHT-SIG-B, and Data Field.
To maintain its backward compatibility with 802.11a/g, the L-STF, L-LTF, and L-SIG in the VHT frame are the same as those in Fig.~\ref{fig:wifi_frame}(a).
VHT-SIG-A and VHT-SIG-B are for similar purpose as the header field (HT-SIG) of 11n and SIG field of 11a. 
In 802.11ac, signal fields are SIG-A and SIG-B. 
They describe channel bandwidth, modulation-coding and indicate whether the frame is for a single user or multiple users. 
These fields are only deployed by the 11ac devices and are ignored by 11a and 11n devices.
VHT-STF has the same function as that of the non-HT STF field. 
It assists the 11ac receiver to detect the repeating pattern.
VHT-LTF consists of a sequence of symbols and is used for demodulating the rest of the frame. 
Its length depends on the number of transmitted streams. 
It could be 1, 2, 4, 6, or 8 symbols. 
It is mainly used for channel estimation purposes.
Data field carries payload data from the upper layers. 
When there are no data from upper layers, the field is referred to as the null data packet (NDP) and is used for measurement and beamforming sounding purposes by the physical layer.

\subsubsection{PHY-Layer Signal Processing Modules} 

Fig.~\ref{fig:wifi_transceiver} shows the PHY-layer signal processing framework of a legacy Wi-Fi transceiver.
On the transmitter side, the data bitstream is first scrambled and then encoded using a convolutional or LDPC encoder.
The coded bits are modulated according to the pre-selected MCS index.
Then, the modulated data and pilot signals are mapped onto the scheduled subcarriers and converted to the time domain using OFDM modulation (IFFT operation).
Following the OFDM modulation, the cyclic prefix (CP) is appended to each OFDM symbol in the time domain.
After that, a preamble is attached to the time-domain signal.
Finally, the output signal samples are up-converted to the desired carrier frequency and transmitted over the air using a radio frequency (RF) front-end module.

Referring to Fig.~\ref{fig:wifi_transceiver} again, on the receiver side, the received radio signal is down-converted to baseband I/Q signals, which are further converted to digital streams by ADC modules.
The start of a packet can be detected by auto-correlating the received signal stream with itself in a distance of one OFDM symbol to identify the two transmitted STF signals within the frame.   
The received STF signals can be used to coarsely estimate the carrier frequency offset, which can then be utilized to correct the offset and improve the timing synchronization accuracy.
Timing synchronization can be done by cross-correlating the received signal and a local copy of the LTF signal at the receiver.
LTF is also used for fine frequency offset correction.
Once the signal is synchronized, it is converted into the frequency domain using the OFDM demodulation, which comprises CP removal and FFT operation.  

The received LTF symbols are used to estimate the wireless channel between the Wi-Fi transmitter and receiver for each subcarrier.
Channel smoothing, which refers to interpolating the estimated channel for each subcarrier using its adjacent estimated subcarriers' channels, is usually used to suppress the impact of noise in the channel estimation process.
The estimated channels are then used to equalize the channel distortion of the received frame in the frequency domain.
The received four pilots are used for residual carrier frequency, and phase offsets correction.
After phase compensation, the received symbols are mapped into their corresponding bits.
This process is called symbol-to-bit mapping. 
Following the symbol-to-bit mapping, convolutional or LDPC decoder and descrambler are applied to recover the transmitted bits.
The recovered bits are fed to the MAC layer for protocol-level interpretation.

\subsection{Jamming Attacks}
\label{subsec:wifi_jamming}

With the primer knowledge provided above, we now dive into the review of existing jamming attacks in WLANs.
In what follows, we first survey the generic jamming attacks proposed for Wi-Fi networks but can also be applied to other types of wireless networks and then review the jamming attacks that delicately target the PHY transmission and MAC protocols of Wi-Fi communications. 


\subsubsection{Generic Jamming Attacks}
While there are many jamming attacks that were originally proposed for Wi-Fi networks, they can also be applied to other types of wireless systems. 
We survey these generic jamming attacks in this part.

\noindent \textbf{Constant Jamming Attacks:}
Constant jamming attacks refer to the scenario where the malicious device broadcasts a powerful signal all the time.
Constant jamming attacks not only destroy legitimate users' packet reception by introducing high-power interference to their data transmissions, but they also prevent them from accessing the channel by continuously occupying it.
In constant jamming attacks, the jammer may target the entire or a fraction of channel bandwidth occupied by legitimate users \cite{grover2014jamming,pelechrinis2010denial}.
In \cite{karhima2004ieee}, Karishma et al. analyzed the performance of legacy Wi-Fi communications under broadband and partial-band constant jamming attacks through theoretical exploration and experimental measurement.
The authors conducted experiments to study the impact of jamming power on Wi-Fi communication performance when the data rate is set to $18$~Mbps.
Their experimental results show that a Wi-Fi receiver fails to decode its received packets under broadband jamming attack (i.e., $100\%$  packet error rate) when the received desired signal power is $4$~dB less than the received jamming signal power (i.e., signal-to-jamming power ratio, abbreviated as SJR, less than $4$~dB).
The theoretical analysis in \cite{karhima2004ieee,jun2007bit} showed that Wi-Fi communication is more resilient to partial-band jamming than broadband jamming attacks.
The experimental results in \cite{karhima2004ieee} showed that, for the jamming signal with bandwidth being one subcarrier spacing (i.e., $312.5$~KHz), Wi-Fi communication fails when $\text{SJR} < -19$~dB.
In \cite{vanhoef2014advanced}, Vanhoef et al. used a commercial Wi-Fi dongle and modified its firmware to implement a constant jamming attack.
To do so, they disabled the CSMA protocol, backoff mechanism, and ACK waiting time.
To enhance the jamming effect, they also removed all interframe spaces and injected many packets for transmissions.   

\noindent \textbf{Reactive Jamming Attacks:}
Reactive jamming attack is also known as channel-aware jamming attack, in which a malicious jammer sends an interfering radio signal when it detects legitimate packets transmitted over the air \cite{cai2013joint}.
Reactive jamming attacks are widely regarded as an energy-efficient attack strategy since the jammer is active only when there are data transmissions in the network.
Reactive jamming attack, however, requires tight timing constraints (e.g., $< 1~\text{OFDM symbols}$, $4~\mu$s) for real-world system implementation because it needs to switch from listening mode to transmitting mode quickly. 
In practice, a jammer may be triggered by either channel energy-sensing or part of a legitimate packet's detection (e.g., preamble detection). 
In \cite{prasad2011jamming}, Prasad~et~al. implemented a reactive jamming attack in legacy Wi-Fi networks using the energy detection capability of cognitive radio devices.
In \cite{yan2014mimo,yan2016jamming}, Yan et al. studied a reactive jamming attack where a jammer sends a jamming signal after detecting the preamble of the transmitted Wi-Fi packets.
By doing so, the jammer is capable of effectively attacking Wi-Fi packet payloads. 
In \cite{schulz2017massive}, Schulz et al. used commercial off-the-shelf (COTS) smartphones to implement an energy-efficient reactive jammer in Wi-Fi networks.
Their proposed scheme is capable of replying ACK packets to the legitimate transmitter to hijack its retransmission protocol, thereby resulting in a complete Wi-Fi packet loss whenever packet error occurs.
In \cite{bayraktaroglu2013performance}, Bayraktaroglu et al. evaluated the performance of Wi-Fi networks under reactive jamming attacks.
Their experimental results showed that reactive jamming could result in a near-zero throughput in real-world Wi-Fi networks. 
In \cite{vanhoef2014advanced}, Vanhoef et al. implemented a reactive jamming attack using a commercial off-the-shelf Wi-Fi dongle.
The device decodes the header of an on-the-air packet to carry out the attack implementation, stops receiving the frame, and launches the jamming signal.

\noindent \textbf{Deceptive Jamming Attacks:}
In deceptive jamming attacks, the malicious jamming device sends meaningful radio signals to a Wi-Fi AP or legitimate Wi-Fi client devices, with the aim of wasting a Wi-Fi network's time, frequency, and/or energy resources and preventing legitimate users from channel access.
In \cite{broustis2009fiji}, Broustis et al. implemented a deceptive jamming attack using a commercial Wi-Fi card.
The results in \cite{broustis2009fiji} showed that a low-power deceptive jammer could easily force a Wi-Fi AP to allocate all the network's resources for processing and replying fake signals issued by a jammer, leaving no resource for the AP to serve the legitimate users in the network.
In \cite{gvozdenovic2020truncate}, Gvozdenovic et al. proposed a deceptive jamming attack on Wi-Fi networks called truncate after preamble (TaP) jamming and evaluated its performance on USRP-based testbed. 
TaP attacker lures legitimate users to wait for a large number of packet transmissions by sending them the packets' preamble and the corresponding signal field header only.

\noindent \textbf{Random and Periodic Jamming Attacks:}
Random jamming attack (a.k.a. memoryless jamming attack) refers to the type of jamming attack where a jammer sends jamming signals for random periods and turns to sleep for the rest of the time.
This type of jamming attack allows the jammer to save more energy compared to a constant jamming attack.
However, it is less effective in its destructiveness compared to constant jamming attack.
Periodic jamming attacks are a variant of random jamming attacks, where the jammer sends periodic pulses of jamming signals.
In \cite{bayraktaroglu2013performance}, the authors investigated the impact of random and periodic jamming attacks on Wi-Fi networks.
Their experimental results showed that the random and periodic jamming attacks' impact became more significant as the duty-cycle of jamming signal increases. 
The experimental results in \cite{bayraktaroglu2013performance} also showed that, for a given network throughput degradation and jamming pulse width, the periodic jamming attack consumes less energy than the random jamming attack.
It is noteworthy that, compared to the random jamming attack, periodic jamming attack bears a higher probability of being detected as it follows a predictable transmission pattern.

\noindent \textbf{Frequency Sweeping Jamming Attacks:}
As discussed earlier, there are multiple channels available for Wi-Fi communications on ISM bands.
For a low-cost jammer, it is constrained by its hardware circuit (e.g., very high ADC sampling rate and broadband power amplifier) in order to attack a large number of channels simultaneously.
Frequency-sweeping jamming attacks were proposed to get around of this constraint, such that a jammer can quickly switch (e.g., in the range of $10~\mu$s) to different channels.
In \cite{bandaru2014investigating}, Bandaru analyzed Wi-Fi networks' performance under frequency-sweeping jamming attacks on $2.4$~GHz, where there are only $3$~non-overlapping $20$~MHz channels.
The preliminary results in \cite{bandaru2014investigating} showed that the sweeping-jammer could decrease the total Wi-Fi network throughput by more than $65 \%$.

\subsubsection{WiFi-Specific Jamming Attacks}
While the above jamming attacks are generic and can apply to any type of wireless network, the following jamming attacks are dedicated to the PHY signal processing and MAC protocols of Wi-Fi networks.

\noindent \textbf{Jamming Attacks on Timing Synchronization:}
As shown in Fig.~\ref{fig:wifi_transceiver}, timing synchronization is a critical component of the Wi-Fi receiver to decode the data packet. 
%
%
Various jamming attacks have been proposed to thwart the signal timing acquisition and disrupt the start-of-packet detection procedure, such as false preamble attack, preamble nulling attack, and preamble warping attack  \cite{la2012jamming, shahriar2014phy,la2016physical}. 
These attacks were sophisticatedly designed to thwart the timing synchronization process at a Wi-Fi receiver.
False preamble attack \cite{la2012jamming,la2016physical}, also known as preamble spoofing, is a simple method devised to falsely manipulate timing synchronization output injecting the same preamble signal as that in legitimate Wi-Fi packets. 
%
By doing so, a Wi-Fi receiver will not be capable of decoding the desired data packet as it will fail in the correlation peak detection. 
Preamble nulling attack \cite{la2012jamming,la2016physical} is another form of timing synchronization attacks.
In this attack, the jammer attempts to nullify the received preamble energy at the Wi-Fi receiver by sending an inverse version of the preamble sequence in the time domain.
Preamble nulling attack, however, requires perfect knowledge of the network timing, so it is hard to be realized in real Wi-Fi networks.
Moreover, preamble nulling attack may have considerable error since the channels are random and unknown at the jammer.
Preamble warping attack  \cite{la2012jamming,la2016physical} designed to disable the STF-based auto-correlation synchronization at a Wi-Fi receiver by transmitting the jamming signal on the subcarriers where STF should have zero data.

\noindent \textbf{Jamming Attacks on Frequency Synchronization:\!}
For a \mbox{Wi-Fi} receiver, carrier frequency offset may cause subcarriers to deviate from mutual orthogonality, resulting in inter-channel interference (ICI) and SNR degradation.
Moreover, carrier frequency offset may introduce an undesired phase deviation for modulated symbols, thereby degrading symbol demodulation performance.
In \cite{shahriar2013performance}, Shahriar et al. argued that, under off-tone jamming attacks, the orthogonality of subcarriers in an OFDM system would be destroyed.
This idea has been used in \cite{zhao2019orthogonality}, where the jammer takes down 802.11ax communications by using $20$--$25$\% of the entire bandwidth to send an unaligned jamming signal. 
In Wi-Fi communications, frequency offset in Wi-Fi communications is estimated by correlating the received preamble signal in the time domain.
Then, the preamble attacks proposed for thwarting timing synchronization can also be used to destroy the frequency offset correction functionalities.
In \cite{la2013phase}, two attacks have been proposed to malfunction the frequency synchronization correction: 
\textit{preamble phase warping attack} and \textit{differential scrambling attack}.
In the preamble phase warping attack, the jammer sends a frequency shifted version of the preamble, causing an error in frequency offset estimation at the Wi-Fi receiver.
Differential scrambling attack targets the coarse frequency correction in Fig.~\ref{fig:wifi_transceiver}, where STF is used to estimate the carrier frequency.
The jammer transmits interfering signals across the subcarriers used in STF, aiming to distort the periodicity pattern of the received preamble required for frequency offset estimation.

\noindent \textbf{Jamming Attacks on Channel Estimation:}
As shown in Fig.~\ref{fig:wifi_transceiver}, channel estimation and channel equalization are essential modules for a Wi-Fi receiver.
Any malfunction in their operations is likely to result in a false frame decoding output.
A Wi-Fi receiver uses the received frequency-domain preamble sequence to estimate the channel frequency response of each subcarrier.
A natural method to attack channel estimation and channel equalization modules is to interfere with the preamble signal.
Per \cite{la2012jamming,clancy2011efficient}, the preamble nulling attack can also be used to reduce the channel estimation process's accuracy.
The simulation results in \cite{clancy2011efficient} showed that, while preamble nulling attacks are highly efficient in terms of active jamming time and power, they are incredibly significant to degrade network performance.
However, it would be hard to implement preamble nulling attacks in real-world scenarios due to the timing and frequency mismatches between the jammer and the legitimate target device.
The impact of synchronization mismatches on preamble nulling attacks has been studied in \cite{shahriar2012performance}.
In \cite{sodagari2012efficient} and \cite{sodagari2015singularity}, Sodagari et al. proposed the singularity of jamming attacks in MIMO-OFDM communication networks such as 802.11n/ac, LTE, and WiMAX, intending to minimize the rank of estimated channel matrix on each subcarrier at the receiver.
Nevertheless, the proposed attack strategies require the global channel state information (CSI) to be available at the jammer to design the jamming signal.

\begin{figure}
	\centering
	\includegraphics[width=3.5in]{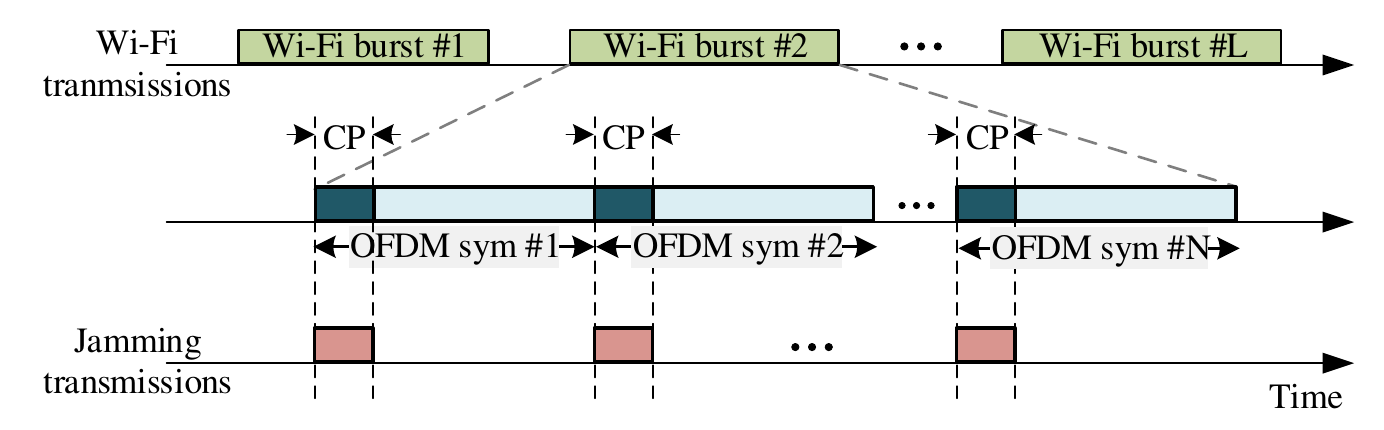}
	\caption{Illustration of a jamming attack targeting on OFDM symbol's cyclic prefix (CP) \cite{scott2011effects}.
	}
	\label{fig:cp_attack}
\end{figure}

\noindent \textbf{Jamming Attacks on Cyclic Prefix (CP):}
%
Since most wireless communication systems employ OFDM modulation at the physical layer and every OFDM symbol has a CP, jamming attacks on OFDM symbols' CP have attracted many research efforts.
In \cite{scott2011effects}, Scott et al. introduced a CP jamming attack, where a jammer targets the CP samples of each transmitted OFDM symbol, as shown in Fig.~\ref{fig:cp_attack}.
The authors showed that the CP jamming attack is an effective and efficient approach to break down any OFDM communications such as \mbox{Wi-Fi}.
The CP corruption can easily lead to a false output of linear channel equalizers (e.g., ZF and MMSE).
Moreover, the authors also showed that the CP jamming attack saves more than $80\%$ energy compared to constant jamming attacks to pull down Wi-Fi transmissions.
However, jamming attack on CP is challenging to implement as it requires jammer to have a precise estimation of the network transmission timing \cite{shahriar2014phy}.

\begin{table*}[]
\centering
\caption{A summary of existing jamming attacks in Wi-Fi networks.}
\begin{tabular}{|l|l|l|l|l|}
\hline
\multicolumn{1}{|c|}{\textbf{Attacks}}                                                         & \multicolumn{1}{c|}{\textbf{Ref.}}                                                                                                                                                  & \multicolumn{1}{c|}{\textbf{Mechanism}}                                                                                                                           & \multicolumn{1}{c|}{\textbf{Strngths}}                                                                     & \multicolumn{1}{c|}{\textbf{Weaknesses}}                                                                           \\ \hline
\multirow{5}{*}{\begin{tabular}[c]{@{}l@{}}Generic jamming attacks\end{tabular}}             & \begin{tabular}[c]{@{}l@{}}\cite{grover2014jamming,pelechrinis2010denial}, \\ \cite{karhima2004ieee,jun2007bit,vanhoef2014advanced}\end{tabular}                                            & Constant jamming attack                                                                                                                                           & Highly effective                                                                                           & Energy inefficient                                                                                                 \\ \cline{2-5} 
                                                                                               & \begin{tabular}[c]{@{}l@{}}\cite{cai2013joint, prasad2011jamming, yan2014mimo,yan2016jamming}, \\ \cite{schulz2017massive, bayraktaroglu2013performance,vanhoef2014advanced}\end{tabular} & Reactive jamming attack                                                                                                                                           & \begin{tabular}[c]{@{}l@{}}Highly effective\\ Energy efficient\end{tabular}                                & Hardware constraints                                                                                               \\ \cline{2-5} 
                                                                                               & \cite{broustis2009fiji,gvozdenovic2020truncate}                                                                                                                                    & Deceptive jamming attack                                                                                                                                          & Energy efficient                                                                                           & Less effective                                                                                                     \\ \cline{2-5} 
                                                                                               & \cite{bayraktaroglu2013performance}                                                                                                                                                & Random and periodic jamming attack                                                                                                                                & Energy efficient                                                                                           & Less effective                                                                                                     \\ \cline{2-5} 
                                                                                               & \cite{bandaru2014investigating}                                                                                                                                                    & Frequency sweeping jamming attack                                                                                                                                 & Highly effective                                                                                           & Energy inefficient                                                                                                 \\ \hline
\begin{tabular}[c]{@{}l@{}}Timing synchronization \\ attacks\end{tabular}                      & \cite{la2012jamming, shahriar2014phy}                                                                                                                                              & \begin{tabular}[c]{@{}l@{}}Preamble jamming attack\\ False preamble timing attack\\ Preamble nulling attack\end{tabular}                                          & \begin{tabular}[c]{@{}l@{}}High Effective\\ Energy-efficient\\ High stealthy\end{tabular}                  & \begin{tabular}[c]{@{}l@{}}Hard to implement\\ Tight timing synchronization required\end{tabular}                  \\ \hline
\multirow{3}{*}{\begin{tabular}[c]{@{}l@{}}Frequency synchronization \\ attacks\end{tabular}}  & \cite{shahriar2013performance,zhao2019orthogonality}                                                                                                                               & Asynchronous off-tone jamming attack                                                                                                                              & \multirow{3}{*}{\begin{tabular}[c]{@{}l@{}}Energy-efficient\\ High stealthy\end{tabular}}                  & \multirow{3}{*}{Less effective}                                                                                    \\ \cline{2-3}
                                                                                               & \multirow{2}{*}{\cite{la2013phase}}                                                                                                                                                & \multirow{2}{*}{\begin{tabular}[c]{@{}l@{}}Phase warping attack\\ Differential scrambling attack\end{tabular}}                                                    &                                                                                                            &                                                                                                                    \\
                                                                                               &                                                                                                                                                                                     &                                                                                                                                                                   &                                                                                                            &                                                                                                                    \\ \hline
\multirow{3}{*}{\begin{tabular}[c]{@{}l@{}}Channel estimation \\ (pilot) attacks\end{tabular}} & \cite{la2012jamming,shahriar2012performance}                                                                                                                                       & Pilot jamming attack                                                                                                                                              & \multirow{3}{*}{\begin{tabular}[c]{@{}l@{}}Energy-efficient\\ High Effective\\ High stealthy\end{tabular}} & \multirow{3}{*}{\begin{tabular}[c]{@{}l@{}}Hard to implement\\ Tight timing synchronization required\end{tabular}} \\ \cline{2-3}
                                                                                               & \cite{clancy2011efficient,shahriar2012performance}                                                                                                                                 & Pilot nulling attack                                                                                                                                              &                                                                                                            &                                                                                                                    \\ \cline{2-3}
                                                                                               & \cite{sodagari2012efficient,sodagari2015singularity}                                                                                                                               & Singularity jamming attack                                                                                                                                        &                                                                                                            &                                                                                                                    \\ \hline
\begin{tabular}[c]{@{}l@{}}Cyclic prefix \\ attacks\end{tabular}                               & \cite{scott2011effects}                                                                                                                                                            & Cyclic prefix (CP) jamming attack                                                                                                                                 & \begin{tabular}[c]{@{}l@{}}Energy-efficient\\ High effective\\ High stealthy\end{tabular}                  & \begin{tabular}[c]{@{}l@{}}Hard to implement\\ Tight timing synchronization required\end{tabular}                  \\ \hline
Beamforming attacks                                                                            & \cite{patwardhan2014jamming}                                                                                                                                                       & NDP jamming attack                                                                                                                                                & Energy-efficient                                                                                           & Applies to 802.11ac/ax and beyond                                                                                          \\ \hline
\multirow{3}{*}{\begin{tabular}[c]{@{}l@{}}MAC layer \\ jamming attacks\end{tabular}}          & \cite{thuente2006intelligent,acharya2004intelligent}                                                                                                                               & \begin{tabular}[c]{@{}l@{}}CTS corruption jamming attack\\ ACK corruption jamming attack\\ Data corruption jamming attack\\ DIFS-wait jamming attack\end{tabular} & \multirow{3}{*}{\begin{tabular}[c]{@{}l@{}}Energy-efficient\\ High Effective\\ High stealthy\end{tabular}} & \multirow{3}{*}{Tight timing synchronization required}                                                             \\ \cline{2-3}
                                                                                               & \cite{negi2005analysis}                                                                                                                                                            & Fake RTS transmissions                                                                                                                                            &                                                                                                            &                                                                                                                    \\ \cline{2-3}
                                                                                               & \cite{vanhoef2014advanced}                                                                                                                                                         & Selfish jamming attack                                                                                                                                            &                                                                                                            &                                                                                                                    \\ \hline
\begin{tabular}[c]{@{}l@{}}Rate adaption \\ algorithm attacks\end{tabular}                     & \cite{noubir2011robustness,orakcal2012rate,orakcal2014jamming}                                                                                                                     & \begin{tabular}[c]{@{}l@{}}Keeping the network throughout \\ below a threshold\end{tabular}                                                                       & \begin{tabular}[c]{@{}l@{}}Energy-efficient\\ High stealthy\end{tabular}                                   & Less effective                                                                                                     \\ \hline
\end{tabular}
\label{tab:wlanjam}
\end{table*}

\noindent \textbf{Jamming Attacks on MU-MIMO Beamforming:}
Given the asymmetry of antenna configurations at an AP and its serving client devices in Wi-Fi networks, recent Wi-Fi technologies (e.g., IEEE 802.11ac and IEEE 802.11ax) support multi-user MIMO (MU-MIMO) transmissions in their downlink, where a multi-antenna AP can simultaneously serve multiple single-antenna (or multi-antenna) users using beamforming technique \cite{7797535}.
To design beamforming precoders (a.k.a. beamforming matrix), a Wi-Fi AP requires to obtain an estimation of the channels between its antennas and all serving users.
Per IEEE 802.11ac standard, the channel estimation procedure in VHT Wi-Fi communications is specified by the following three steps: 
First, the AP broadcasts a sounding packet to the users. 
Second, each user estimates its channel using the received sounding packet.
Third, each user reports its channel estimation results to the AP.

\begin{figure}
	\centering
	\includegraphics[width=3.5in]{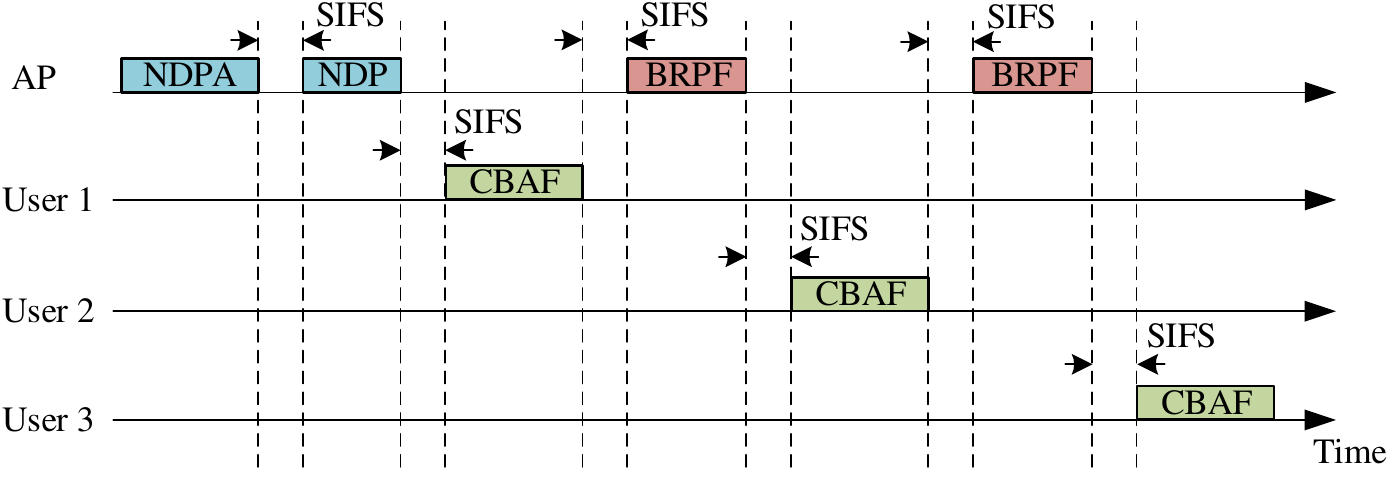}
	\caption{The beamforming sounding protocol in 802.11 VHT Wi-Fi networks.}
	\label{fig:bf_pro}
\end{figure}

Fig.~\ref{fig:bf_pro} shows the beamforming sounding protocol in VHT Wi-Fi networks.
The AP issues a null data packet announcement (NDPA) in order to reserve the channel for channel sounding and beamforming processes.
Following the NDPA signaling, the AP broadcasts a null data packet (NDP) as the sounding packet.
The users use the preamble transmitted within the NDP to estimate the channel frequency response on each subcarrier.
Then, the Givens Rotations technique is generally used to decrease the channel report overhead, where a series of angles are sent back to the AP as the compressed beamforming action frame (CBAF), rather than the original estimated channel matrices.
The AP uses beamforming report poll frame (BRPF) to manage the report transmissions among users.

In \cite{patwardhan2014jamming}, Patwardhan et al. studied the VHT Wi-Fi beamforming vulnerabilities.
They have built a prototype of a radio jammer using a USRP-based testbed that jams the NDP transmissions such that the users will no longer be able to estimate their channels and then report false CBAFs.
Their experimental results showed that, in the presence of the NDP jamming attack, less than $7\%$ of packets could be successfully beamformed in MU-MIMO transmission.

\noindent \textbf{Jamming Attacks on MAC Protocols:}
A series of MAC-layer jamming attacks, also called intelligent jamming attacks, have been proposed in \cite{thuente2006intelligent,acharya2004intelligent}, aiming to degrade Wi-Fi communications' performance.
The main focus of intelligent jamming attacks is on corrupting the control packets such as CTS and ACK packets used by Wi-Fi MAC protocols.
For CTS attack, the jammer listens to the RTS packet transmitted by an active node, waits a SIFS time slot from the end of RTS, and jams the CTS packet. 
Failing to decode the CTS packet can simply stop data communication.
A similar idea was proposed to attack ACK packet transmissions.
As the transmitter cannot receive the ACK packet, it retransmits the data packet.
Retransmission continues until the TCP limit is reached or an abort is issued to the application. 
An intelligent jamming attack can also target the data packet where the jammer senses the RTS and CTS and sends the jamming signal following a SIFS time slot.

Per \cite{thuente2006intelligent,acharya2004intelligent}, DIFS wait jamming is another form of MAC-layer attack, in which the jammer continuously monitors the channel traffic and sends a short pulse jamming signal when it senses the channel idle for a DIFS period, aiming to cause an interference for the next transmission.
%
%
Also, per \cite{negi2005analysis}, MAC-layer jamming attacks can be designed to keep the medium busy, preventing other nodes from accessing the channel by sending fake RTS packet to reserve the channel for the longest possible duration.
In \cite{vanhoef2014advanced}, Vanhoef et al. implemented a selfish jamming attack in Wi-Fi networks using a cheap commercial Wi-Fi dongle.
The dongle's firmware was particularly modified to disable the backoff mechanism and shrink the SIFS time window to implement the attack.

\noindent \textbf{Algorithm Attacks on Rate Adaptation:}
In Wi-Fi networks, rate adaptation algorithms (RAAs) were mainly designed to make a proper modulation and coding scheme (MCS) selection for data modulation.
RRAs can be considered as a defense mechanism to overcome lossy channels in the presence of low-power interference and jamming signals.
However, the pattern designed for RRAs can be targeted by jammer to degrade the network throughput below a certain threshold.
RAAs change the transmission MCS based on the statistical information of the successful and failed decoded packets.
The Automatic Rate Fallback (ARF) \cite{biaz2008rate}, SampleRate \cite{he2011theoretic}, and ONOE \cite{he2012performance} are the main RAAs using in commercial Wi-Fi devices.

In \cite{noubir2011robustness}, Noubir et al. investigated the RAAs' vulnerabilities against periodic jamming attacks.
In \cite{orakcal2012rate} and \cite{orakcal2014jamming}, Orakcal et al. evaluated the performance of ARF and SampleRate RAAs under reactive jamming attacks. 
The simulation results showed that, in order to keep the throughput below a certain threshold in Wi-Fi point-to-point communications, higher RoJ is required in ARF RAA compared to the SampleRate RAA, where the RoJ is defined as the ratio of the number of jammed packets to the total number of transmitted packets.
This reveals that SampleRate RAA is more vulnerable to jamming attacks.

\subsubsection{A Summary of Jamming Attacks}

Table~\ref{tab:wlanjam} summarizes existing jamming attacks in WLANs. 
We hope such a table will facilitate the audience's reading and offer a high-level picture of different jamming attacks.

\subsection{Anti-Jamming Techniques}
\label{sec:wifi_anti}
In this subsection, we review existing anti-jamming countermeasures proposed to eliminate or alleviate the impacts of jamming threats in WLANs.
In what follows, we categorize the existing anti-jamming techniques into the following classes:
channel hopping, MIMO-based jamming mitigation, coding protection, rate adaptation, and power control. 
We note that, given the destructiveness of jamming attacks and the complex nature of WLANs, there are no generic solutions that can tackle all types of jamming attacks.

\noindent
\subsubsection{Channel Hopping Techniques}
Channel hopping is a low-complexity technique to improve the reliability of wireless communications under intentional or unintentional interference.
Channel hopping has already been implemented in Bluetooth communications to enhance its reliability against undesired interfering signals and jamming attacks. 
In \cite{navda2007using}, Navda et al. proposed to use channel hopping to protect Wi-Fi networks from jamming attacks.
They implemented a channel hopping scheme for Wi-Fi networks in a real-world environment.
The reactive jamming attack can decrease Wi-Fi network throughput by $80$\% based on their experimental results.
It was also shown that, by using the channel hopping technique, $60$\% Wi-Fi network throughput could be achieved in the presence of reactive jamming attacks when compared to the case without jamming attack.
In \cite{jeung2011adaptive}, Jeung et al. used two concepts of window dwelling and a deception mechanism to secure WLANs against reactive jamming attacks. 
The window dwelling refers to adjusting the Wi-Fi packets' transmission time based on the jammer's capability. 
Their proposed deception mechanism leverages an adaptive channel hopping mechanism in which the jammer is cheated to attack inactive channels.

\subsubsection{Spectrum Spreading Technique}

Spectrum spreading is a classical wireless technique that has been used in several real-world wireless systems such as 3G cellular, ZigBee, and 802.11b. 
It is well known that it is resilient to narrowband interference and narrowband jamming attack. 
802.11b employs DSSS to enhance link reliability against undesired interference and jamming attacks.
It uses an 11-bit Barker sequence for $1$~Mbps and $2$~Mbps data rates, and an 8-bit complementary code keying (CCK) for $5.5$~Mbps and $11$~Mbps data rates.
In \cite{karhima2004ieee}, Karishma et al. evaluated the resiliency of DSSS in 802.11b networks against broadband, constant jamming attacks through simulation and experiments.
Their simulation results show that the packet error rate hits $100\%$ when SJR $ < -3$~dB for $1$~Mbps data rate, when SJR $< 0$~dB for $2$~Mbps data rate, when SJR $< 2$~dB for $5.5$~Mbps data rate, and when SJR $< 5$~dB for $11$~Mbps data rate.
Their experimental results show that an 802.11b Wi-Fi receiver fails to decode its received packets when received SJR~$ < -7$~dB for $1$~Mbps data rate, when SJR~$< -4$~dB for $2$~Mbps data rate, when SJR~$< -1$~dB for $5.5$~Mbps data rate, and when SJR $< 2$~dB for $11$~Mbps data rate.
In addition, \cite{harjula2011performance} evaluated the performance of $11$~Mbps 802.11b DSSS communications under periodic and frequency sweeping jamming attacks. 
The results show that 802.11b is more resilient against periodic and frequency sweeping jamming attacks compared to OFDM 802.11g.
%

\subsubsection{MIMO-based Jamming Mitigation Techniques}
Recently, MIMO-based jamming mitigation techniques emerge as a promising approach to salvage wireless communications in the face of jamming attacks. 
In \cite{yan2014mimo,yan2016jamming}, Yan et al. proposed a jamming-resilient wireless communication scheme using MIMO technology to cope with the reactive jamming attacks in OFDM-based Wi-Fi networks.
The proposed anti-jamming scheme employs a MIMO-based interference mitigation technique to decode the data packets in the face of jamming signal by projecting the mixed received signals into the subspace orthogonal to the subspace spanned by jamming signals. 
The projected signal can be decoded using existing channel equalizers such as zero-forcing technique.
However, this anti-jamming technique requires the knowledge of channel state information of both the desired user and jammer.
Conventionally, a user's channel can be estimated in this case because the reactive jammer starts transmitting jamming signals in the aftermath of detecting the preamble of a legitimate packet.
Therefore, the user's received preamble signal is not jammed.
Moreover, it is shown in \cite{yan2014mimo} that the complete knowledge of the jamming channel is not necessary, and the jammer's channel ratio (i.e., jammer's signal direction) suffices.
Based on this observation, the authors further proposed inserting known pilots in the frame and using the estimated user's channel to extract the jammer's channel ratio.
In \cite{shen2014mcr}, a similar idea called multi-channel ratio (MCR) decoding was proposed for MIMO communications to defend against constant jamming attacks.
In the proposed MCR scheme, the jammer's channel ratio is first estimated by the received signals at each antenna when the legitimate transmitter stays silent.
The jammer's channel ratio and the preamble in the transmitted frame are then used to estimate the projected channel component, which are later deployed to decode the desired signal.

While it is not easy to estimate channel in the presence of an unknown jamming signal, research efforts have been invested in circumventing this challenge.
In \cite{zeng2017enabling}, Zeng et al. proposed a practical anti-jamming solution for wireless MIMO networks to enable legitimate communications in the presence of multiple high-power and broadband radio jamming attacks. 
They evaluated their proposed scheme using real-world implementations in a Wi-Fi network.
Their scheme benefits from two fundamental techniques: A jamming-resilient synchronization module and a blind jamming mitigation equalizer.
The proposed blind jamming mitigation module is a low-complex linear spatial filter capable of mitigating the jamming signals from unknown jammers and recovering the desired signals from legitimate users.  
Unlike the existing jamming mitigation algorithms that rely on the availability of accurate jamming channel ratio, the algorithm does not need any channel information for jamming mitigation and signal recovery.
Besides, a jamming-resilient synchronization algorithm was also crafted to carry out packet time and frequency recovery in the presence of a strong jamming signal.
The proposed synchronization algorithm consists of three steps.
First, it alleviates the received time-domain signal using a spatial projection-based filter.
Second, the conventional synchronization techniques were deployed to estimate the start of frame and carrier frequency offset.
Third, the received frames by each antenna were synchronized using the estimated frequency offset. 
The proposed scheme was validated and evaluated in a real-world implementation using GNURadio-USRP2. 
It was shown that the receiver could successfully decode the desired Wi-Fi signal in the presence of $20$~dB stronger than the signals of interest.

\begin{table*}[]
\caption{{A summary of anti-jamming techniques for WLANs.}}
\centering
\begin{tabular}{|l|l|l|l|}
\hline
\multicolumn{1}{|c|}{\textbf{Anti-jamming technique}}                                                    & \multicolumn{1}{c|}{\textbf{Ref.}}                              & \multicolumn{1}{c|}{\textbf{Mechanism}}                                                                                          & \multicolumn{1}{c|}{\textbf{Application Scenario}}                                                                     \\ \hline
\multirow{2}{*}{\begin{tabular}[c]{@{}l@{}}Channel hopping \\ techniques\end{tabular}}                   & \cite{navda2007using}                                          & Channel hopping scheme for Wi-Fi networks                                                                                        & All jamming attacks on a channel                                                                                       \\ \cline{2-4} 
                                                                                                         & \cite{jeung2011adaptive}                                       & Window dwelling and adaptive channel hopping                                                                                     & Reactive jamming attack on a channel                                                                                   \\ \hline
\multirow{2}{*}{DSSS techniques}                                                                         & \multirow{2}{*}{\cite{karhima2004ieee,harjula2011performance}} & \multirow{2}{*}{802.11b performance evaluation}                                                                                  & \multirow{2}{*}{\begin{tabular}[c]{@{}l@{}}Constant, periodic, and \\ frequency-sweeping jamming attacks\end{tabular}} \\
                                                                                                         &                                                                 &                                                                                                                                  &                                                                                                                        \\ \hline
\multirow{3}{*}{\begin{tabular}[c]{@{}l@{}}MIMO-based \\ techniques\end{tabular}}                        & \cite{yan2014mimo,yan2016jamming}                              & \begin{tabular}[c]{@{}l@{}}Mixed received signals projection onto the subspace \\ orthogonal to the jamming signal.\end{tabular} & Reactive jamming attack                                                                                                \\ \cline{2-4} 
                                                                                                         & \cite{shen2014mcr}                                             & Multi-channel ratio (MCR) decoding                                                                                              & Constant jamming attack                                                                                                \\ \cline{2-4} 
                                                                                                         & \cite{zeng2017enabling}                                        & \begin{tabular}[c]{@{}l@{}}Blind jamming mitigation and jamming-resilient\\ synchronization\end{tabular}                        & Constant jamming attack                                                                                                \\ \hline
Coding techniques                                                                                        & \cite{noubir2003low,lin2005link}                               & LDPC and Reed-Solomon code schemes' analysis                                                                                     & Low-power random jamming attack                                                                                        \\ \hline
\multirow{4}{*}{\begin{tabular}[c]{@{}l@{}}Rate adaptation and \\ power control techniques\end{tabular}} & \cite{pelechrinis2009ares,pelechrinis2011measurement}          & Rate adaptation and power control mechanism evaluation                                                                           & Random and periodic jamming attacks                                                                                    \\ \cline{2-4} 
                                                                                                         & \cite{orakcal2012jamming}                                      & Randomized rate adaptation algorithm                                                                                           & Reactive jamming attacks                                                                                               \\ \cline{2-4} 
                                                                                                         & \cite{broustis2009fiji}                                        & Packet fragmentation                                                                                                            & Low-power random jamming attack                                                                                        \\ \cline{2-4} 
                                                                                                         & \cite{garcia2010detecting}                                     & Cell breathing and load balancing concepts                                                                                      & Low-power constant jamming attack                                                                                      \\ \hline
Detection mechanisms                                                                                     & \cite{punal2014machine}                                        & Multi-factor learning-based algorithm                                                                                            & Constant and reactive jamming attacks                                                                                  \\ \hline
\end{tabular}
\label{tab:wlanantijam}
\end{table*}

\subsubsection{Coding Techniques}
Channel coding techniques are originally designed to improve the communication reliability in unreliable channels.
In \cite{noubir2003low} and \cite{lin2005link}, the performance of low-density parity codes (LDPC) and Reed-Solomon codes were analyzed for different packet sizes under noise (pulse) jamming attacks with low duty cycle.
It was shown that, for long size packets (e.g., a few thousand bits), LDPC coding scheme is a suitable choice as it can achieve throughput close to its theoretical Shannon limit while bearing a low decoding complexity.

\subsubsection{Rate Adaptation and Power Control Techniques}
Rate adaptation and power control mechanisms are proposed to combat jamming attacks, provided that wireless devices have sufficient power supply and the jamming signal's power is limited.
In \cite{noubir2011robustness}, a series of rate adaptation algorithms (RAAs) were proposed to provide reliable and efficient communication scheme for Wi-Fi networks.
Based on channel conditions, RAAs set a data rate such that the network can achieve the highest possible throughput.
Despite the differences among existing RAAs, all RAAs trace the rate of successful packet transmissions and may increase or decrease the data rate accordingly. 
A power control mechanism is another technique that can be used to improve wireless communication performance over poor quality links caused by interference and jamming signals.
However, the power control mechanisms are highly subjected to the limit of power budget available at the transmitter side.
Clearly, rate adaptation and power control techniques will not work in the presence of high power constant jamming attacks.

In \cite{pelechrinis2009ares,pelechrinis2011measurement}, Pelechrinis et al. studied the performance of these two techniques (rate adaptation and power control) in jamming mitigation for legacy Wi-Fi communications via real-world experiments.
%
It was shown that the rate adaptation mechanism is generally effective in lossy channels where the desired signal is corrupted by low-power interference and jamming signals.
When low transmission data rates are adopted, the jamming signal can be alleviated by increasing the transmit power.
Nevertheless, power control is ineffective in jamming mitigation at high data rates.
In \cite{orakcal2012jamming}, a randomized RAA was proposed to enhance rate adaptation capability against jamming attacks.
The jammer attack was designed to keep the network throughput under a certain threshold, as explained earlier in RAA attacks.
The main idea of this scheme lies in an unpredictable rate selection mechanism.
When a packet is successfully transmitted, the algorithm randomly switches to another data rate with a uniform distribution.
The proposed scheme shows higher reliability against this class of attacks.
The results in \cite{orakcal2012jamming} show that a jammer aiming to pull down network throughput below 1~Mbps will need to transmit a periodic jamming signal with $3\times$ more energy in order to achieve the same performance when legacy ARF algorithm applies.

In \cite{broustis2009fiji}, an alternative approach was proposed for RAAs to cope with low-power jamming attacks using packet fragmentation.
Although the smaller-sized packet transmissions induce more considerable overhead to the network, it can improve communications reliability under periodic and noise jamming attacks by reducing each packet's probability of being jammed.
In \cite{garcia2010detecting}, Garcia et al. borrowed the concept of cell breathing in cellular networks and deployed it in dense WLANs for jamming mitigation purposes.
Here, cell breathing refers to the dynamic power control for adjusting an AP's transmission range.
That is, an AP decreases its transmission range when bearing a high load and increases its transmission range when bearing a light load. 
Meanwhile, load balancing was proposed as a complementary technique to cell breathing.
For a WLAN with cell breathing capability, the jamming attack can be treated as a case with a high load imposed on target APs \cite{garcia2010detecting}.


\subsubsection{Jamming Detection Mechanisms}

In \cite{punal2014machine}, Pu{\~n}al et al. proposed a learning-based jamming detection scheme for Wi-Fi communications.
The authors used the parameters of noise power, the time ratio of channel being busy, the time interval between two frames, the peak-to-peak signal strength, and the packet delivery ratio as the training dataset, and used the random forest algorithm for classification. 
The performance of the proposed scheme was evaluated under constant and reactive jamming attacks. 
The simulation results show that the proposed scheme could detect the presence of jammer with $98.4\%$ accuracy for constant jamming and with $94.3\%$ accuracy for reactive jamming.

\subsubsection{A Summary of Anti-Jamming Techniques}
Table~\ref{tab:wlanantijam} summarizes existing anti-jamming techniques designed for WLANs.

\section{Jamming and Anti-Jamming Attacks in Cellular Networks}
\label{sec:cellular}

\begin{figure}
	\centering
	\includegraphics[width=3.5in]{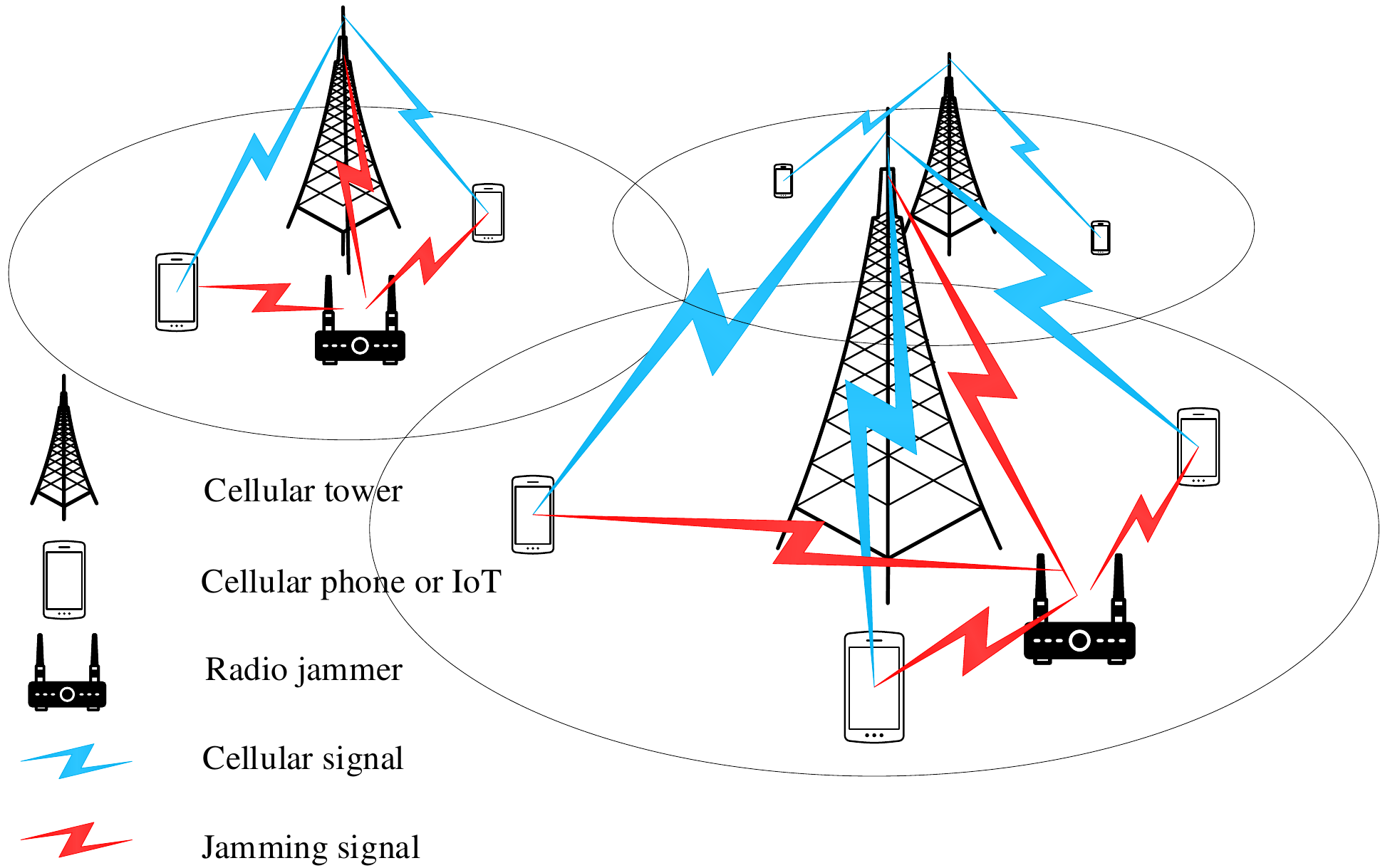}
	\caption{Jamming attack in a cellular network.}
	\label{fig:cellular_jammer}
\end{figure}

Although cellular networks have been evolving for more than four decades, existing cellular wireless communications are still vulnerable to jamming attacks.
The vulnerability can be mainly attributed to the lack of practical yet efficient anti-jamming techniques at the wireless PHY/MAC layer that are capable of securing radio packet transmissions in the presence of jamming signals. 
The vulnerability also underscores the critical need for an in-depth understanding of jamming attacks and for more research efforts on the design of efficient anti-jamming techniques. 
In this section, we consider a cellular network under jamming attacks, as shown in Fig.~\ref{fig:cellular_jammer}.
We first provide a primer of cellular networks, focusing on long-term evolution (LTE) systems.
Then, we conduct an in-depth review on existing jamming attacks and anti-jamming strategies at the PHY/MAC layers of cellular networks.

\subsection{A Primer of Cellular Networks}
\label{subsec:lte_background}

Cellular networks have evolved from the first generation toward the fifth-generation (5G). 
While 5G is still under construction, we focus our overview on 4G LTE/LTE-advanced cellular networks. 
Generally speaking, the jamming attacks in 4G LTE networks can also apply to 5G networks as they share the same wireless technologies at the PHY/MAC layers. 
4G LTE/LTE-advanced has been widely adopted by mobile network operators to provide wide-band, high throughput, and extended coverage services for mobile devices.
Due to its success in mobile networking, the LTE framework is now known as the primary reference scheme for future cellular networks such as 5G. 
LTE supports channel bandwidth from $1.4$~MHz to $20$~MHz in licensed frequency spectrum and targets $100$~Mbps peak data rate for downlink transmissions and $50$~Mbps peak data rate for uplink transmissions.
LTE was designed to support both TDD and FDD transmission schemes for further spectrum flexibility.
It uses OFDM modulation scheme in downlink and SC-FDMA (DFTS-OFDM) in uplink transmissions.

In what follows, we will overview the PHY and MAC layers of LTE, including its downlink/uplink time-frequency resource grid, the uplink/downlink transceiver structure, and the random access procedure.

\begin{figure}
	\centering
	\includegraphics[width=3.5in]{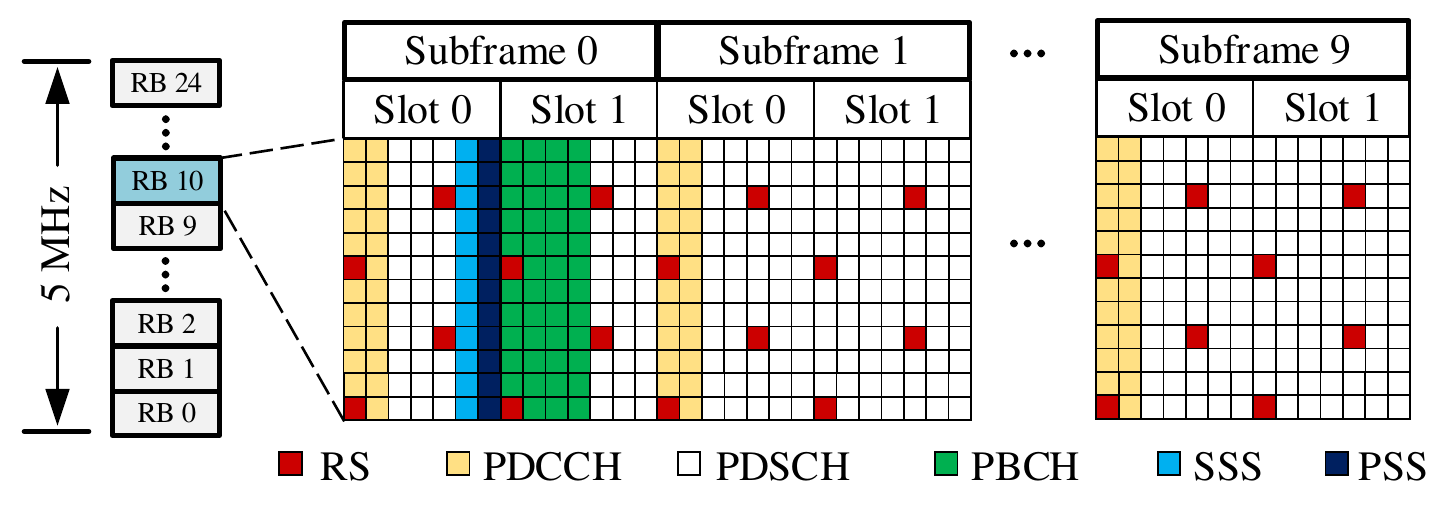}
	\caption{LTE downlink resource grid \cite{dahlman20073g}.}
	\label{fig:lte_dl}
\end{figure}

\noindent
\textbf{LTE Downlink Resource Grid:}
Fig.~\ref{fig:lte_dl} shows a portion of LTE downlink resource grid for $5$~MHz channel bandwidth.
The frame is of 10~ms time duration in the time domain and includes ten equally-sized 1~ms subframes.
Each subframe consists of two time slots, each composed of seven (or six) OFDM symbols.
In the frequency domain, a generic subcarrier spacing is set to 15~KHz.
Every 12 consecutive subcarriers (180 KHz) in one time slot are grouped as one physical resource block (PRB).
Depending on the channel bandwidth (i.e., FFT size), the frame may have $6 \leq \text{PRB} \leq 110$.
The LTE downlink resource grid shown in Fig.~\ref{fig:lte_dl} carries multiple physical channels and signals for different purposes \cite{access2009physical}, which we elaborate as follows.

\begin{itemize}
\item
\textit{Synchronization signals} consist of primary synchronization signal (PSS) and secondary synchronization signal (SSS), both of which are used for UE frame timing synchronization and cell ID detection.

\item
\textit{Reference signals} (a.k.a. pilot signals) are used for channel estimation and channel equalization.
	There are 504 predefined reference signal sequences in LTE, each corresponding to a 504 physical-layer cell identity.
	Different reference signal sequences are used in neighbor cells.
	
\item
\textit{Physical downlink shared channel (PDSCH)} is the primary physical downlink channel and is used to carry user data.
The main part of the system information, known as system information blocks (SIBs), required for random access procedure is also transmitted using PDSCH.  

\item
\textit{Physical downlink control channel (PDCCH)} is used to transmit downlink control information (DCI), which carries downlink scheduling decisions and power control commands.
	
\item
\textit{Physical broadcast channel (PBCH)} 
carries the system information called master information block (MIB), including downlink transmission's bandwidth, PHICH configuration, and the number of transmit antennas.
PBCH is always transmitted within the first $4$~OFDM symbols of the second slot in subframe $0$ and mapped to the six resource blocks (i.e., 72 subcarriers) centered around DC subcarrier. 

\end{itemize}

\noindent
\textbf{LTE Uplink Resource Grid:}
To conserve mobile devices' energy consumption, LTE employs single carrier frequency division multiple access (SC-FDMA) for uplink data transmission.
Compared to OFDM modulation, SC-FDMA modulation renders a better PAPR performance and provides a better battery lifetime for mobile devices.
Fig.~\ref{fig:lte_ul} shows the resource grid for uplink transmission.
Most of the physical transport channels and the signal processing blocks in an LTE transceiver are common for uplink and downlink.
In what follows, we focus only on their differences:

\begin{figure}
	\centering
	\includegraphics[width=3.45in]{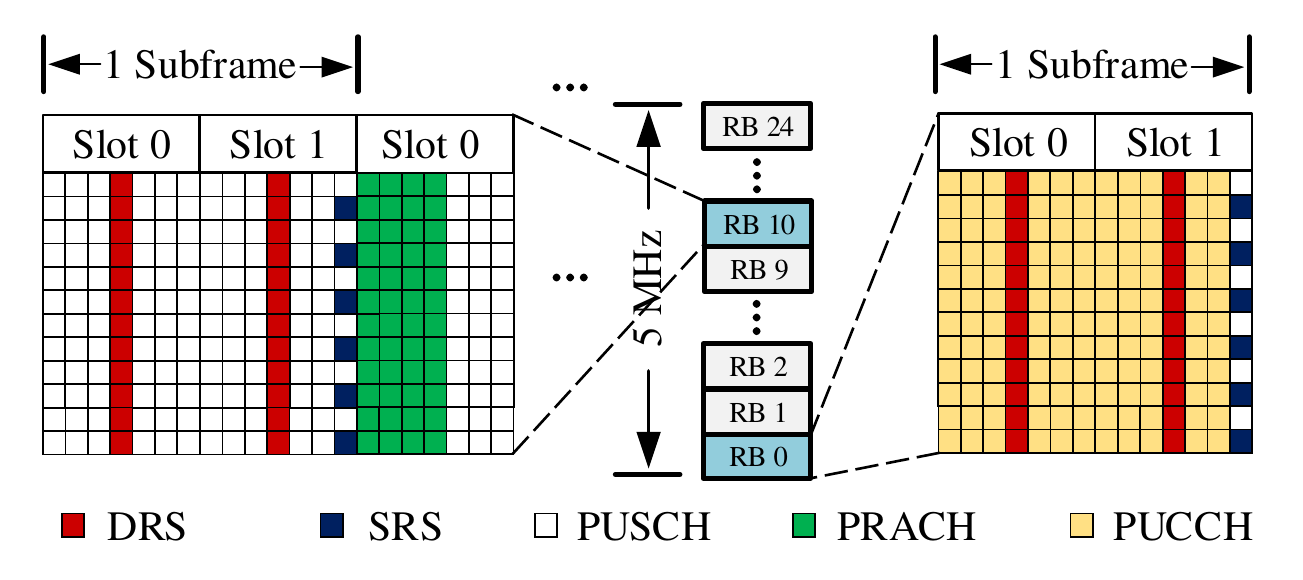}
	\caption{LTE uplink resource grid.}
	\label{fig:lte_ul}
\end{figure}

\begin{itemize}
\item
\textit{Demodulation reference signals (DRS)} are used for uplink channel estimation.

\item
\textit{Sounding reference signals (SRS)} are transmitted for the core network to estimate channel quality for different frequencies in the uplink transmissions.

\item
\textit{Physical uplink shared channel (PUSCH)} is the primary physical uplink channel used for data transmission and UE-specific higher layer information.

\item
\textit{Physical uplink control channel (PUCCH)} is used to acknowledge the downlink transmission.
It is also used to report the channel state information for downlink channel-dependent transmission and request time and frequency resources required for uplink transmission.

\item
\textit{Physical random access channel (PRACH)} is used by the UE for the initial radio link access. 

\end{itemize}


\begin{figure*}
	\centering
	\begin{subfigure}{7in}
		\centering
		\includegraphics[width=7in]{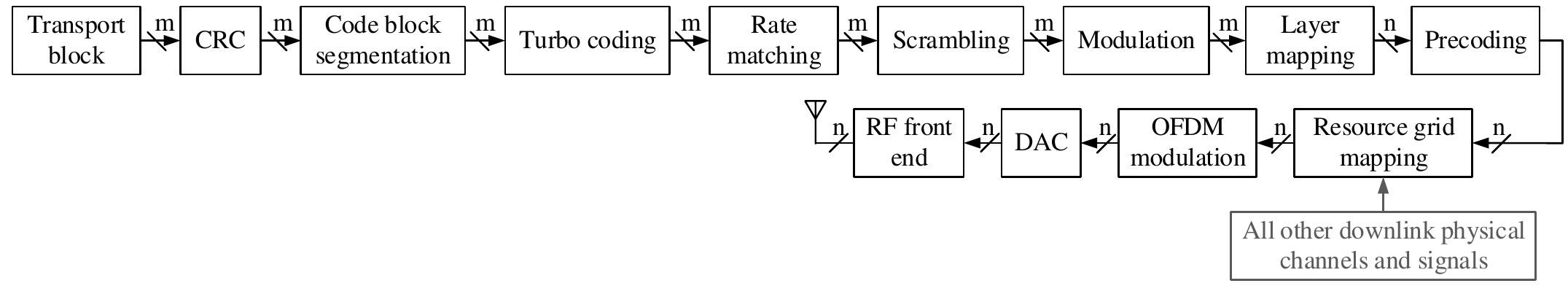}
		\caption{}
		\vspace{.2in}
	\end{subfigure}
	\begin{subfigure}{7in}
		\centering
		\includegraphics[width=7in]{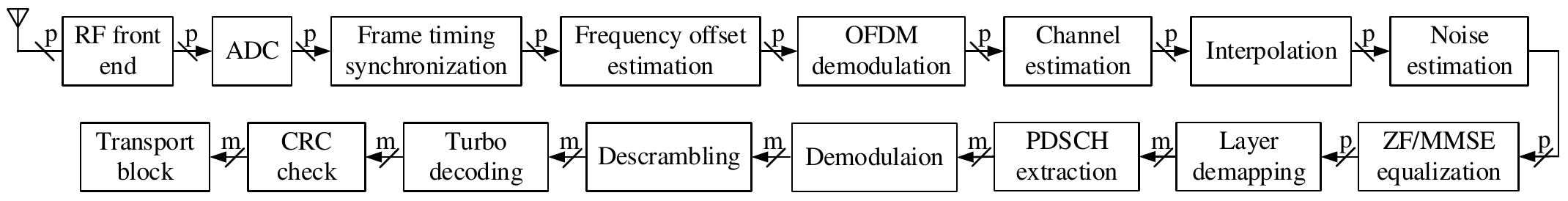}
		\caption{}
	\end{subfigure}
	\caption{(a) The schematic diagram of downlink LTE PDSCH signal processing. (b) The schematic diagram of a LTE receiver to decode PDSCH signal. ($n \le 8$, $m \le 2$, and $p \le 4$)}
	\label{fig:lte_trx}
\end{figure*}

\noindent 
\textbf{LTE Downlink Transceiver Structure:}
Fig.~\ref{fig:lte_trx}(a) shows the signal processing block diagram used for PDSCH transmission. 
PDSCH is the main downlink physical channel, which carries user data and system information.
The transport block(s) to be transmitted are delivered from the MAC layer.
Legacy LTE can support up to two transport blocks in parallel for downlink transmission.
For each transport block, the signal processing chain consists of CRC attachment, code block segmentation, channel coding, rate matching, bit-level scrambling, data modulation, antenna mapping, resource block mapping, OFDM modulation, and carrier up-conversion.
CRC is attached for error detection in received packets at the receiver side.
Code block segmentation segments an over-lengthed code block into small-size fragments matched to the given block sizes defined for Turbo encoder.
It is particularly applied when the transmitted code block exceeds 6,144~bits.
Turbo coding is used for error correction.
Rate matching is applied to select the exact number of bits required for each packet transmission.
The scrambled codewords are mapped into corresponding complex symbol blocks.
Legacy LTE can support QPSK, 16QAM, 64QAM, and 256QAM, corresponding to two, four, six, and eight bits per symbol, respectively.
The modulated codeword(s) are mapped into different predefined antenna port(s) for downlink transmission. 
Antenna port configuration can be set such that it realizes different multi-antenna schemes, including spatial multiplexing, transmit diversity, or beamforming.
Codeword(s) can be transmitted on up to eight antenna ports using spatial multiplexing.
For transmit diversity, only one codeword is mapped into two or four antenna ports.
The symbols are assigned to the corresponding PDSCH resource units, as illustrated in Fig.~\ref{fig:lte_dl}.
The other physical channels and signals are then added to the resource grids.
The resource grids on each antenna port are injected through the OFDM modulation and then up-converted to the carrier frequency and transmitted over the air.

Fig.~\ref{fig:lte_trx}(b) shows the signal processing block diagram of a receiver for downlink frame reception.
It comprises synchronization, OFDM demodulation, channel estimation and equalization, and data extraction blocks.
We elaborate on them as follows.


%
\begin{itemize}
\item
\textit{Synchronization:} 
PSS and SSS are used for detecting the start of a frame and searching for the cell ID.
Particularly, the received signal is cross-correlated with PSS to find the PSS and SSS positions.
A SSS cross-correlation is later performed to find the cell ID.
LTE uses the CP, which is appended to the end of every OFDM symbol, to estimate carrier frequency offset in the time domain.

\item
\textit{OFDM Demodulation:} 
OFDM demodulation is performed to transform the received signals from the time domain to the frequency domain, where the resource grid is constructed for further process. 

\item
\textit{Channel Estimation:} 
Channel estimation is performed by leveraging the reference signals embedded in the resource grid. 
Specifically, it is done by the following three steps: 
i) the received reference signals are extracted from the received resource grid;
ii) least square or other method is used to estimate the channel frequency responses at the reference positions in the resource grid;
and
iii) the estimated channels are interpolated using an averaging window, which can apply to the time domain, the frequency domain, or both of the resource grid.

\item
\textit{Channel equalization:} The estimated channel is used to equalize the received packet.
Minimum mean square error (MMSE) or zero-forcing (ZF) are the most two equalizers used in cellular systems. 
For the MMSE equalizer, it is also required to take the noise power into account as well. 
The noise power is estimated by calculating the variance between original and interpolated channel coefficients on the pilot subcarriers.
Compared to the MMSE equalizer, the ZF equalizer does not require the knowledge of noise power. 
It tends to offer the same performance as MMSE equalizer in the high SNR scenario.

\end{itemize}

PDSCH is extracted from the received resource grid and demodulated into bits. The received bits are then fed into the reverse process of PDSCH in order to recover the transport data.

\noindent
\textbf{LTE Uplink Transceiver Structure:}
In the uplink, the scrambled codewords are mapped into QPSK, 16QAM, or 64QAM modulated blocks.
The block symbols are divided into sets of $M$ modulated symbols, where each set is fed into DFT operation with the length of $M$.
Finally, physical channels and reference signals are mapped into resource blocks, as illustrated in Fig.~\ref{fig:lte_ul}, and fed into the OFDM modulator.
The uplink receiver structure is similar to that of PDSCH decoding in the downlink except that the frequency and symbol synchronizations to the cell and frame timing of the cell are performed in the cell search procedure.

\begin{figure}
	\centering
	\includegraphics[width=2.6in]{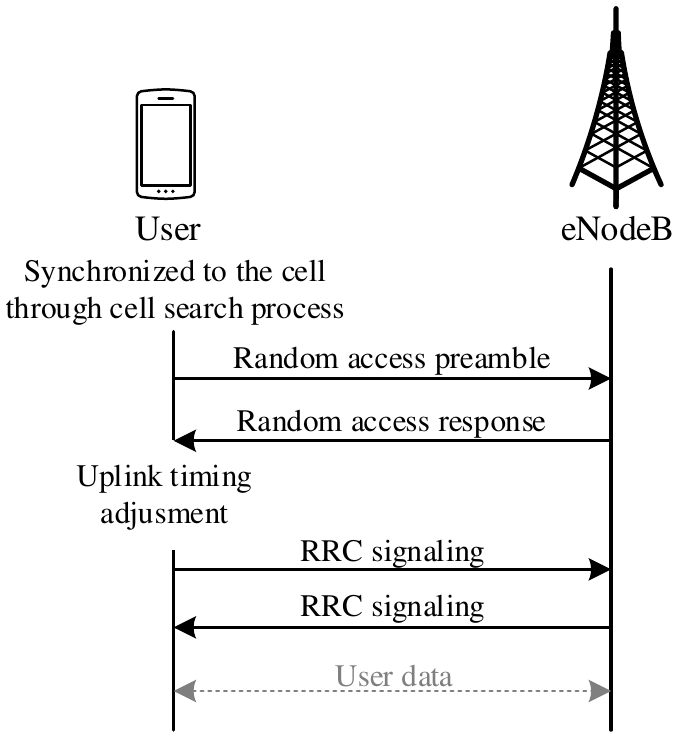}
	\caption{Illustration of random access process in cellular networks.}
	\label{fig:lte_ra}
\end{figure}

\noindent
\textbf{LTE Random Access Process:} Before an LTE user can initiate the random access procedure with the network, it has to synchronize with a cell in the network and successfully receive and decode the cell system information.
LTE users can acquire the cell's frame timing and determine the cell ID using PSS and SSS synchronization signals transmitted within the downlink resource grid, as illustrated in Fig.~\ref{fig:lte_dl}.
PDSCH and PBCH carry system information in the downlink resource grid.
Once the system information is successfully decoded, the user can communicate with the network throughout the random access procedure.
The basic steps in the random access procedure have been illustrated in
Fig.~\ref{fig:lte_ra}.
In the first step, the LTE user transmits the random access preamble (i.e., PRACH in uplink resource grid as illustrated in Fig.~\ref{fig:lte_ul}) for uplink synchronization.
It allows the eNodeB to estimate the transmission timing of the user.
In the second step, the eNodeB responds to the preamble transmission by issuing an advanced timing command to adjust uplink timing transmission based on the delay estimated in the first step.
Moreover, the uplink resources required for the user in the third step are provided in this step.
In the third step, the user transmits the mobile-terminal identity to the network using the UL-SCH resources assigned in the second step.
In the fourth and final step, the network transmits a contention-resolution message to the user using DL-SCH.

\subsection{Jamming Attacks}

With the above knowledge of LTE networks, this section dives into the review of malicious jamming attacks in cellular networks.
A jammer may attempt to disrupt cellular wireless communications either using generic jamming attacks (see Section~\ref{subsec:wifi_jamming}) or using cellular-specific jamming attack strategies.

In \cite{romero2019lte}, Romero et al. studied the performance of LTE uplink transmission under a commercial frequency-sweeping jamming attack.
The authors conducted experiments to evaluate the sensitivity of uplink reference signal when the jammer sweeps the 20~MHz channel within $T$ microseconds, where $T \in [1,~ 200]$.
The EVM of uplink demodulation reference signal was measured at the receiver and used as the performance metric. 
The experimental results show that, for a given signal-to-jamming ratio (SJR), a jammer with $T \in [20,~ 40]$ is least destructive (yielding highest EVM), while a jammer with $T \in [160,~ 200]$ is most destructive (yielding lowest EVM).

In \cite{zorn2012smart} and \cite{zorn2011power}, Zorn et al. proposed an intelligent jamming attack for WCDMA cellular networks, with the aim of forcing a victim user to switch from WCDMA to GSM service.
To do so, the jammer sends interfering signal to degrade the SNR of WCDMA cell primary common pilots.
The authors argued that, if the SNR of WCDMA CPCPCH is below a certain threshold, the user will leave the WCDMA to use GSM service.
Through experiments, the authors show that $37$~dBm jamming power is sufficient to force a user leave WCDMA to join GSM service. 


As we discussed earlier in section \ref{subsec:lte_background}, the PHY layer of LTE is made up of several physical signals and channels that carry specific information throughout the downlink and uplink resource grid to provide reliable and interference-free communications between eNodeB and users within a cell.
In what follows, we focus on cellular-specific jamming attacks that target on PHY-layer downlink/uplink signaling and channels of cellular networks.

\subsubsection{Jamming Attacks on Synchronization}
In cellular networks, synchronization signals, PSS and SSS, are critical for the cell search process, through which a user can obtain the frequency synchronization to a cell, the frame timing of the cell, and the physical identity of the cell.
An LTE user must perform a cell search before initializing the random access procedure, and the cell search is also performed for cell reselection and handover.
In FDD LTE networks, synchronization signals are transmitted within the last two OFDM symbols of the first time slot of subframe 0 and subframe 5, as illustrated in Fig.~\ref{fig:lte_dl}.
%
Once a user decodes PSS, it will find the cell's timing, the two positions of SSS, and partial information of cell identity.
Then, SSS is used to acquire frame timing (start of packet) and determine cell identity.
Once cell identity is obtained, the user becomes aware of the reference signals and their positions within the resource grid used in downlink transmission.
This information allows the user to perform channel estimation and extract the system information by decoding PBCH and PDSCH.

Clearly, the synchronization process is critical in cellular networks. 
If a user fails to detect the synchronization signals, it cannot conduct the cell search process and cannot access it.
The synchronization process is, however, vulnerable to jamming attacks. 
In \cite{krenz2015jamming}, Krenz et al. studied jamming attack on synchronization signals and PBCH.
The authors studied the LTE network under such jamming attacks where the jammer interferes with the bandwidth portions occupied by synchronization signals.
%
%
In \cite{lichtman2016lte} and \cite{lichtman2013vulnerability}, Litchman et al. investigated the vulnerabilities of LTE networks to synchronization signals spoofing attack, where the jammer intentionally sends fake PSS and SSS signals to lure the LTE user.
The authors claim that the synchronization signals spoofing attack is an efficient denial-of-service attack in cellular networks by the following two arguments.
First, synchronization signals occupy a small fraction of the downlink resource grid (e.g., $< 0.7\%$ of the total resource grid in $5$~MHz channel bandwidth) to carry information. 
That means the jammer entails very low air transmissions to jam the synchronization signals.
Second, per  \cite{access2009physical}, once the LTE user receives the fake synchronization signals, it will decode the PBCH to acquire the master information block (MIB).
If the user fails to receive MIB, it believes the cell is out of service and selects the most robust neighboring cell in the same channel, thereby degrading the network performance.

\subsubsection{Jamming Attacks on PDCCH and PUCCH}

PDCCH and PUCCH are critical control channels in cellular networks, and their vulnerability to jamming attacks have been investigated. 
In \cite{aziz2015resilience}, Aziz et al. considered PDCCH and PUCCH as potential physical channels that a smart jammer may target to interrupt since they carry critical control information on downlink and uplink resource allocations.
Before a jammer can attack the PDCCH, it requires to decode the physical control format indicator channel (PCFICH) that determines the position of the PDCCH resource elements within the downlink frame.
In \cite{kakar2014analysis}, Kakar et al. studied PCFICH jamming attacks. 
Control format indicator (CFI) is a two-digit binary data encoded into $32$ bits codeword and modulated into $16$~QPSK symbols and mapped into $16$ sparse resource elements within the downlink resource grid.
It was argued in \cite{kakar2014analysis} that the jamming attack on PCFICH is an efficient and effective jamming strategy in LTE networks because PCFICH occupies only a small fraction of the downlink resource grid and carries vital information on PDCCH resource allocations.
That means the user will no longer be able to decode the PDCCH and suffer from a denial of service when jamming the PCFICH.
In \cite{lichtman2014detection}, Lichtman et al. analyzed the physical uplink control channel (PUCCH) vulnerabilities against jamming attacks.
The authors argued that a jammer could simply attack PUCCH only by learning the LTE channel bandwidth since PUCCH is transmitted on the uplink resource grid's edge.
That means the PUCCH highly susceptible to jamming attacks as their locations within the resource grid is fix and predictable for a malicious attacker.

\subsubsection{Jamming Attacks on PDSCH and PUSCH}
PDSCH and PUSCH carry user data and upper-layer network information and dominate the major available resources in downlink and uplink transmissions, respectively.
In \cite{lichtman2016lte}, Litchman et al. investigated the vulnerability of PDSCH and PUSCH under jamming attacks.
However, the jamming attacks on PDSCH and PUSCH require the synchronization to the cell and a prior knowledge of control information and cell ID.
In \cite{girke2019towards}, Girke et al. implemented a PUSCH jamming attack using srsLTE testbed for a smart grid infrastructure and evaluated uplink throughput for different jamming gains.
The results in \cite{girke2019towards} show that the total number of received packets reduces approximately by $90\%$ under jamming signal $35$~dB stronger than PUSCH signal.

\begin{table*}[]
\centering
\caption{{A summary of existing jamming attacks for cellular networks.}}
\begin{tabular}{|l|l|l|l|l|}
\hline
\multicolumn{1}{|c|}{\textbf{Attacks}}                                                                 & \multicolumn{1}{c|}{\textbf{Ref.}}                                                                                                  & \multicolumn{1}{c|}{\textbf{Mechanism}}                                                                                         & \multicolumn{1}{c|}{\textbf{Strength}}                                                                        & \multicolumn{1}{c|}{\textbf{Weakness}}                                                     \\ \hline
\begin{tabular}[c]{@{}l@{}}Generic jamming \\ attacks\end{tabular}                                     & \cite{romero2019lte}                                                                                                               & Frequency-sweeping jamming attack                                                                                               & \begin{tabular}[c]{@{}l@{}}Easy to implement\\ High effective\end{tabular}                                    & \begin{tabular}[c]{@{}l@{}}Energy-inefficient\\ Less stealthy\end{tabular}                 \\ \hline
\begin{tabular}[c]{@{}l@{}}WCDMA CPCPCH\\ jamming attacks\end{tabular}                                 & \cite{zorn2012smart,zorn2011power}                                                                                                 & \begin{tabular}[c]{@{}l@{}}Forcing a user to leave WCDMA RAN and \\ switch to GSM by interfering the CPCPCH signal\end{tabular} & \begin{tabular}[c]{@{}l@{}}Energy-efficient\\ High stealthy\end{tabular}                                      & \begin{tabular}[c]{@{}l@{}}Cell synchronization required\\ Less effective\end{tabular}     \\ \hline
\multirow{2}{*}{\begin{tabular}[c]{@{}l@{}}Synchronization \\ signals jamming \\ attacks\end{tabular}} & \cite{krenz2015jamming}                                                                                                            & PSS and SSS corruption jamming attack                                                                                           & \multirow{2}{*}{\begin{tabular}[c]{@{}l@{}}Energy-efficient\\ High effective\\ High stealthy\end{tabular}}    & \multirow{2}{*}{Tight timing constraint}                                                   \\ \cline{2-3}
                                                                                                       & \begin{tabular}[c]{@{}l@{}}\cite{lichtman2016lte,lichtman2013vulnerability},\\ \cite{la2012jamming, shahriar2014phy}\end{tabular} & synchronization signals' spoofing attack                                                                                        &                                                                                                               &                                                                                            \\ \hline
\multirow{3}{*}{\begin{tabular}[c]{@{}l@{}}PDCCH/PUCCH \\ jamming attacks\end{tabular}}                & \cite{aziz2015resilience}                                                                                                          & Downlink control information (DCI) jamming attack                                                                               & \multirow{3}{*}{\begin{tabular}[c]{@{}l@{}}Energy-efficient\\ High effective\\ High stealthy\end{tabular}}    & \multirow{3}{*}{cell synchronization required}                                             \\ \cline{2-3}
                                                                                                       & \cite{kakar2014analysis}                                                                                                           & Control format indicator (CFI) jamming attack                                                                                   &                                                                                                               &                                                                                            \\ \cline{2-3}
                                                                                                       & \cite{lichtman2014detection}                                                                                                       & Uplink control channel attack                                                                                                   &                                                                                                               &                                                                                            \\ \hline
\begin{tabular}[c]{@{}l@{}}PDSCH/PUSCH \\ jamming attacks\end{tabular}                                 & \cite{lichtman2016lte,girke2019towards}                                                                                            & \begin{tabular}[c]{@{}l@{}}User data corruption jamming attack\\ System information block (SIB) jamming attack\end{tabular}     & High effective                                                                                                & \begin{tabular}[c]{@{}l@{}}Energy-inefficient\\ cell synchronization required\end{tabular} \\ \hline
\begin{tabular}[c]{@{}l@{}}PBCH jamming \\ attacks\end{tabular}                                        & \cite{krenz2015jamming,lichtman2016lte,lichtman2013vulnerability}                                                                  & Master information block (MIB) jamming attack                                                                                   & \begin{tabular}[c]{@{}l@{}}Energy-efficient\\ High effective\\ High stealthy\end{tabular}                     & cell synchronization required                                                              \\ \hline
\begin{tabular}[c]{@{}l@{}}PHICH jamming\\ attacks\end{tabular}                                        & \cite{lichtman2016lte}                                                                                                             & Hybrid-ARQ acknowledgement bit jamming attack                                                                                   & \begin{tabular}[c]{@{}l@{}} energy-efficient\\ High effective\\ High stealthy\end{tabular}                & cell synchronization required                                                              \\ \hline
\multirow{3}{*}{\begin{tabular}[c]{@{}l@{}}Reference signal \\ jamming attacks\end{tabular}}           & \cite{aziz2015resilience,lichtman2016lte,lichtman2013vulnerability}                                                                & Reference signal jamming attack                                                                                                 & \multirow{3}{*}{\begin{tabular}[c]{@{}l@{}}Energy-efficient\\ High Effective\\ High stealthy\end{tabular}} & \multirow{3}{*}{cell synchronization required}                                              \\ \cline{2-3}
                                                                                                       & \cite{clancy2011efficient}                                                                                                         & Reference signal nulling attack                                                                                                 &                                                                                                               &                                                                                            \\ \cline{2-3}
                                                                                                       & \cite{sodagari2012efficient,sodagari2015singularity}                                                                               & \begin{tabular}[c]{@{}l@{}}Singularity jamming attack in MIMO-OFDM \\ communications\end{tabular}                               &                                                                                                               &                                                                                            \\ \hline
\begin{tabular}[c]{@{}l@{}}Random access \\ attacks\end{tabular}                                       & \cite{aziz2014resilience,arjoune2020smart,mavoungou2016survey}                                                                     & PRACH, handover and link re-establishing jamming attack                                                                         & \begin{tabular}[c]{@{}l@{}}Energy-efficient\\ High Effective\\ High stealthy\end{tabular}                     & cell synchronization required                                                              \\ \hline
\end{tabular}
\label{tab:celljam}
\end{table*}

\subsubsection{Jamming Attacks on PBCH}
PBCH is used to carry MIB information in downlink required for LTE users to initial random access process.
MIB conveys the information on downlink cell bandwidth, PHICH configuration, and system frame number (SFN) required at the user side for packet reception and data extraction.
PBCH is transmitted only within the first subframe and mapped to the central 72 subcarriers regardless of channel bandwidth.
In \cite{lichtman2016lte} and \cite{lichtman2013vulnerability}, PBCH jamming attack was introduced as an effective and efficient adversary attack to the LTE PHY-layer communications.
This is because the PBCH carries essential system information for the user and occupies a limited portion of the resource elements (i.e., $<0.7\%$) in the downlink resource grid \cite{lichtman20185g}.
PBCH jamming attack prevents LTE users from performing their random access process.
This may lead LTE users to switch to neighbor cells \cite{lichtman20185g}.
In \cite{krenz2015jamming}, Krenz et al. evaluated the network's performance under the PBCH jamming attack using a real-world system implementation.
Their experimental results show that the LTE communications can be blocked when the jamming signal is $3$~dB stronger than the desired signal at LTE receivers. 

\subsubsection{Jamming Attacks on PHICH}
Physical hybrid automatic repeat request (ARQ) indicator channel (PHICH) is used to carry hybrid-ARQ ACKs and NACKs in response to PUSCH transmissions.
The hybrid-ARQ acknowledgment is a single bit of information (i.e.,`1' stands for ACK and `0' stands for NACK).
The bit is further repeated three times, modulated by BPSK, and spread with an orthogonal sequence to minimize the error probability of the acknowledgment detection.
PHICH occupies a small portion of the downlink resource grid (e.g., $\leq 0.3$\% in 10 MHz channel bandwidth).
In \cite{lichtman2016lte}, Lichtman et al. introduced a jamming attack on PHICH, where a jammer attacks the logic of ACK/NACK bit in order to degrade the network performance by wasting the resources for false retransmission requests.

\subsubsection{Jamming Attacks on Reference Signal}
Reference signals are transmitted within the frame for channel estimation purposes.
Downlink reference signals (pilots) are generated using a pseudo-random sequence in the frequency domain, followed by a quadrature phase shift keying (QPSK) modulation scheme.
The reference signals in the LTE downlink resource grid are transmitted on a subset of predefined subcarriers.
The estimated channel responses for the subcarriers carrying reference signals are later interpolated for the entire bandwidth.
%
%

In \cite{aziz2015resilience,lichtman2016lte,lichtman2013vulnerability}, the authors introduced potential jamming attack on cell-specific reference signals.
The LTE user under reference signals jamming attack will fail to demodulate the physical downlink channels transmitted within the downlink resource grid.
Moreover, it will lose its initial synchronization to the cell and fail to perform the handover process \cite{aziz2015resilience}.
However, a jamming attack on reference signals requires prior knowledge of the cell identity to determine the resource grid's reference signals' position.

In \cite{clancy2011efficient}, Clancy et al. proposed a reference signal nulling attack (a.k.a. pilot nulling attack).
This attack attempts to force the received energy at the pilot OFDM samples (i.e., reference signal resource elements) to zero, thereby disabling channel estimation capability at cellular networks. 
In \cite{sodagari2012efficient} and \cite{sodagari2015singularity}, Sodagari et al. studied pilot jamming attack in MIMO-OFDM communications, where their main goal is to design jamming signal so that the estimated channel matrix at a cellular receiver is rank-deficient and, as a result, the channel matrix will no longer be invertible.
This prevents the cellular receiver from correctly equalizing the received resource grid.

%
%
%

\subsubsection{Jamming Attacks on Random Access}

An LTE user can establish a radio connection to a cellular base station by using a random access procedure, provided that it correctly performs cell search as explained in Section \ref{subsec:lte_background}.
In \cite{aziz2014resilience,arjoune2020smart,mavoungou2016survey}, jamming attacks on PRACH were introduced as one of the critical PHY-layer vulnerabilities in LTE networks.
The jamming attacks on random access channels will cause DoS by preventing LTE users from connecting to the network or reestablishing the link in a handover process.
However, neither theoretical analysis nor experimental measurements were presented for PRACH jamming attacks.



\subsubsection{A Summary of Jamming Attacks}
Table~\ref{tab:celljam} presents a summary of existing jamming attacks that were delicately designed for cellular networks. 
The table also outlines the primary mechanism, strength, and weakness of each jamming attack.

\subsection{Anti-Jamming Techniques}

In the presence of security threats from existing and potential jamming attacks, researchers have been studying anti-jamming strategies to thwart jamming threats and secure cellular wireless services. 
In what follows, we survey existing anti-jamming strategies and present a table to summarize the state-of-the-art jamming defense mechanisms.

\subsubsection{MIMO-based Jamming Mitigation Techniques}
The anti-jamming strategies, such as the ones originally designed for WLANs in \cite{yan2014mimo,yan2016jamming,shen2014mcr,zeng2017enabling}, can also be applied to secure wireless communications in cellular networks.
The interference cancellation capability of MIMO communications can be enhanced when the number of antennas installed on the devices tends to be large (i.e., massive MIMO technology).
However, massive MIMO is very likely deployed in cellular base stations (eNodeB) as it requires a high power budget and a large space to accommodate massive number of antennas.
Therefore, massive MIMO techniques are exploited to mitigate jamming attacks in the cellular uplink transmissions.

In \cite{do2017jamming}, a massive MIMO jamming-resilient receiver was designed to combat constant broadband jamming signal in the uplink transmission of cellular networks. 
The basic idea behind their design is to reserve a portion of pilots in a frame so that these unused pilots can be leveraged to estimate the jammer's channel.
At the same time, a legitimate user can also estimate its desired channel in the presence of jamming signal.
This can be done using the pilots in a frame based on the large number law originated from the massive number of antennas on BS.
With the estimated channel information, a linear spatial filter can be designed at the BS to mitigate the jamming signal and recover the legitimate signal.

In \cite{vinogradova2016detection}, Vinogradova et al. proposed to use the received signal projection onto the estimated signal subspace to nullify the jamming signal.
The main challenge in this technique is to find the correct user's signal subspace.
The eigenvector corresponding to the legitimate user eigenvalue in eigenvalue decomposition of the received covariance matrix can be used for signal projection.
It is known that the user's eigenvalues can be expressed as its transmitted power when the legitimate user's and jammer's power are distinct.
In this case, the eigenvalues corresponding to the legitimate users can be selected by an exhaustive search over all the calculated eigenvalues.

\subsubsection{Spectrum Spreading Techniques}
Spectrum spreading is a classic anti-jamming technique, which has been used in 3G CDMA, WCDMA, TDS-CDMA cellular systems. 
It is particularly effective in coping with narrowband jamming signals.
At the PHY layer of WCDMA communications, the bit-streams are spread using orthogonal channelization codes called orthogonal variable spreading factor (OVSF) codes. 
The length of OVSF codes are determined by a spreading factor (SF) that varies in the range from $4$ to $512$ for downlink and from $4$ to $256$ for uplink. 
The chip rate of WCDMA communications is $3.84$~Mcps, and the bit rate can be adjusted by selecting different SF values.

In \cite{pinola2012experimental}, Pinola et al. studied the performance of WCDMA PHY layer under broadband jamming attacks by conducting experimental measurements.
The authors reported the results for different services provided by WCDMA. 
For the uplink services, the acceptable quality can be achieved when SJR $\ge -22$~dB for voice service, SJR $\ge -13$~dB for text message service, and SJR $\ge -16$~dB for data service with SF $= 32$ (i.e., $64$~kbps data rate).  
For the downlink services, the acceptable quality can be achieved when SJR $\ge -26$~dB for voice service, SJR $\ge -14$~dB for text message service, and SJR $\ge -18$~dB for data service with SF $= 32$.

\begin{table*}[]
\centering
\caption{A summary of existing anti-jamming techniques for cellular networks.}
\begin{tabular}{|l|l|l|l|}
\hline
\multicolumn{1}{|c|}{\textbf{\begin{tabular}[c]{@{}c@{}}Anti-jamming   technique\end{tabular}}} & \multicolumn{1}{c|}{\textbf{Ref.}} & \multicolumn{1}{c|}{\textbf{Mechanism}}                                                                                                                                                     & \multicolumn{1}{c|}{\textbf{Application scenario}}                            \\ \hline
\multirow{2}{*}{{\begin{tabular}[c]{@{}l@{}}MIMO-based   techniques\end{tabular}}}       & \cite{do2017jamming}               & \begin{tabular}[c]{@{}l@{}}Massive MIMO jamming-resilient receiver: jammer's channel estimation, \\ and spatial linear filter design to mitigate jamming signal.\end{tabular}               & Constant jamming attack                                                       \\ \cline{2-4} 
     & \cite{vinogradova2016detection}    & Jamming signal nullification using signal projection techniques                                                                                                                             & Constant jamming attack                                                       \\ \hline
{\begin{tabular}[c]{@{}l@{}}Spreading spectrum techniques\end{tabular}}                 & \cite{pinola2012experimental}      & \begin{tabular}[c]{@{}l@{}}Evaluating of WCDMA-based cellular RAN against jamming attacks\end{tabular}                                                                                    & Constant jamming attack                                                       \\ \hline
{Alternative eNodeB}                                                                      & \cite{makarevitch2006jamming}      & Rerouting the user's traffic via alternative eNodeBs.                                                                                                                                       & DoS jamming attacks                                                           \\ \hline
{Coding techniques}                                                                       & \cite{jover2014enhancing}          & \begin{tabular}[c]{@{}l@{}}Spread spectrum for PBCH modulation, scrambling of PRB allocation for \\  PUCCH transmission,   distributed encryption scheme for PDCCH coding.\end{tabular} & \begin{tabular}[c]{@{}l@{}}PBCH and PDCCH/PUCCH\\ jamming attack\end{tabular} \\ \hline
{\begin{tabular}[c]{@{}l@{}}Dynamic resource   allocation\end{tabular}}                  & \cite{lichtman2014detection}       & Dynamic resource allocation for PUCCH transmissions.                                                                                                                                        & PUCCH jamming attack                                                          \\ \hline
{\begin{tabular}[c]{@{}l@{}}Jamming detection  mechanisms\end{tabular}}                  & \cite{arjoune2020novel}            & Machine learning algorithms using PER, RSS, and PDR features                                                                                                                                & Constant jamming attack                                                       \\ \hline
\end{tabular}
\label{tab:cellantijam}
\end{table*}

\subsubsection{Multiple Base Stations Schemes}
In addition to MIMO-based jamming mitigation, another approach for coping with jamming attacks in cellular networks is rerouteing the traffic using alternative eNodeB. 
When current serving eNodeB is under jamming attacks and out of service, a user can re-connect to another eNodeB if available \cite{makarevitch2006jamming}.
This mechanism has already been used in LTE networks.
When a user fails to decode the MIB of its current serving eNodeB, it will search for the strongest neighboring eNodeB for a new connection \cite{access2009physical}.

\subsubsection{Coding and Scrambling Techniques}

Coding and scrambling techniques have also been used to protect wireless communications against jamming attacks in cellular networks. 
In \cite{jover2014enhancing}, Jover et al. focus on the security enhancement for the physical channels of LTE networks, including PBCH, PUCCH, and PDCCH.
These physical channels carry crucial information and must be well protected against malicious jamming adversary.
To protect the information in these physical channels, the authors proposed using spectrum spreading for PBCH modulation, scrambling the PRB allocation for PUCCH transmissions, and distributed encryption scheme for PDCCH coding.

\subsubsection{Dynamic Resource Allocation Schemes}
The static resource allocation for the physical channel such as PUCCH is regarded as the main vulnerability of the LTE framework that can be used by a malicious attacker to interrupt the legitimate transmissions, thereby causing DoS to the network.
The PUCCH is transmitted on the edges of the system bandwidth, as illustrated in Fig.~\ref{fig:lte_ul}.
It carries HARQ acknowledgments in response to PDSCH transmissions.
When the acknowledgment signals cannot be correctly decoded, the eNodeB will need to retransmit the downlink data packets, thereby imposing traffic congestion on the network and wasting the resources.
In \cite{lichtman2014detection}, Lichtman et al. studied the vulnerability of PUCCH and proposed a dynamic resource allocation scheme for PUCCH transmission by incorporating a jamming detection mechanism.
In the jamming detection process, the energy of the received PUCCH signal is continuously monitored and compared with the signal strength of other physical channels to identify any unexpected received energy behavior. 
Moreover, the number of consecutive errors on PUCCH decoding is tracked to detect the jamming attack on PUCCH.
The key idea of the proposed dynamic resource allocation scheme is to assign pseudo-random resource elements or a diverse combination of resource blocks for PUCCH transmissions in the uplink frame. 

\subsubsection{Jamming Detection Mechanisms}
 
In \cite{arjoune2020novel}, Arjoune et al. studied the performance of existing machine learning methods in jamming detection for 5G communications.
Particularly, the authors evaluated the accuracy of neural networks, super vector machine (SVM), and random forest algorithms.
They generated the database using packet error rate, packet delivery ratio, and the received signal strength features.
The simulation results show that the random forest algorithms achieve higher accuracy ($95.7\%$) in jamming detection compared to other algorithms.

\subsubsection{A Summary of Anti-Jamming Techniques}

To facilitate the reading for the audience, we summarize existing cellular-specific anti-jamming techniques as well as their mechanisms in Table~\ref{tab:cellantijam}.

\section{Jamming and Anti-Jamming Attacks in Cognitive Radio Networks (CRNs)}
\label{sec:crn}	
\begin{figure}
	\centering
	\includegraphics[width=3.5in]{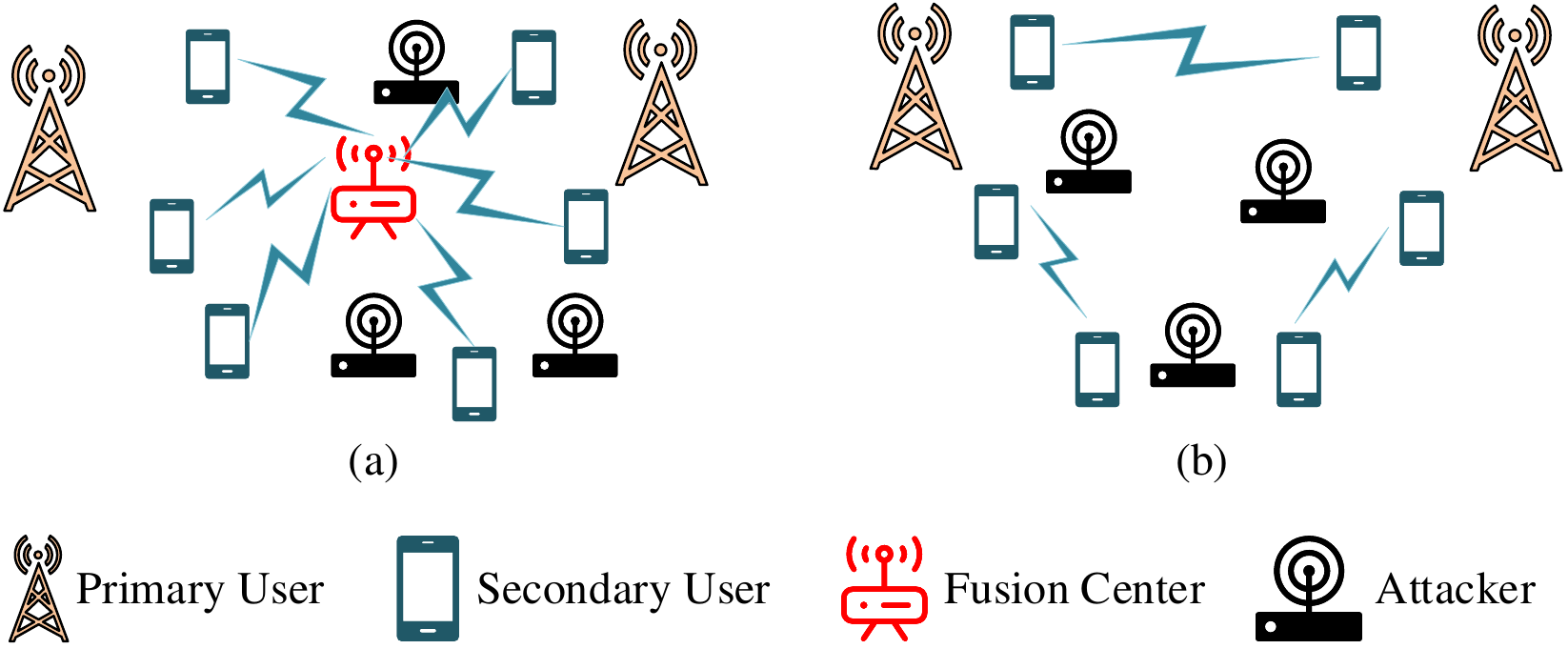}
	\caption{(a) Security attack in a cooperative sensing cognitive radio network, (b) Security attack in a non-centralized cognitive radio network.}
	\label{fig:ccrn_attack}
\end{figure}

In this section, we survey the jamming and anti-jamming strategies in cognitive radio networks. 
Following the previous section structure, we first provide an introduction of CRN and then review the jamming and anti-jamming attacks in literature. 
Fig.~\ref{fig:ccrn_attack} describes a general scheme of security attacks on centralized and distributed CRNs.

\subsection{Background}

Cognitive radio network (CRN) is a technique proposed to alleviate the spectrum shortage problem by enabling unlicensed users to coexist with incumbent users in licensed spectrum bands without inducing interference to incumbent communications.
Thus, a CRN has two types of users:
primary users (PUs), which are also known as licensed users or incumbent users, and secondary users (SUs), which are also known as unlicensed users or cognitive users.
In order to realize spectrum access for cognitive users, spectrum sensing plays a crucial role.
Cooperative spectrum sensing (CSS) refers to combining the sensing results reported from multiple secondary users (SUs) and coming up with a near-optimal carrier sensing output to improve the spectrum utilization.

The main components in CSS are signal detection, hypothesis testing, and data fusion \cite{wu2013spatial}.
Signal detection in CSS is independently carried out by active SUs to sense the medium and record the raw information on possible PU activities.
Signal detection can be implemented using matched filtering, energy detection, or feature detection techniques.
While matched filtering and feature techniques use PU's signal shaping and packet format to detect the presence of its transmissions, the energy detection method does not need to have prior knowledge of the PU's protocols, making it easy to implement.
However, the energy detection method will not be able to differentiate between the PU's transmitted signal and unknown noise and interference sources, resulting in a decrease of the sensing accuracy \cite{cabric2004implementation}.
After signal detection, each user uses hypothesis analysis to reach to a binary decision whether the medium is busy or not.
The hypothesis analysis can be performed using the Bayesian test, the Neyman-Pearson test, and the sequence probability ratio test \cite{varshney2012distributed}.
Once individual SUs have made their decisions, data fusion is performed by leveraging the reports received from the SUs to turn them into a solid output.
Data fusion can be done centralized or decentralized, corresponding to the network setting.

In centralized CCS, each SU's detection results are transmitted to a common control entity known as fusion center (FC).
The FC performs the data fusion processing and broadcasts the final results to all the SUs collaborated in the sensing phase.
Unlike centralized fusion, which relies on a fusion center to collect the data and make the fusion, for decentralized fusion in cognitive radio ad-hoc networks (CRAHN), no dedicated central entity exists to perform data fusion, and instead, each sensor exchanges its sensing output with its neighbors and iteratively fuses the sensing outputs from its neighbors \cite{zhang2015byzantine}.

\subsection{Jamming Attacks}

In what follows, we survey the jamming attacks uniquely designed for CRNs.

\subsubsection{Primary User Emulation (PUE) Attacks}
In conventional CRNs, secondary users rely on spectrum sensing to realize spectrum access opportunistically.
This implies that secondary users need to vacate the channel upon an incumbent signal is detected.
In this case, secondary users must re-sense the spectrum again for other idle channels.
This process is referred to as spectrum hand-off.
Spectrum hand-off can result in performance degradation for secondary users as their available time for spectrum access is wasted for the spectrum sensing process.
In general, this inherent concept of cognitive radio networks may be targeted by an adversary attacker to prevent secondary users access the channel.
The primary user emulation attack refers to the scenario where the attacker mimics the primary user signal's interface (transmits fake primary user signal), so the secondary users treat the attacker as the primary user and vacate the channel \cite{das2013primary}.

An attacker may launch the PUE attack as a selfish attack or a malicious attack:

\begin{itemize}
\item
\textit{Selfish attack:}
In \cite{jo2013selfish}, Jo et al. studied a selfish PUE attack in which the attacker utilizes a PUE attack to prevent legitimate secondary users from accessing the spectrum while it can exclusively acquire the spectrum resources.
This means that a non-legitimate secondary user launches a PUE attack to maximize its utilization from the spectrum channels.

\item
\textit{Malicious Attack:}
The main objective of the malicious PUE attack is to cause denial of service to the secondary networks \cite{anand2008analytical}.
In \cite{chen2011cooperative}, Chen et al. investigated the malicious PUE attack in centralized cooperative spectrum sensing CRNs.
In \cite{chen2009modeling}, the authors proposed an advanced malicious PUE attack design in which the attacker first estimates the primary user transmit power and design the incumbent signal such that the legitimate secondary users receive the fake signal with the same power as the received primary user signal.
The proposed design allows the attacker to stay undetectable.
\end{itemize}

\begin{table*}[]
\caption{{A summary of jamming attacks in CRNs.}}
\centering
\begin{tabular}{|l|l|l|}
\hline
\multicolumn{1}{|c|}{\textbf{Attacks}}                                           & \multicolumn{1}{c|}{\textbf{Ref.}}                                                         & \multicolumn{1}{c|}{\textbf{Description}}                                                                                                                               \\ \hline
\multirow{3}{*}{PUE attacks}                                                     & \cite{jo2013selfish}                                                                      & Used PUE attack to vacate the spectrum for its exclusive usage.                                                                                                         \\ \cline{2-3} 
& \cite{chen2011cooperative}                                                                & Used malicious PUE attack to cause DoS in centralized cooperative spectrum sensing CRNs.                                                                                \\ \cline{2-3} 
& \cite{chen2009modeling}                                                                   & Designed an advance malicious PUE attack by taking the PU's receive power into account.                                                                                 \\ \hline
\multirow{8}{*}{\begin{tabular}[c]{@{}l@{}}False-report \\ attacks\end{tabular}} & \cite{fatemieh2011using}                                                                  & \begin{tabular}[c]{@{}l@{}}Introduced selfish false-report attack where the attacker sends fake channel busy reports \\ to exclusively access the channel.\end{tabular} \\ \cline{2-3} 
& \cite{chen2016joint,zhang2014performance}                                                 & Proposed an attack to report a value opposite to its local sensing decision to degrade the FC performance.                                                              \\ \cline{2-3} 
& \cite{fatemieh2010secure}                                                                 & Designed an attack to report the channel as unused when it is busy to cause interference to the primary network.                                                        \\ \cline{2-3} 
& \cite{chen2008robust,wang2010phy,abdelhakim2011cooperative}                               & Studied false-report attack in centralized cooperative spectrum sensing.                                                                                                \\ \cline{2-3} 
& \cite{akyildiz2009crahns,li2009distributed,bazerque2009distributed,ding2012decentralized} & Studied false-report attack in decentralized cooperative spectrum sensing networks.                                                                                     \\ \cline{2-3} 
& \cite{kailkhura2013distributed}                                                           & Proposed spectrum sensing protocol-aware false report attack.                                                                                                           \\ \cline{2-3} 
& \cite{penna2011detecting}                                                                 & Proposed probabilistic false report attack.                                                                                                                             \\ \cline{2-3} 
& \cite{rawat2010collaborative}                                                             & Studied the impact of attack population on false-report attack performance.                                                                                             \\ \hline
\begin{tabular}[c]{@{}l@{}}Common control \\ channel attacks\end{tabular}        & \cite{bian2006mac}                                                                        & Evaluated the impact of jamming attack on common control channel.                                                                                                       \\ \hline
\begin{tabular}[c]{@{}l@{}}Learning-based \\ jamming attacks\end{tabular}        & \cite{erpek2018deep}                                                                      & Proposed to use users' ACK reports to train a deep learning classifier for the attack initializing.                                                                     \\ \hline
\end{tabular}
\label{tab:crn_table_jam}
\end{table*}

\begin{table*}[]
\caption{{A summary of anti-jamming techniques for CRNs.}}
\centering
\begin{tabular}{|l|l|l|}
\hline
& \multicolumn{1}{c|}{\textbf{Ref.}} & \multicolumn{1}{c|}{\textbf{Description}}                                                                                                                       \\ \hline
\multirow{6}{*}{\begin{tabular}[c]{@{}l@{}}Anti-jamming \\ techniques\end{tabular}} & \cite{li2009dogfight}             & Proposed random channel hopping against PUE attack.                                                                                                             \\ \cline{2-3} 
& \cite{chang2017jamming}           & Introduce Tri-CH, an anti-jamming channel hopping algorithm.                                                                                                    \\ \cline{2-3} 
& \cite{wu2011anti}                 & Used zero-sum game modeling and a channel hopping defense strategy.                                                                                             \\ \cline{2-3} 
& \cite{lo2012multiagent}           & \begin{tabular}[c]{@{}l@{}}Proposed JRCC game modeling; Cooperation for control channel allocation and adaptive rate learning for primary user.\end{tabular} \\ \cline{2-3} 
& \cite{wang2011anti}               & Proposed a game theoretical approach to model the CRN under jamming attack.                                                                                     \\ \cline{2-3} 
& \cite{asterjadhi2010jenna}        & Propose JENNA;  Random channel hopping and network coding for control packets sharing.                                                                          \\ \hline
\end{tabular}
\label{tab:crn_table_ant}
\end{table*}

\subsubsection{False-Report Attacks}
In cooperative spectrum sensing, secondary users send their local channel sensing results to a fusion center that analyzes the reports and makes a global decision on CRN's channel access.
It is well known that centralized cooperative spectrum sensing can improve CRN interference management by synthesizing the diverse reports from multiple secondary nodes.

False-report Attack, also known as spectrum sensing data falsification (SSDF) attack or Byzantine Attack, is one of the main MAC-layer security threats on cooperative spectrum sensing in CRNs.
False-report attack refers to the scenario where the malicious network sends misleading channel sensing reports to the fusion center to cause miss spectrum decision \cite{zhang2015byzantine}.
Per \cite{fatemieh2011using}, false-report attack is mainly conducted in the following two forms;

\begin{itemize}
\item 
\textit{Malicious Attack:}
In this class, the malicious network injects false local sensing results to degrade the performance of decision-making in the fusion center.
In \cite{chen2016joint} and \cite{zhang2014performance}, the authors considered a malicious false-report attack model in which the attacker performs channel sensing and local decision and then sends a random value with its opposite decision output distribution.
In \cite{fatemieh2010secure}, Fatemieh et al. introduced \textit{"vandalism"} attack where the malicious attackers report the channel as unused when it is busy in order to impose interference to the primary network transmissions.

\item 
\textit{Selfish Attacks:}
Selfish false-report attack refers to the scenario in which the attacker intends to vacate the spectrum for its exclusive usage by reporting fake channel busy information to the fusion center \cite{fatemieh2011using}.
\end{itemize}

False report attack is also investigated from the network setting perspective in the literature;
In \cite{chen2008robust,wang2010phy,abdelhakim2011cooperative}, the authors focused on the false-report attack in centralized cooperative spectrum sensing where a fusion center generates the final decision on available channel detection.

In \cite{akyildiz2009crahns,li2009distributed,bazerque2009distributed,ding2012decentralized}, the authors considered false-report attack in decentralized cooperative spectrum sensing networks, where the channel access decision is made through iterative information exchange among cognitive users.
Compared to centralized cooperative spectrum sensing, the authors argued that decentralized cooperative spectrum sensing is more susceptible to false-report attack as there is no common control unit to receive all the users' local channel sensing and perform the global decision.
Studies in \cite{kailkhura2013distributed} showed that the false-report attack would be more destructive if the attacker acquires a prior-knowledge of the cooperative spectrum sensing protocols and fusion techniques used in the fusion center before it starts to launch the attack.

In \cite{penna2011detecting}, Penna et al. evaluated the performance of the cooperative spectrum sensing CRNs under statistical false-report attack in which the attackers are assumed to have certain attack probability, instead of following a predefined attack strategy.
It was shown that the statistical attacks are more challenging to be detected compared to non-probabilistic false-report attacks.
%
In \cite{rawat2010collaborative}, Rawat et al. demonstrated the impact of attack population on false-report attack performance.
The authors used Kullback-Leibler divergence (KLD) as a measurement metric to validate the detection performance.
It was shown that the attack would be more destructive as the number of malicious nodes increases.

\subsubsection{Jamming Attacks on Secondary Network and Common Control Channel}
The common control channel is a medium through which the secondary nodes share their channel sensing reports.
Common control channel attack refers to the overwhelming secondary network common control channel by fake MAC control frames.
The attacker network launches the common control channel attack in order to cause denial of service to the secondary network.
In \cite{bian2006mac}, Bian et al. argued that the common control channel attack benefits from the following two features:
First, it is hard to be detected as it injects the MAC control frames identical to the secondary networks' protocols.
Second, it is an energy-efficient attack as it requires to send a small number of control packets to saturate the common control channel.

\subsubsection{Learning-based Jamming Attacks}
In \cite{erpek2018deep}, the authors proposed a deep learning-based jamming attack on cognitive radio networks.
The proposed scheme takes advantage of users' ACK reports to train a deep learning classifier to predict whether an ACK transmission would occur through the medium.
The performance of the proposed jamming attack was compared with a random jamming attack and a sensing-based reactive jamming attack.
The simulation results show that the proposed deep-learning jamming attack could degrade the network's throughput to $0.05$~packet/slot.
As a comparison, the authors also show that the network's throughput is $0.38$~packet/slot under the random jamming attack and $0.14$~packet/slot under the sensing-based jamming attack.

\subsubsection{A Summary of Existing Jamming Attacks}
Table~\ref{tab:crn_table_jam} is a summary of existing attacks that were dedicatedly designed for CRNs.

\subsection{Anti-Jamming Techniques}
Generally speaking, the anti-jamming strategies presented in Section~\ref{sec:wifi_anti} (e.g., MIMO-based jamming mitigation) can also be applied to thwarting jamming attacks in CRNs.
Table~\ref{tab:crn_table_ant} summarizes existing anti-jamming strategies uniquely designed for CRNs. 
In what follows, we elaborate on these anti-jamming strategies.

Per \cite{li2009dogfight,zhu2010stochastic,wu2011anti}, game-theoretical modeling is a well-known technique used in CRNs to express the interaction among different parties contributing to the network.
Particularly, the stochastic zero-sum game is widely adopted to model the interaction between secondary users and adversary jamming attackers since they have opposite objectives.

In \cite{li2009dogfight}, Li et al. proposed that secondary users randomly hop over the multiple channels to countermeasure PUE jamming attacks.
For the secondary network, each user needs to search the optimal tradeoff between choosing good channels and avoiding jamming signals.
In \cite{chang2017jamming}, Chang et al. introduced Tri-CH, an anti-jamming channel hopping algorithm for cognitive radio networks.
In \cite{wu2011anti}, the interaction between the secondary network and attackers is formulated as a zero-sum game, and a channel-hopping defense strategy is proposed using the Markov decision process.

In \cite{lo2012multiagent}, Lo et al. proposed JRCC, a jamming resilient control channel game that models the strategies chosen by cognitive users and an attacker under the impact of primary user activity.
JRCC uses user cooperation for control channel allocation and adaptive rate learning for primary user using the Win-or-Learn-Fast scheme.
The optimal control channel allocation strategy for secondary users is derived using multiagent reinforcement learning (MARL).

In \cite{wang2011anti}, a game-theoretical approach is picked up to model the cognitive radio users' interactions in the presence of a jamming attack.
Secondary users at each stage update their strategies by observing the channel quality and availability and the attackers' strategy from the status of jammed channels.
The strategies defined for cognitive users consider the number of channels they can reserve to transmit control and data messages and how they can switch between the different channels.
A minimax-Q learning method is used to find the optimal anti-jamming channel selection strategy for cognitive users.

In \cite{asterjadhi2010jenna}, Asterjadhi et al. propose a scheme called JENNA, jamming evasive network-coding neighbor-discovery algorithm, to secure cognitive radio networks against PUE jamming attack.
JENNA combines random channel hopping and network coding to share the control packets to neighbor users.
The proposed neighbor-discovery algorithm is fully distrusted and scalable.

\section{Jamming and Anti-Jamming Attacks in ZigBee Networks}
\label{sec:zigbee}

In this section, we survey existing jamming attacks and anti-jamming techniques in ZigBee networks. 
Following the same structure of previous sections, we first offer a primer of ZigBee communication and then explore existing jamming and anti-jamming strategies that were uniquely designed for ZigBee networks.

\subsection{A Primer of ZigBee Communication}

ZigBee is a key technology for low-power, low-data-rate, and short-range wireless communication services such as home automation, medical data collection, and industrial equipment control \cite{ieee2011ieee}.
With the proliferation of low-power IoT devices, ZigBee becomes increasingly important and emerges as a crucial component of wireless networking infrastructure in smart home and smart city environments. 
ZigBee was developed based on the IEEE 802.15.4 standard.
It operates in the unlicensed spectrum band from $2.4$ to $2.4835$~GHz worldwide, $902$ to $928$~MHz in North America and Australia, and $868$ to $868.6$~MHz in Europe.
On these unlicensed bands, sixteen $5$~MHz spaced channels are available for ZigBee communications.
At the PHY layer, ZigBee uses offset quadrature phase-shift keying (OQPSK) modulation scheme and direct-sequence spread spectrum (DSSS) coding for data transmission.
A typical data rate of ZigBee transmission is 250~kbit/s, corresponding to 2~Mchip/s.

\begin{figure}
	\centering
	\includegraphics[width=3.5in]{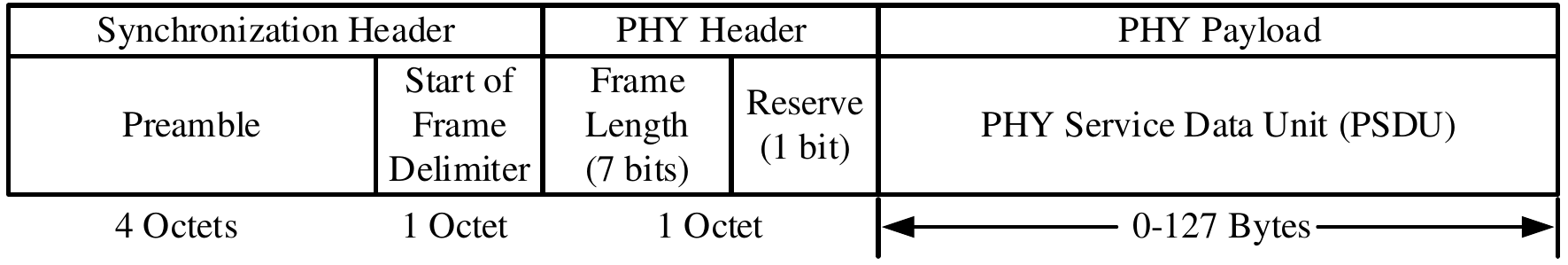}
	\caption{ZigBee frame structure.}
	\label{fig:zigbee_frame}
\end{figure}

Fig.~\ref{fig:zigbee_frame} shows the frame structure of ZigBee communication.
The frame consists of a synchronization header, PHY header, and data payload.
Synchronization header comprises a preamble and a start of frame delimiter (SFD).
The preamble field is typically used for chip and symbol timing, frame synchronization, and carrier frequency and phase synchronizations.
The preamble length is four predefined Octets (32 bits) that all are binary zeros.
The SFD field is a predefined Octet, which is used to indicate the end of SHR.
Following the SFD is the PHY header, which carries frame length information.
PHY payload carries user payload and user-specific information from the upper layers.

\begin{figure}
	\centering
	\begin{subfigure}{3.5in}
		\centering
		\includegraphics[width=3.5in]{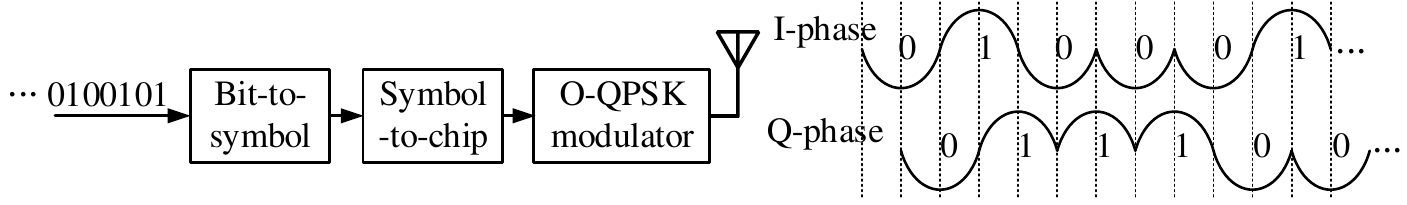}
		\caption{ZigBee transmitter structure.}
	\end{subfigure}
	\begin{subfigure}{3.5in}
		\centering
		\includegraphics[width=3.5in]{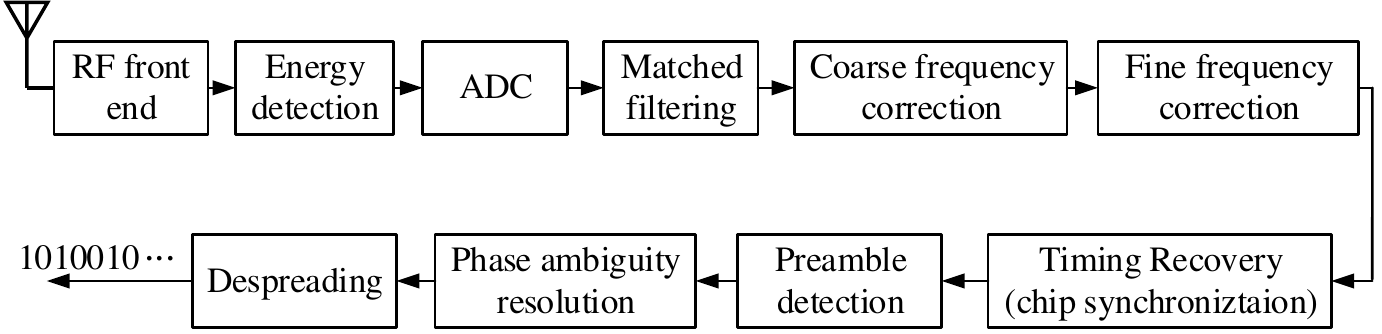}
		\caption{ZigBee receiver structure.}
	\end{subfigure}
	\caption{The signal processing block diagram of a ZigBee transceiver.}
	\label{fig:zigbee_trx}
\end{figure}

Fig.~\ref{fig:zigbee_trx} shows the block diagram of the baseband signal processing in a ZigBee transceiver.
Referring to Fig.~\ref{fig:zigbee_trx}(a), at a ZigBee transmitter, every $4$ bits of the data for transmission are mapped to a predefined $32$-chip pseudo-random noise (PN) sequence, followed by a half-sine pulse shaping process.
The chips of the PN sequence are modulated onto carrier frequency by using the OQPSK modulation scheme.
%
Referring to Fig.~\ref{fig:zigbee_trx}(b), at a ZigBee receiver, the main signal processing block chain comprises coarse and fine frequency correction, timing recovery (chip synchronization), preamble detection, phase ambiguity resolution, and despreading.
The RF front-end module first demodulates the received signal from carrier radio frequency to baseband.
The received chip sequences are passed through a matched filter to boost the received SNR.
Then, an FFT-based algorithm is typically used for coarse frequency offset estimation.
What follows is a closed-loop PLL-based algorithm for fine carrier frequency, and phase offsets correction.
Timing recovery (chip synchronization) can be performed using classic methods such as zero-crossing or Mueller-Muller error detection.
Following the timing recovery is the preamble detection, which can compensate for the phase ambiguity generated by the fine frequency compensation module.
Finally, the decoded chips are despreaded to recover the original transmitted bits.

\begin{figure}
	\centering
	\includegraphics[width=3in]{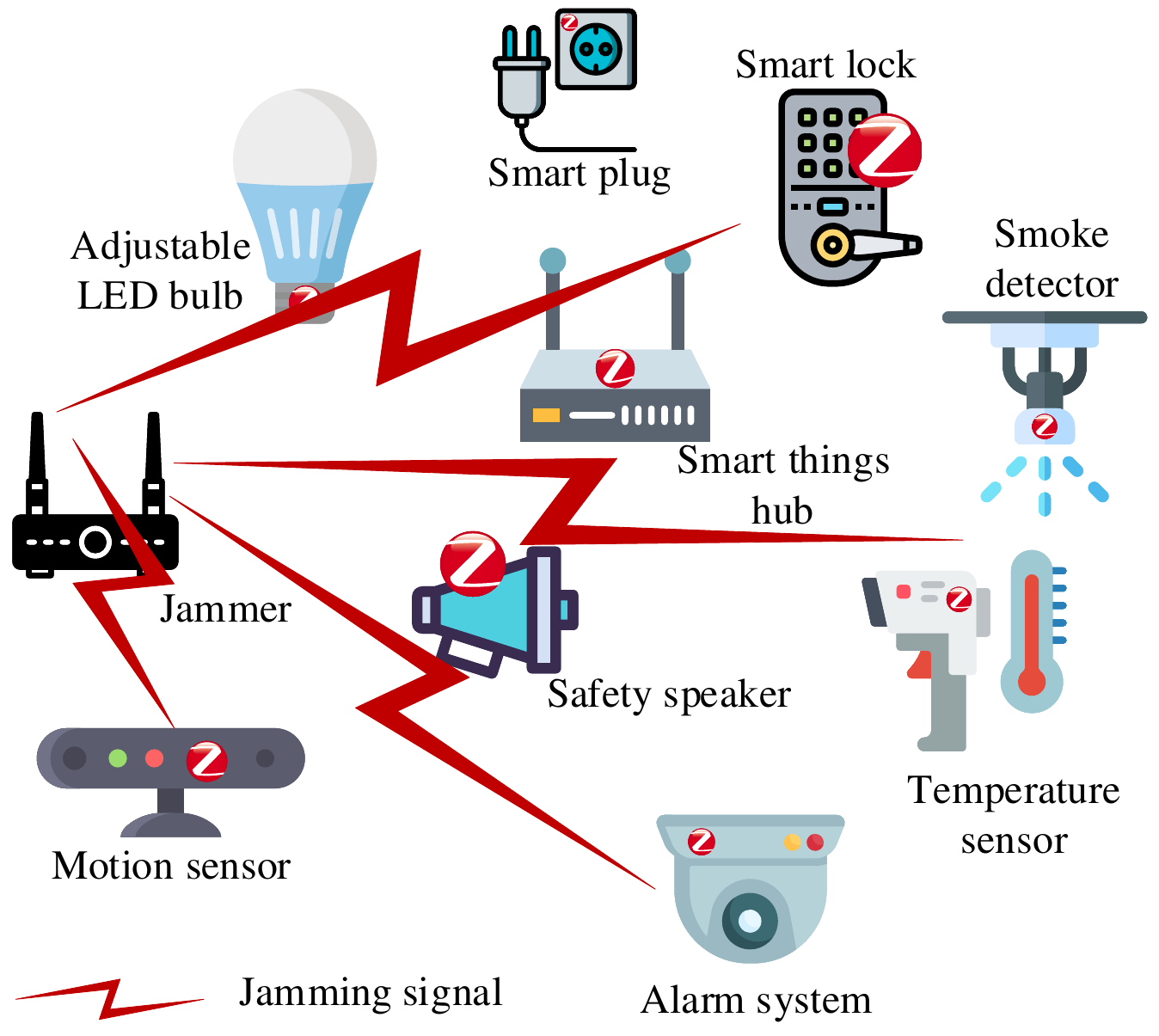}
	\caption{Illustration of jamming attacks in a ZigBee network.}
	\label{fig:zigbee_net}
\end{figure}

\begin{table*}[]
\caption{{A summary of jamming attacks and anti-jamming strategies for ZigBee networks.}} 
\centering
\begin{tabular}{|l|l|l|}
\hline
                                                                                    & \multicolumn{1}{c|}{\textbf{Ref.}}          & \multicolumn{1}{c|}{\textbf{Description}}                                                                                                                                    \\ \hline
\multirow{4}{*}{\begin{tabular}[c]{@{}l@{}}Jamming \\ attacks\end{tabular}}         & \cite{rewienski2018investigation}          & Studied the performance of constant jamming attack on ZigBee communications.                                                                                                 \\ \cline{2-3} 
                                                                                    & \cite{wilhelm2011short}                    & Implemented reactive jamming attack in ZigBee communications using real-world implementation.                                                                                \\ \cline{2-3} 
                                                                                    & \cite{cao2016ghost}                        & Designed the energy depletion attack on ZigBee networks.                                                                                                                     \\ \cline{2-3} 
                                                                                    & \cite{chi2020countering}                   & Proposed cross technology jamming attack on ZigBee communications                                                                                                            \\ \hline
\multirow{9}{*}{\begin{tabular}[c]{@{}l@{}}Anti-jamming \\ techniques\end{tabular}} & \cite{pirayesh2020securing}                & Proposed a MIMO-based receiver using machine learning to cancel the jamming and decode the desired signal.                                                                   \\ \cline{2-3} 
                                                                                    & \cite{fang2010study}                       & Evaluated conventional DSSS performance against interference and jamming signals.                                                                                            \\ \cline{2-3} 
                                                                                    & \cite{liu2010randomized}                   & Introduced randomized differential DSSS (RD-DSSS) scheme.                                                                                                                    \\ \cline{2-3} 
                                                                                    & \cite{heo2017dodge,heo2018mitigating}      & Proposed an anti-jamming scheme (Dodge-Jam) based on channel hopping and frame segmentation.                                                                                 \\ \cline{2-3} 
                                                                                    & \cite{wood2007deejam}                      & \begin{tabular}[c]{@{}l@{}}Proposed a MAC-layer anti-jamming scheme based on frame masking, channel hopping, \\ packet fragmentation, and fragment replication.\end{tabular} \\ \cline{2-3} 
                                                                                    & \cite{debruhl2011digital}                  & Designed digital filter to reject the frequency components of periodically cycling jamming attacks.                                                                          \\ \cline{2-3} 
                                                                                    & \cite{liu2012bittrickle, fang2015wireless} & Proposed to manage the reaction time of the reactive jammer and use the unjammed time slots to transmit data.                                                                \\ \cline{2-3} 
                                                                                    & \cite{spuhler2014detection}                & \begin{tabular}[c]{@{}l@{}}Proposed jamming attack detection based on extracting statistics from jamming-free \\ symbols of DSSS synchronizer.\end{tabular}                  \\ \cline{2-3} 
                                                                                    & \cite{chi2020countering}                   & Proposed a multi-staged cross-technology jamming attack detection.                                                                                                           \\ \hline
\end{tabular}
\label{tab:zigbee}
\end{table*}

\subsection{Jamming Attacks}

Since ZigBee is a wireless communication system, the generic jamming attack strategies (e.g., constant jamming, reactive jamming, deceptive jamming, random jamming, and frequency-sweeping jamming) presented in Section~\ref{subsec:wifi_jamming} can also be applied to ZigBee networks.
Here, we focus on the jamming attack strategies that were delicatedly designed for ZigBee networks, as shown in Fig.~\ref{fig:zigbee_net}.
Table~\ref{tab:zigbee} presents the existing ZigBee-specific jamming attacks in the literature.
In what follows, we elaborate on these jamming attacks.

In \cite{rewienski2018investigation}, the authors studied the impact of constant radio jamming attack on connectivity of an IEEE 802.15.4 (ZigBee) network.
They evaluated the destructiveness of jamming attack in an indoor ZigBee environment via experiments.
In their experiments, the ZigBee network was configured in a tree topology, and a commercial ZigBee module with modified firmware was used as the radio jammer.
The authors reported the number of nodes affected by a jamming attack when the jammer was located at different locations. 

In \cite{wilhelm2011short}, the authors studied reactive radio jamming attacks in ZigBee networks, focusing on improving the effectiveness and practicality of reactive jamming attacks.
The reactive jammer first detects ZigBee packets in the air by searching for the PHY header in ZigBee frames shown in Fig.~\ref{fig:zigbee_frame}, and then sends a short jamming signal to corrupt the detected ZigBee packets.
The results show that the jamming signal of more than $26~\mu$s time duration suffices to bring the packets reception rate of a ZigBee receiver down to zero.
The authors also built a prototype of the proposed jamming attack on a USRP testbed and evaluated its performance in an indoor environment. 
Their experimental results show that the prototyped jammer blocks more $96\%$ packets in all scenarios.

In \cite{cao2016ghost}, Cao et al. presented an energy depletion attack targeting on ZigBee networks.
Through sending fake packets, the proposed attack intends to keep the ZigBee receiver busy and waste its physical resources (e.g., CPU).
The authors evaluated the cost of such an attack in terms of energy consumption, time requirement, and computational cost of processing the fake messages.
It was shown that the energy depletion attack can serve as a DoS attack in ZigBee networks by depleting the airtime resource and leaving no airtime for serving legitimate users.

In \cite{chi2020countering}, Chi et al. proposed a cross-technology jamming attack on ZigBee communications, where a Wi-Fi dongle was used as the malicious jammer to interrupt ZigBee communication using Wi-Fi signal.
The firmware of Wi-Fi dongle was modified to disable its carrier sensing and set the SIFS and DIFS time windows to zero, such that the Wi-Fi dongle can continuously transmit packets.
The reasons for using Wi-Fi cross-technology to attack ZigBee communications are three-fold:
i) Wi-Fi dongle is cheap and easy to be driven as a constant jammer;
ii) a WiFi-based jammer is easy to detect as its traffic tends to be considered legitimate;
iii) Wi-Fi bandwidth ($20$~MHz) is larger than ZigBee bandwidth ($5$~MHz), making it possible to pollute several ZigBee channels at the same time.

\subsection{Anti-Jamming Techniques}

Anti-jamming strategies such as MIMO-based jamming mitigation and spectrum spreading techniques can also be applied to ZigBee communications against jamming attacks. 
Particularly, ZigBee employs DSSS at its PHY layer, which can enhance the link reliability against jamming and interference signals.
Table~\ref{tab:zigbee} presents existing anti-jamming attacks that were delicately designed for ZigBee networks. 
We elaborate on these works in the following.

As a performance baseline, Fang et al. \cite{fang2010study} studied the bit error rate (BER) performance of DSSS in ZigBee communications.
Their theoretical analysis and simulation show that, in the AWGN channel, a ZigBee receiver renders 
$\mbox{BER} = 10^{-1}$ when $\mbox{SNR} = 3~\mbox{dB}$, 
$\mbox{BER} = 10^{-2}$ when $\mbox{SNR} = 6~\mbox{dB}$,
and 
$\mbox{BER} = 10^{-3}$ when $\mbox{SNR} = 8~\mbox{dB}$.
These theoretical results provide a reference for the study of ZigBee communications in the presence of jamming attacks.

In \cite{pirayesh2020securing}, Pirayesh et al. proposed a MIMO-based jamming-resilient receiver to secure ZigBee communications against constant jamming attack. 
They employed an online learning approach for a multi-antenna ZigBee receiver to mitigate unknown jamming signal and recover ZigBee signal.
Specifically, the proposed scheme uses the preamble field of a ZigBee frame, as shown in Fig.~\ref{fig:zigbee_frame}, to train a neural network, which then is used for jamming mitigation and signal recovery.
A prototype of the proposed ZigBee receiver was built using a USRP SDR testbed.
Their experimental results show that the proposed ZigBee receiver achieves $100\%$ packet reception rate in the presence of jamming signal that is $20$~dB stronger than ZigBee signal. 
Moreover, it was shown that the proposed ZigBee receiver yields an average of $26.7$~dB jamming mitigation gain compared to commercial off-the-shelf ZigBee receivers.

In \cite{liu2010randomized}, a randomized differential DSSS (RD-DSSS) scheme was proposed to salvage ZigBee communication in the face of a reactive jamming attack.
RD-DSSS decreases the probability of being jammed using the correlation of unpredictable spreading codes.
In \cite{heo2018mitigating} and \cite{heo2017dodge}, a scheme called Dodge-Jam was proposed to defend IEEE 802.15.4 ACK frame transmission against reactive jamming attacks.
Dodge-Jam relies on two main techniques: channel hopping and frame segmentation.
Particularly, frame segmentation is done by splitting the original frame into multiple small blocks.
These small blocks are shifted in order when retransmission is required.
In this scenario, the receiver can recover the transmitted frame after a couple of retransmission.
In \cite{wood2007deejam}, a MAC-layer protocol called \mbox{DEEJAM} was proposed to reduce the impact of jamming attack in ZigBee communications.
\mbox{DEEJAM} offers four different countermeasures, namely frame masking, channel hopping, packet fragmentation, and redundant encoding, to defend against different jamming attacks.
Specifically, frame masking refers to using a confidential start of frame delimiter (SFD) symbols by the ZigBee transmitter and receiver when a jammer is designed to use the SFD detection for its transmission initialization.
Channel hopping was proposed to defend against reactive jamming attacks.
Packet fragmentation was proposed to defend against scan jamming.
Redundant encoding (e.g., fragment replication) was designed to defend against noise jamming.

In \cite{debruhl2011digital}, the impact of a periodically cycling jamming attack on ZigBee communication was studied, and a digital filter was designed to reject the frequency components of the jamming signal.
In \cite{liu2012bittrickle} and \cite{fang2015wireless}, an anti-jamming technique was proposed to defend against high-power broadband reactive jamming attacks for low data rate wireless networks such as ZigBee.
The proposed technique undertakes reactive jammers' reaction time and uses the unjammed time slots to transmit data.
In \cite{spuhler2014detection}, Spuhler et al. studied a reactive jamming attack and its detection method in ZigBee networks. 
The key idea behind their design is to extract statistics from the jamming-free symbols of the DSSS synchronizer to discern jammed packets from those lost due to bad channel conditions.  
This detection method, however, focuses only on jamming attacks without considering jamming defense mechanism.
%
In \cite{chi2020countering}, Chi et al. proposed a detection mechanism to cope with cross-technology jamming attacks.
Their proposed detection technique consists of several steps, including multi-stage channel sensing, sweeping channel, and tracking the number of consecutive failed packets. 
Once the number of failed packets exceeds a certain threshold, the ZigBee device transmits its packets even if the channel is still busy, letting the signal recovery be made at the receiver side.

\section{Jamming and Anti-Jamming Attacks in Bluetooth Networks}
\label{sec:bluetooth}

\begin{figure}
	\centering
	\includegraphics[width=3in]{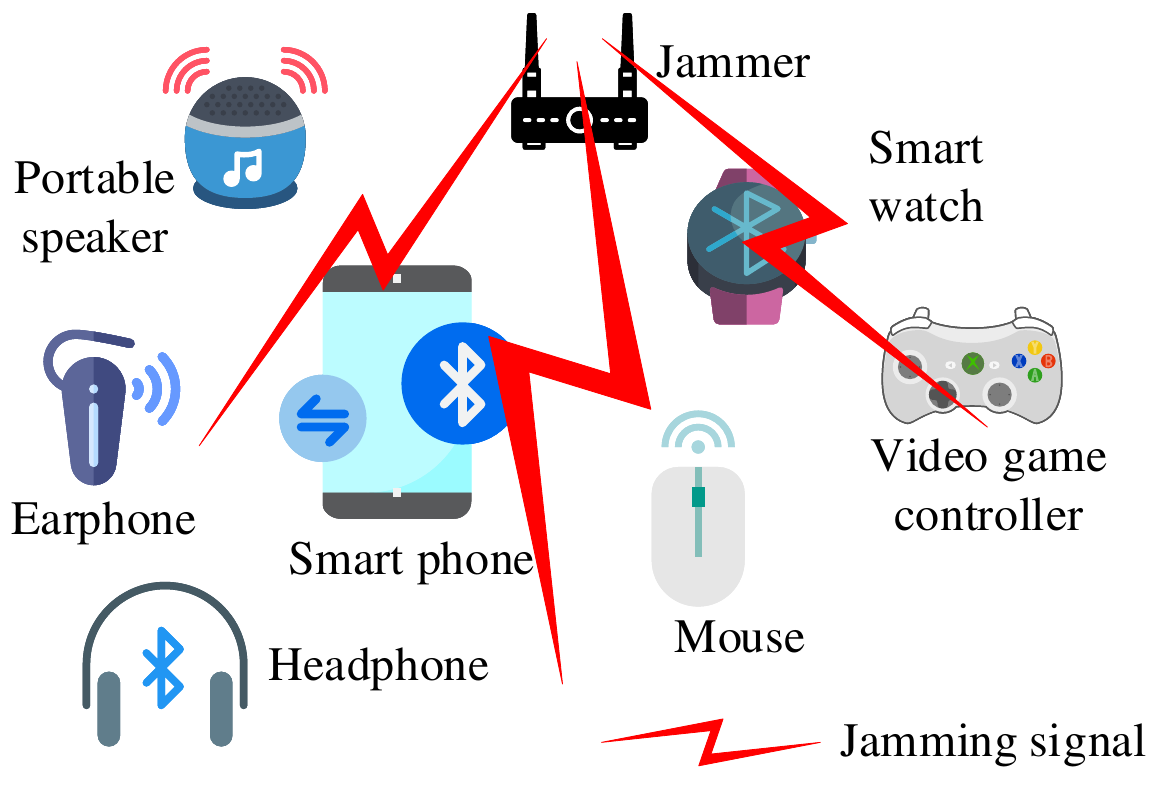}
	\caption{Jamming attack in a Bluetooth network.}
	\label{fig:bl_jamming}
\end{figure}

In this section, we survey existing jamming and anti-jamming attacks in a Bluetooth network, as shown in Fig.~\ref{fig:bl_jamming}. 
By the same token, we first offer a primer of the PHY and MAC layers of Bluetooth communication and then review the existing jamming/anti-jamming techniques that were dedicatedly designed for Bluetooth networks.

\subsection{A Primer of Bluetooth Communication}
\begin{figure}
	\centering
	\includegraphics[width=3.5in]{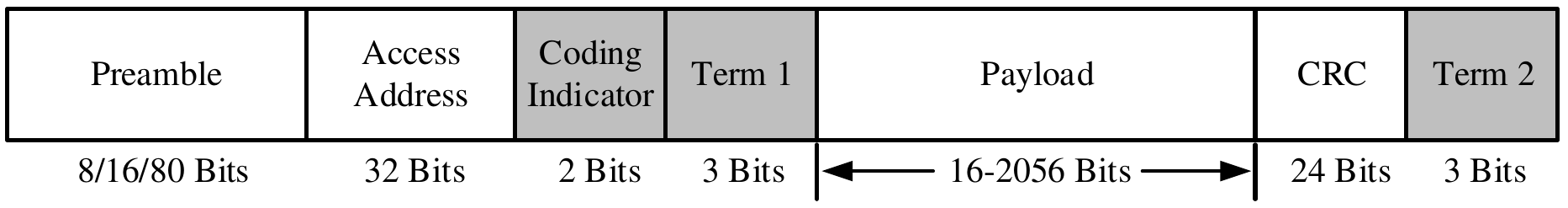}
	\caption{Bluetooth low energy frame structure.}
	\label{fig:bluetooth_frame}
\end{figure}

Bluetooth is a wireless technology that has been deployed for real-world applications for many years. 
It was initially designed for short-range device to device (D2D) communication and then evolved toward many other communication purposes in the IoT applications.
A Bluetooth network, also known as a piconet, is typically composed of one master device and up to seven slave devices. 
A Bluetooth device can serve as a master node in only one piconet, but it can be connected to multiple piconets as a slave.
In response to the low power constraints in IoT networks, a new concept of Bluetooth Low Energy (BLE) was recently introduced to offer a more energy-efficient and flexible communication scheme compared to classic Bluetooth technology.   
BLE operates in unlicensed 2.4GHz--2.4835GHz ISM frequency bands, where there are $40$ channels of $2$~MHz bandwidth available for its communications.
Unlike Wi-Fi channels, the BLE channels are not overlapping.
In its latest version (e.g., BLE v5), BLE can support $1$~Msym/s and $2$~Msym/s symbol rates, where $1$~Msym/s symbol rate is the mandatory modulation scheme while $2$~Msym/s symbol rate is optional.
BLE at 1~Msym/s modulation may use packet frames with coded or uncoded data.
Fig.~\ref{fig:bluetooth_frame} shows a general packet structure for BLE at 1~Msym/s transmission.
As can be seen, a BLE packet consists of a preamble, access address, payload, and CRC.
The coding indicator and termination fields are further transmitted within the packet for the coded frame format, as shown in Fig.~\ref{fig:bluetooth_frame}.

\begin{figure}
	\begin{subfigure}{3.5in}
		\includegraphics[width=3.5in]{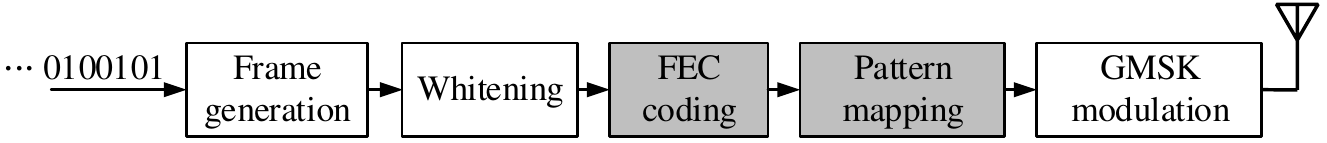}
		\caption{BLE transmitter structure.}
	\end{subfigure}
	\begin{subfigure}{3.5in}
		\includegraphics[width=3.5in]{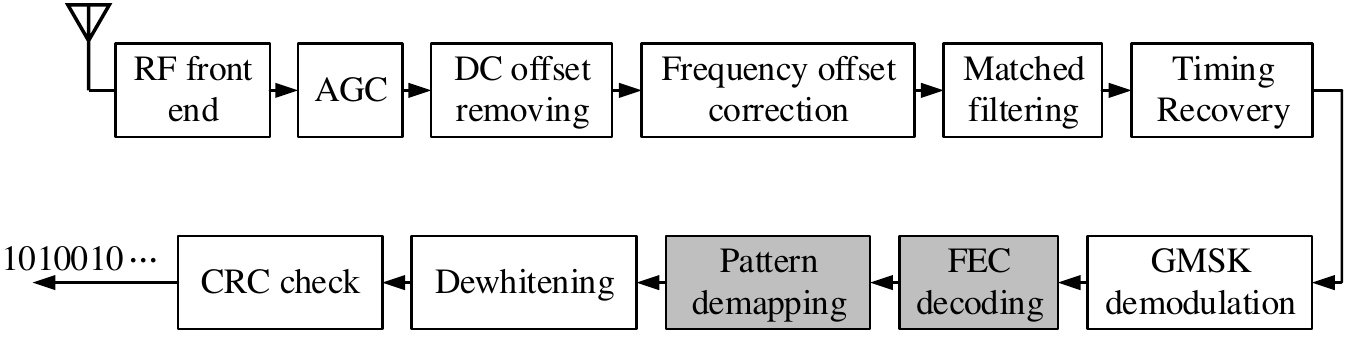}
		\caption{BLE receiver structure.}
	\end{subfigure}
	\caption{Signal processing block diagram of Bluetooth communication.}
	\label{fig:bt_transceiver}
\end{figure}

Fig.~\ref{fig:bt_transceiver} shows the block diagram of baseband signal processing in a Bluetooth device. 
At the transmitter side, the generated frame is fed into a whitening (scrambling) operation to avoid transmitting long sequences consisting of consecutive zeros or ones.    
FEC and pattern mapping are used when the coded frame format is transmitted.
The scrambled data is encoded using a $1/2$ rate binary convolutional FEC encoder.
Pattern mapping might be used to map each encoded bit to 4-bits symbol to decrease the error decoding probability.
The final bits to be sent are modulated on the carrier frequency using Gaussian minimum shift keying (GMSK).
The data rate of uncoded packet format transmission is $1$~Mbps.
The data rate of coded packet format is $500$~kbps.
After applying pattern mapping to the encoded bits, the data rate is $125$~kbps.

Referring Fig.~\ref{fig:bt_transceiver}(b), at the receiver side, the received signal power is adjusted using an AGC block.
Following AGC, the DC offset of the received signal is removed.
A coarse frequency offset correction algorithm is employed to estimate and compensate the carrier frequency mismatch between the transmitter and receiver clocks.
The received signal is later passed through a Gaussian matched filter to reduce the noise. 
In the aftermath of matched filtering, the frame timing synchronization is performed based on preamble detection.
Then, the received frame is demodulated using the GMSK demodulation block.
FEC decoding and pattern de-mapping are used when the coded packet frames are transmitted. 
Finally, the decoded bits are de-whitened and the CRC check is performed to check the correctness of the decoded bits.

FHSS is a key technology of Bluetooth communications as it enhances the reliability of Bluetooth link in the presence of intentional or unintentional interference.
With FHSS, Bluetooth packets are transmitted by rapidly hopping over a set of predefined channels, following a mutually-agreed hopping pattern.
Conventional Bluetooth has $79$ distinct and separate channels, each of $1$~MHz bandwidth.
The new standard, Bluetooth Low Energy (BLE), has $40$ distinct channels, each of $2$~MHz bandwidth.
The hopping pattern is determined by a pseudo-random spreading code, which is shared with the receiver to keep transmitter and receiver synchronized during the channel hopping.
The spreading code must be shared after a pre-shared key establishment process to secure the data communication.

\subsection{Jamming Attacks}

In the original design of Bluetooth communication systems, frequency hopping spread spectrum (FHSS) technology was adopted to avoid interference so as to improve the communication reliability.
The inherent FHSS technology provides Bluetooth systems with the capability of coping with jamming attacks to some extent.
For this reason, the study of jamming attacks in Bluetooth networks is highly overlooked, and the investigation of Bluetooth-specific jamming attacks is very limited in the literature.

However, the FHSS technology is effective only for narrowband, low-power jamming attacks; and it becomes ineffective when the bandwidth of high-power jamming signal spans over all possible Bluetooth channels, i.e., $2.4-2.4835$~GHz ISM band. 
In fact, such a jamming attack can be easily launched in practice, thanks to the advancement of SDR technology. 
In light of powerful jamming attacks, Bluetooth networks are vulnerable to generic jamming attacks (e.g., constant, reactive, and deceptive jamming attacks, see Section~\ref{subsec:wifi_jamming}) as much as other wireless communication systems.

In the context of Bluetooth networks, although many works have been done to investigate the underlying security vulnerabilities of Bluetooth applications (see, e.g.,  \cite{haataja2008man,gehrmann2004bluetooth,hager2003analysis,jakobsson2001security}), the research effort on investigating jamming attacks for Bluetooth communications remains very limited.
In \cite{koppel2013bluetooth}, K{\"o}ppel et al. built an intelligent jamming attack that synchronously tracks Bluetooth communications and sends the jamming signal.
The proposed algorithm estimates the frequency hopping sequence of Bluetooth communications by 
i) decoding the master's upper address part (UAP) and the lower address part (LAP), and 
ii) determining the communication's clock.
The jammer uses the estimated clock to synchronize itself with the Bluetooth devices.
The experimental results demonstrate that a jammer is capable of completely blocking Bluetooth communications if the jammer can obtain a correct estimate of frequency hopping sequence and clock.

\subsection{Anti-jamming Techniques}

As mentioned before, FHSS is the key technology used at the physical layer of Bluetooth communication, which has a certain capability of taming interference and jamming signals.
The pre-shared key establishment, however, is itself a challenging constraint in the presence of jamming attacks.

In Bluetooth communication, FHSS relies on a pre-shared key at a pair of devices to determine their frequency hopping pattern.
The key establishment procedure (for reaching the consent on frequency hopping pattern at transmitter and receiver) exposes Bluetooth networks to adversarial jamming attackers.
In \cite{strasser2009efficient,lee2010randomized,popper2010anti,liu2010usd}, uncoordinated FHSS schemes were proposed to secure the key establishment procedure for Bluetooth communication in the presence of jamming attacks.
These schemes employ a randomized FHSS technique, where a pair of Bluetooth devices randomly switch over multiple channels before being able to initiate the communication.
Once the two Bluetooth devices come across on the same channel, they exchange the hopping and spreading keys.
To fasten the convergence of the procedure, one approach is to let Bluetooth transmitter hop over the channels much faster than Bluetooth receiver (e.g., $20$ times faster).

In \cite{xiao2011jamming}, Xiao et al. proposed a collaborative broadcast scheme to eliminate the need for pre-shared key exchanging in an uncoordinated FHSS technique.
The authors considered a star-topology network, where a master node intends to broadcast a message to its multiple slave nodes.
The key idea behind this scheme is to broadcast messages in all possible channels and use the nodes having already acquired the message to serve as relays to forward the messages to other nodes.  
It is assumed that the jammer cannot block all available channels; otherwise, the proposed algorithm will not work.
Multiple channel selection schemes, including random, sweeping-channel, and static, were considered for relaying broadcast messages.
Packet reception rate and cooperation gain were investigated to evaluate the performance of the proposed scheme.
Experimental results confirm a significant improvement of jamming resiliency. 
Moreover, it was shown that, for a certain jamming probability, when there is a small number of slave nodes in the network (e.g., less than $50$), sweeping channel selection achieves a higher packet reception rate.
But for a large number of nodes, static channel selection outperforms other channel selection methods.


%

\section{Jamming and Anti-Jamming Attacks in LoRa Networks}
\label{sec:lora}
\subsection{A Primer of LoRaWANs}

LoRa is a low power wide area networking (LPWAN) technology.
It recently becomes popular thanks to its promising features such as open-source development, easy deployment, flexibility, security, cost effectiveness, and energy efficiency.
LoRa has been widely used in diverse IoT applications such as smart parking, smart lighting, waste management services, life span monitoring of civil structures, and air and noise pollution monitoring.
Compared to other low-power wireless technologies (e.g., ZigBee and Bluetooth), LoRa offers a large coverage ($5$km--$15$km).
LoRa is a lightweight communication system with low-complexity signal processing and low-complexity MAC protocols.
It consumes only $120$--$150$~mW power in its transmission mode.
The lifetime of a LoRa device varies in the range from $2$ to $5$ years, depending on its in-use duty cycle \cite{sundaram2019survey}.

LoRa uses chirp spread spectrum (CSS) modulation for its data transmission.
It supports a data rate from $980$~bps to $21.9$~kbps and spreading factors (SF) from $6$ to $12$.
In the CSS modulation, data is carried by frequency modulated chirp pulses.
The CSS modulation scheme appears to be resilient to both interference and Doppler shift, making it particularly suitable for long-range mobile applications.

LoRa operates in sub-GHz ISM bandwidth in North America.
On $902.3$MHz--$914.9$~MHz spectrum, $64$ channels are defined for LoRa, each of $125$~kHz bandwidth.
On $1.6$MHz spectrum, $8$ channels are defined for LoRa's uplink transmission, each of $500$kHz bandwidth.
On $923.3$MHz--$927.5$MHz spectrum, $8$ channels are defined for LoRa's downlink transmission, each of $500$kHz bandwidth.
Typically, LoRa operates in a star-network topology, where one or multiple LoRa devices are served by a central gateway connected to a network server.

\begin{figure}
	\centering
	\includegraphics[width=3.5in]{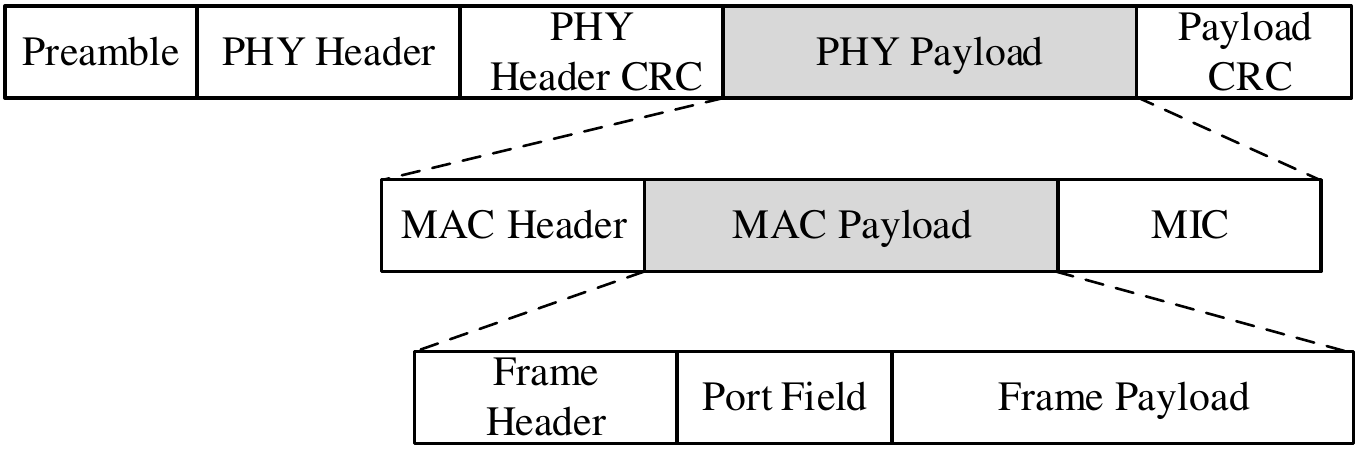}
	\caption{The structure of LoRa frame.}
	\label{fig:lora_frame}
\end{figure}

\noindent
\textbf{Packet Structure:}
Fig.~\ref{fig:lora_frame} shows the structure of a LoRa packet.
The preamble comprises a sequence of upchirps and a sequence of downchirps. 
The number of upchirps in the preamble depends on the data rate and spreading factor. 
The downchirps in the preamble is used for packet detection and clock synchronization. 

The PHY header and its CRC fields are optional and transmitted to indicate the length of PHY payload.
The PHY payload consists of MAC header, MAC payload, and message integrity code (MIC).
Depending on the selected data rate, the maximum user payload size varies in the range of $11$ to $242$ bytes.
MAC header is 1-byte data, specifying MAC message type (MType) such as uplink/downlink data, join request, and accept request.
LoRa supports both ``\textit{confirmed}'' data transmission and ``\textit{unconfirmed}'' data transmission.
The former requires a receiver to acknowledge the frame reception, while the latter requires no ACK feedback.
The MAC payload consists of a frame header (FHDR), an optional port field (FPort), and frame payload, as shown in  Fig.~\ref{fig:lora_frame}.
The frame header (FHDR) mainly carries the end-device address, and the frame payload carries the application-specific end-device data.


LoRa supports bi-directional communications.
That is, when a LoRa device sends an uplink packet, it waits for two window time slots for downlink data.
This means that the downlink transmission requires a LoRa device to send the uplink packets first.
In a LoRaWAN, the ALOHA scheme is used as the medium control protocol for LoRa devices to access the channel for their uplink transmissions.

\begin{figure}
	\begin{subfigure}{3.5in}
		\includegraphics[width=3.5in]{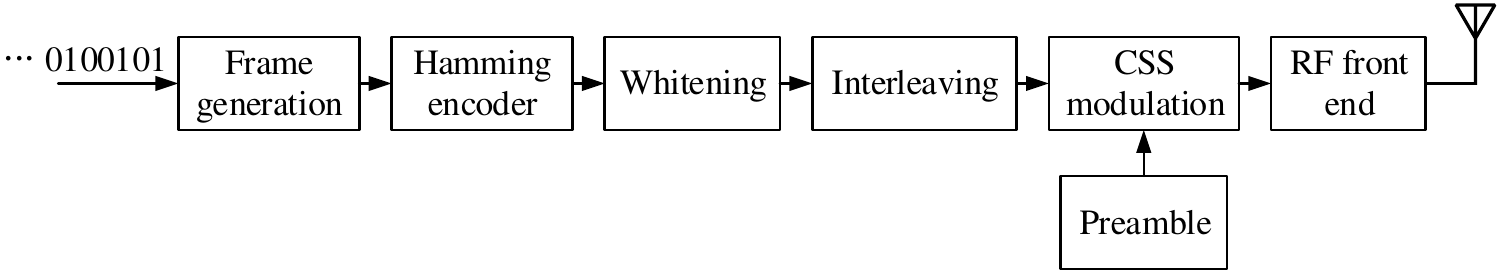}
		\caption{The LoRa transmitter structure.}
	\end{subfigure}
	\begin{subfigure}{3.5in}
		\includegraphics[width=3.5in]{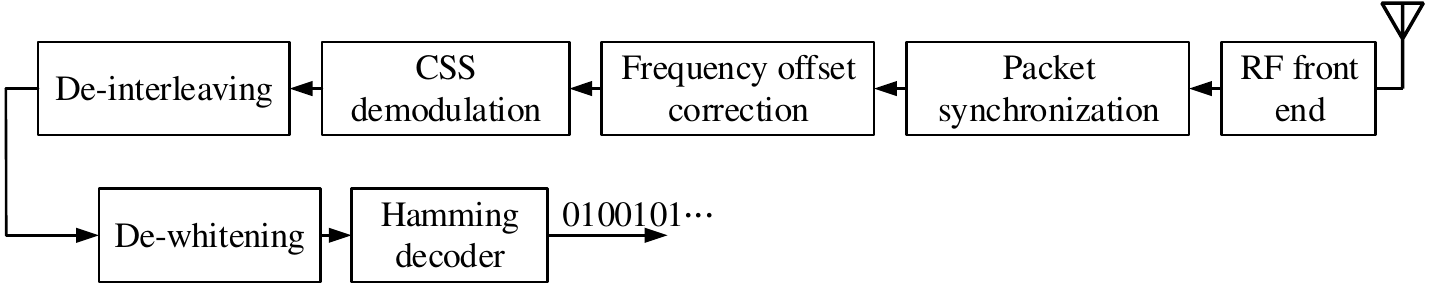}
		\caption{The LoRa receiver structure.}
	\end{subfigure}
	\caption{The schematic diagram of LoRa transceiver.}
	\label{fig:lora_transceiver}
\end{figure}

\noindent
\textbf{Transceiver Structure:}
Fig.~\ref{fig:lora_transceiver} shows the schematic diagram of a LoRa transceiver structure. 
At the transmitter side, the bits to be transmitted are first encoded using a Hamming encoder.
The encoded bits are fed into the whitening module to avoid transmitting a long sequence of consecutive zeros/ones.
Following the whitening module, the interleaving module is applied to the scrambled bits. 
The preamble sequence is attached to the processed bits, and all are modulated on the carrier frequency using the CSS modulation.
At the receiver side, the received frame is synchronized using the preamble detection followed by the carrier frequency offset correction.
The received frame is demodulated using CSS demodulation block and fed into the deinterleaving block, dewhitening block, and hamming decoder, sequentially.

\subsection{Jamming Attacks}
 
In this subsection, we overview the jamming attacks on LoRa communications. 
In \cite{aras2017selective}, Aras et al. investigated the vulnerabilities of LoRa communications under triggered jamming attack and selective jamming attack.
The authors showed that, despite using spreading spectrum scheme at low data rate, LoRaWAN is still highly susceptible to jamming attacks.

\begin{figure}
	\centering
	\includegraphics[width=3.5in]{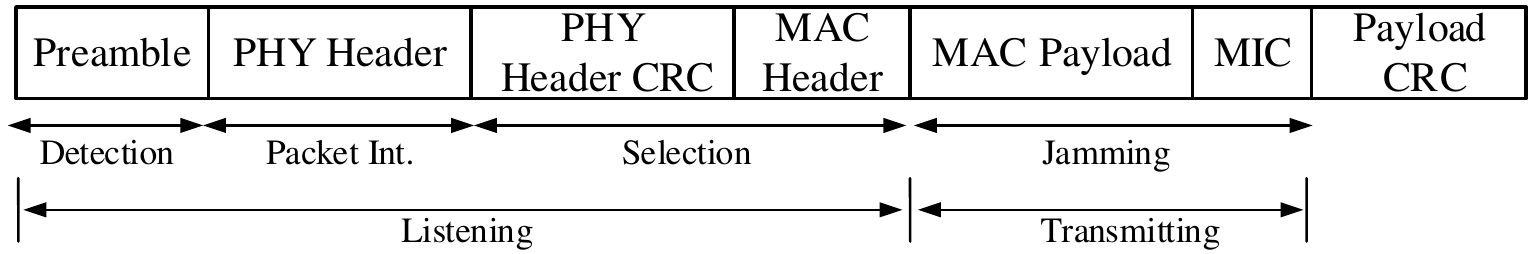}
	\caption{Illustration of selective jamming attack on LoRa packet transmissions \cite{aras2017selective}.}
	\label{fig:lora_sel_jam}
\end{figure}

\begin{itemize}
\item
\textit{Triggered jamming:} 
Triggered jamming attack is similar to reactive jamming attack.
Once the jammer detects the preamble transmitted by a LoRaWAN device, it will broadcast jamming signal.
\cite{aras2017selective} conducted experiments to evaluate the performance of LoRaWAN under  triggered jamming attack, where a commodity LoRa end-device is used as triggered jammer.
The experimental results show that the packet reception rate of a legitimate LoRa device drops to $0.5\%$ under the triggered jamming attack.

\item
\textit{Selective jamming:} 
Fig.~\ref{fig:lora_sel_jam} shows the basic idea of selective jamming attack studied in \cite{aras2017selective}, where a jammer sends jamming signal upon decoding the MAC header and the end-device address.
The selective jamming attack can block a particular device's communications in a LoRaWAN while causing no interference to other devices in the network.
The authors ran experiments to evaluate the performance of selective jamming attack in a LoRaWAN.
They used two commodity LoRa devices and programmed the jammer such that it targets one of their communications.
Experimental results show that the packet reception rate at the victim LoRa device drops to $1.3\%$ under the selective jamming attack. 
\end{itemize}

In \cite{huang2019experimental}, Huang et al. built a prototype of a reactive jammer against LoRaWAN communications using a commodity LoRa module.
The proposed algorithm jointly use the LoRa preamble detection and the signal strength indicator (RSSI) for efficiently detecting LoRa channel activity. 
The authors evaluated the performance of the proposed jammer in LoRaWAN via conducting real-world experiments. 
They considered the impact of the SF and the jammer's bandwidth on signal detection.
The packet delivery rate was used to measure the performance of the victim LoRa device under the proposed jamming attack.
Their results demonstrate that a jammer can achieve more than $90\%$ accuracy in LoRa packet detection when the jammer's SF and bandwidth are matched with the legitimate LoRa signal.
Moreover, the results show that, for most scenarios, when jamming signal is $3$~dB stronger than LoRa signal, a LoRa receiver fails to decode the received packets and the packet delivery rate drops to zero.

In \cite{butun2018analysis}, Butun et al. studied the security challenges regarding the LoRaWAN communications.
A LoRaWAN v1.1 architecture comprising end-devices, gateways, network servers, and application servers was considered. 
At the PHY layer, the authors investigated the replay attack on join-request signaling in LoRaWANs.
The attacker requires to jointly sniff the join-request signal and interferes with it at a LoRa gateway.
The attacker also interrupts the second join-request attempt and simultaneously sends the stored join-request signal captured in the first device's attempt.
It was argued that the gateway tends to accept the join-request signal spoofed by the attacker as it has never been used before. 
In this scenario, the LoRa device will no longer be synchronized to its serving gateway as their in-use join-requests are not matched.

\subsection{Anti-jamming Techniques}

The spreading spectrum technique is the main approach used in the LoRa technology to secure its communications against jamming attacks and unintentional interference signals.
As mentioned earlier, LoRa uses the CSS modulation scheme for its data transmissions, in which every bit to be transmitted is mapped into the sequence of $2^{\text{SF}}$ chips and modulated onto the chirp waveform.

Per \cite{semtech2015an1200}, a LoRa receiver is capable of recovering the received packets (i.e., yielding zero packet error rate) with the RSSI as low as $-121$~dBm when working at $125$~KHz bandwidth and SF = $8$. 
Comparing to conventional FSK modulation scheme, it is shown that, for a given data rate of $1.2$~Kbps, a LoRa receiver achieves 7~dB to 10~dB lower receiving sensitivity.
Moreover, a LoRa receiver achieves up to $15$~dB gain in the packet reception ratio compared to conventional FSK receivers.

In \cite{danish2018network}, Danish et al. proposed a jamming detection mechanism in LoRaWANs using Kullback Leibler divergence (KLD) and Hamming distance (HD) algorithms.
The KLD-based jamming detection scheme uses the likelihood of jamming-free received signal's probability distribution and the received signal's probability distribution under jamming attack.
Particularly, the authors used join request transmissions to determine a mass function of the LoRa device signaling.
If the similarity of the received signal's distribution and the acquired mass function is below a certain threshold, the presence of an undesired interference signal or a jamming signal will be claimed. 
Similarly, in the Hamming distance scheme, the algorithm finds the average Hamming distance between the received signal and the training signal. 
If the calculated distance diverges from a certain threshold, the algorithm claims the presence of jamming attacks.
The authors evaluated the performance of their proposed schemes via a real-world system implementation.
The results show $98\%$ and $88\%$ accuracy in jamming detection for the KLD-based and HD-based algorithms, respectively.

\section{Jamming and Anti-Jamming Attacks in Vehicular Wireless Networks}
\label{sec:vanet}


\begin{figure}
	\centering
	\includegraphics[width=3.5in]{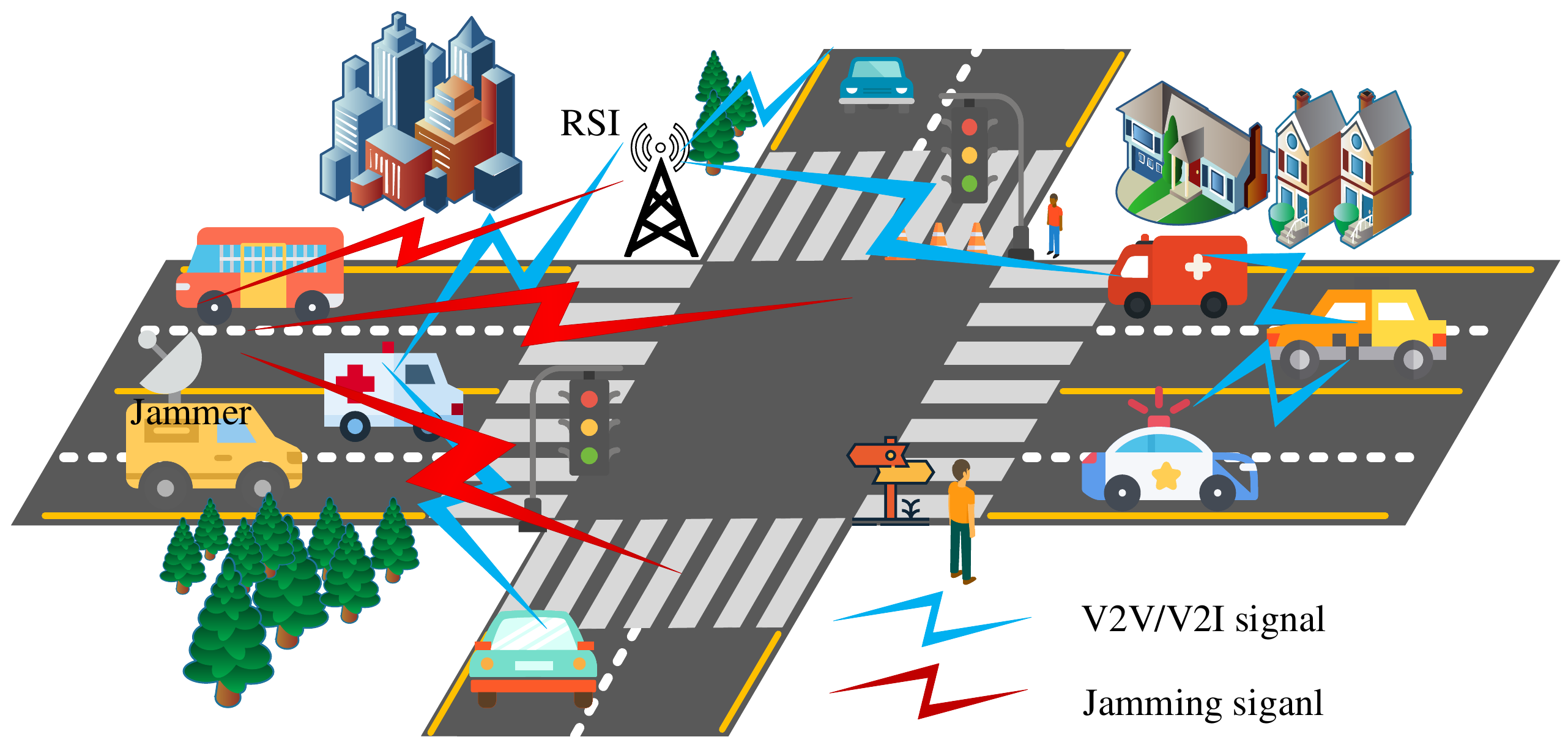}
	\caption{Illustration of jamming attacks in a vehicular network.}
	\label{fig:vanet_jammer}
\end{figure}

\begin{figure}
	\centering
	\includegraphics[width=3.5in]{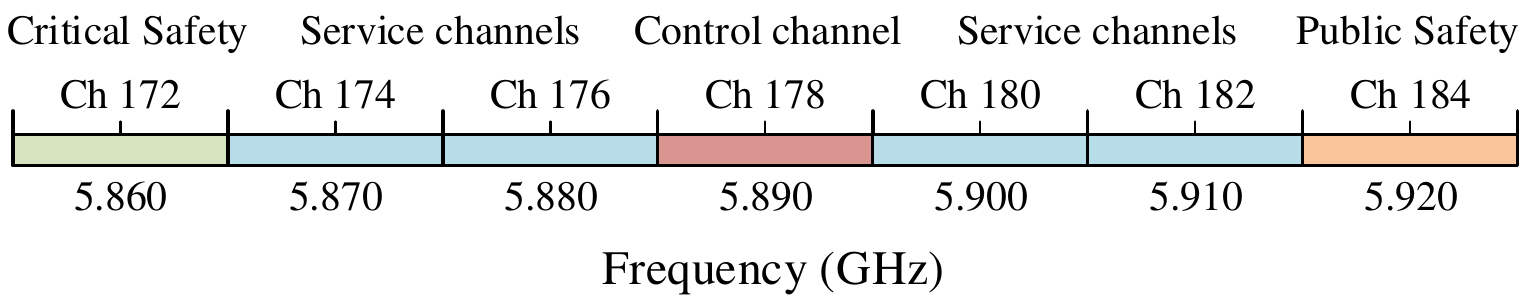}
	\caption{{The frequency channels allocated for DSRC.}}
	\label{fig:dsrc_ch}
\end{figure}

\begin{figure}[t]
	\centering
	\includegraphics[width=3.3in]{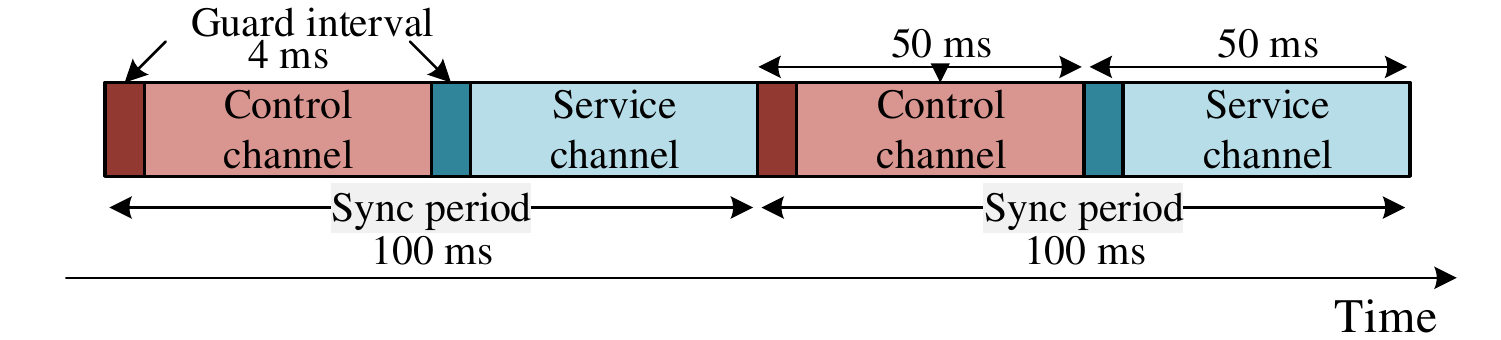}
	\caption{{The time intervals in 802.11p communications.}}
	\label{fig:dsrc_time}
\end{figure}

In this section, we survey the jamming attacks and anti-jamming mechanisms in vehicular wireless communication networks, including on-ground vehicular ad-hoc network (VENET) and in-air unmanned aerial vehicular (UAV) network. 
In what follows, we first provide an introduction of vehicular wireless communication networks and then review the jamming/anti-jamming strategies uniquely designed for vehicular wireless networks.

\subsection{A Primer of Vehicular Wireless Networks}

\noindent
\textbf{On-Ground VANET:}
Every year thousands of deaths occur in the U.S. due to traffic accidents, about 60 percent of which could be avoided by vehicular communication technologies. 
To reduce vehicular fatalities and improve transportation efficiency, the U.S. Department of Transportation (USDOT) launched the Connected Vehicle Program that works with state transportation agencies, car manufacturers, and private sectors to design advanced wireless technologies for vehicular communications.
It is a key component of network infrastructure to realize the vision of an intelligent transportation system (ITS) by enabling efficient vehicle-to-vehicle (V2V) communications and/or vehicle-to-infrastructure (V2I) communications.
Compared to stationary and semi-stationary wireless networks such as Wi-Fi networks, VANETs face two unique challenges in their design and deployment.
First, vehicles are of high speed.
The high speed of the vehicles continuously changes the network topology in terms of both the vehicles' positions and the number of connected vehicles.
Second, VANET is a delay-sensitive communication network. 
Reliable and timely packet delivery is of paramount importance for applications such as collision avoidance in real-world transportation systems.


Dedicated short-range communications (DSRC) is a communication system that enables wireless connectivity for VANETs.
At the PHY and MAC layers, DSRC deploys IEEE 802.11p standard.
802.11p uses the same frame format as legacy Wi-Fi (Fig.~\ref{fig:wifi_frame}(a)) for its data transmission.
However, the channel bandwidth is reduced to $10$~MHz to provide higher link reliability against channel impairments caused by users' mobility.
DSRC operates in the $5.9$~GHz frequency band, where seven distinct channels of $10$~MHz are available for the users' access, as shown in Fig.~\ref{fig:dsrc_ch}.
The center channel (control channel) and the two on-edge channels are used to carry safety messages such as critical information on road crashes or traffic congestion, while the service channels can be used for both safety messages and infotainment data such as video streaming, roadside advertising, and road map and parking-related information.
In the time domain, the resources are separated into equally-sized time intervals of $100$~ms, known as sync period, as shown in Fig.~\ref{fig:dsrc_time}.
Every sync period is divided into two $50$~ms intervals.
Every user needs to listen to the control channel within the first $50$~ms interval to receive the safety messages as well as obtain the information required for accessing the available services.
In the second $50$~ms interval, the user will switch to the intended service channel to send/receive infotainment data.

In a VANET, the vehicles, also known as on-board units (OBUs), and roadside units (RSUs) can communicate in three different network settings:
i) An RSU, similar to a Wi-Fi AP, serves a set of OBUs having the same basic service set (BSS) ID.
ii) The units with the same BSSID communicate with each other within an ad-hoc mode, where there is no dedicated central unit.
iii) The out-of-network units can hear each other, in which the units use a so-called wildcard BSSID to communicate with each other. 
The wildcard BSSID allows the units from different networks to communicate in a critical moment.
While the first two network settings are already deployed in Wi-Fi communications, the third mode is specifically designed for VANETs, as they require all units to be able to receive critical and safety messages.

At the MAC layer, similar to legacy Wi-Fi, 802.11p uses carrier sensing for channel access.
However, on top of the 802.11p MAC mechanism, DSRC takes advantage of an enhanced MAC-layer technique, known as enhanced distributed channel access (EDCA) in IEEE 1609.4, as a prioritized medium access mechanism to enable critical message exchanging.

\noindent
\textbf{In-Air UAV Network:}
With the significant advancement of robotic and battery technologies in the past decades, UAV communication networks have drawn an increasing amount of research interest in the community and enabled a wide spectrum of applications such as photography, film-making, newsgathering, agricultural monitoring, crime scene investigations, border surveillance, armed attacks, infrastructure inspection, search and rescue missions, disaster response, and package delivery.
Depending on the applications, UVAs of different sizes can be deployed in the network.
Small UAVs are generally used to form swarms, while large UAVs are likely used in critical military or civilian missions.
The speed of UAVs in different scenarios may vary from $0$~m/s to $100$~m/s, depending on the applications \cite{fotouhi2019survey}.

Most UAVs operate in the unlicensed $2.4$~GHz and $5.8$~GHz ISM bands.
Depending on the application requirements, UAV networks can be configured to operate in one of the following network topologies:
i) star topology, where each UAV directly communicates with a ground control station;
and 
ii) mesh network topology, in which UAVs communicate with each other as an ad-hoc network, and a small number of them may communicate with ground control station \cite{gupta2015survey}.
Moreover, heavier UAVs can be designed to take advantage of satellite communication (SATCOM) for routing.
UAVs are typically pre-programmed for this service to follow a flight route using global navigation satellite system (GNSS) signals.
In recent years, many standards have been drafted to address the challenges regarding UAV communications.
Cellular networks are currently considered as a promising infrastructure for UAV activities due to their wide coverage.  
3GPP in its release 15 have enhanced the LTE capability to support UAV communications \cite{atis2017unmanned}.

\subsection{Jamming Attacks}

Given the openness nature of wireless medium for vehicular communications, both VENET and UAV networks are vulnerable to generic jamming attacks (see Section~\ref{subsec:wifi_jamming}), such as constant jamming, reactive, deceptive, and frequency-sweeping jamming attacks.
In this subsection, we focus on the existing jamming attacks that were dedicatedly designed for vehicular wireless networks. 
Table~\ref{tab:vehicle} presents a summary of existing jamming attacks in vehicular networks. 

\noindent
\textbf{VANET-Specific Jamming Attacks:}
Jamming attacks are of particular importance in VANETs as the connection loss caused by the jamming attacks may lead to car collision and road fatality.
A jamming attacker can adopt different strategies (e.g., constant, reactive, and deceptive jamming) to cause the loss of wireless connection for V2V and V2I communications in VANETs.
Fig.~\ref{fig:vanet_jammer} shows an instance of jamming attack on a VANET, where the destructiveness of existing jamming attack strategies has been studied in literature.

\begin{table*}[]
\caption{{A summary of jamming attacks and anti-jamming strategies for vehicular wireless networks.}}
\centering
\begin{tabular}{|ll|l|l|}
\hline
                                                                                                          &                         & \multicolumn{1}{c|}{\textbf{Ref.}}            & \multicolumn{1}{c|}{\textbf{Description}}                                                                                                                               \\ \hline
\multicolumn{1}{|l|}{\multirow{5}{*}{\begin{tabular}[c]{@{}l@{}}Jamming \\ attacks\end{tabular}}}         & \multirow{3}{*}{VANETs} & \cite{azogu2013new}                          & Studied constant jamming attacks’ impact on IEEE 802.11p-based V2V communications.                                                                                      \\ \cline{3-4} 
\multicolumn{1}{|l|}{}                                                                                    &                         & \cite{punal2012vanets,punal2014experimental} & \begin{tabular}[c]{@{}l@{}}Evaluated the V2V communications' performance under constant, periodic, \\ and reactive jamming attacks\end{tabular}                         \\ \cline{3-4} 
\multicolumn{1}{|l|}{}                                                                                    &                         & \cite{sumra2011behavior,sumra2014effects}    & Introduced data integrity attack.                                                                                                                                       \\ \cline{2-4} 
\multicolumn{1}{|l|}{}                                                                                    & \multirow{2}{*}{UAVs}   & \cite{hartmann2013vulnerability}             & Investigated jamming attack on UAVs' satellite communications.                                                                                                          \\ \cline{3-4} 
\multicolumn{1}{|l|}{}                                                                                    &                         & \cite{rudinskas2009security}                 & Introduced jamming attack on UAVs' control command.                                                                                                                     \\ \hline
\multicolumn{1}{|l|}{\multirow{7}{*}{\begin{tabular}[c]{@{}l@{}}Anti-jamming \\ techniques\end{tabular}}} & \multirow{4}{*}{VANETs} & \cite{malla2013security}                     & Investigated frequency hopping technique for IEEE 802.11p communications.                                                                                               \\ \cline{3-4} 
\multicolumn{1}{|l|}{}                                                                                    &                         & \cite{lu2017anti,xiao2018uav}                & \begin{tabular}[c]{@{}l@{}}Proposed to use UAV communications to reroute the traffic from jammed regions\\ to alternative RSUs.\end{tabular}                            \\ \cline{3-4} 
\multicolumn{1}{|l|}{}                                                                                    &                         & \cite{karagiannis2018jamming}                & Proposed to use an unsupervised learning algorithm for detecting mobile jammers.                                                                                        \\ \cline{3-4} 
\multicolumn{1}{|l|}{}                                                                                    &                         & \cite{kumar2019delimitated}                  & Deployed a machine learning-based approach to localize the jammers.                                                                                                     \\ \cline{2-4} 
\multicolumn{1}{|l|}{}                                                                                    & \multirow{3}{*}{UAVs}   & \cite{lv2017anti}                            & Used power control game modeling for UAV communications under jamming attack.                                                                                           \\ \cline{3-4} 
\multicolumn{1}{|l|}{}                                                                                    &                         & \cite{xu2017anti,xu2018one,xu2019joint}      & Used power control game modelling for UAV ad-hoc network under jamming attack.                                                                                          \\ \cline{3-4} 
\multicolumn{1}{|l|}{}                                                                                    &                         & \cite{peng2019anti}                          & \begin{tabular}[c]{@{}l@{}}Used multi-parameter of frequency, motion ,and antenna spatial domains to optimally \\ compromise the impact of jamming attack.\end{tabular} \\ \hline
\end{tabular}
\label{tab:vehicle}
\end{table*}

In \cite{azogu2013new}, Azogu et al. studied the impact of jamming attacks on IEEE 802.11p-based V2X communications, where the jammers were designed to dynamically switch to in-use channels.
The authors conducted experiments in a congested city area using $100$ vehicles and $20$ RSI, and placed multiple radio jammers with the range of $100$~m in the proximity of RSIs.
They used two 802.11p channels for V2X communications and packet send ratio (PSR) as the evaluation metric, where the PSR is the number of packets sent per total number of queued messages for transmission.
The experimental results show that $10$ radio jammers can degrade the network PSR to $0.4$.

In \cite{punal2012vanets} and \cite{punal2014experimental}, Punal et al. evaluated the achievable throughput of V2V communications under constant, periodic, and reactive jamming attacks using real-world software-defined radio implementation.
The authors conducted several real-world experiments in both indoor and outdoor environments.
Particularly, their experiments in an outdoor scenario show that the packet delivery rates in 802.11p communications drop to zero when SNJR is less than $9$~dB under constant jamming attack, when SNJR is less than $55$~dB under periodic jamming attack with the duty cycle of $86\%$ and $74~\mu$s period, and when SNJR is less than $12$~dB under reactive jamming attack with the packet detection duration of $40~\mu$s and jamming signal duration of $500~\mu$s.

In \cite{sumra2011behavior} and \cite{sumra2014effects}, Sumra et al. introduced a series of data integrity attacks, in which an attacker aims to reduce the accuracy of information travel through a vehicular network.
Their study's basic idea is that once an attacker receives a packet, it attempts to alter the packet content and relay the false packet to a roadside unit or other vehicles.

\begin{figure}
	\centering
	\includegraphics[width=2.7in]{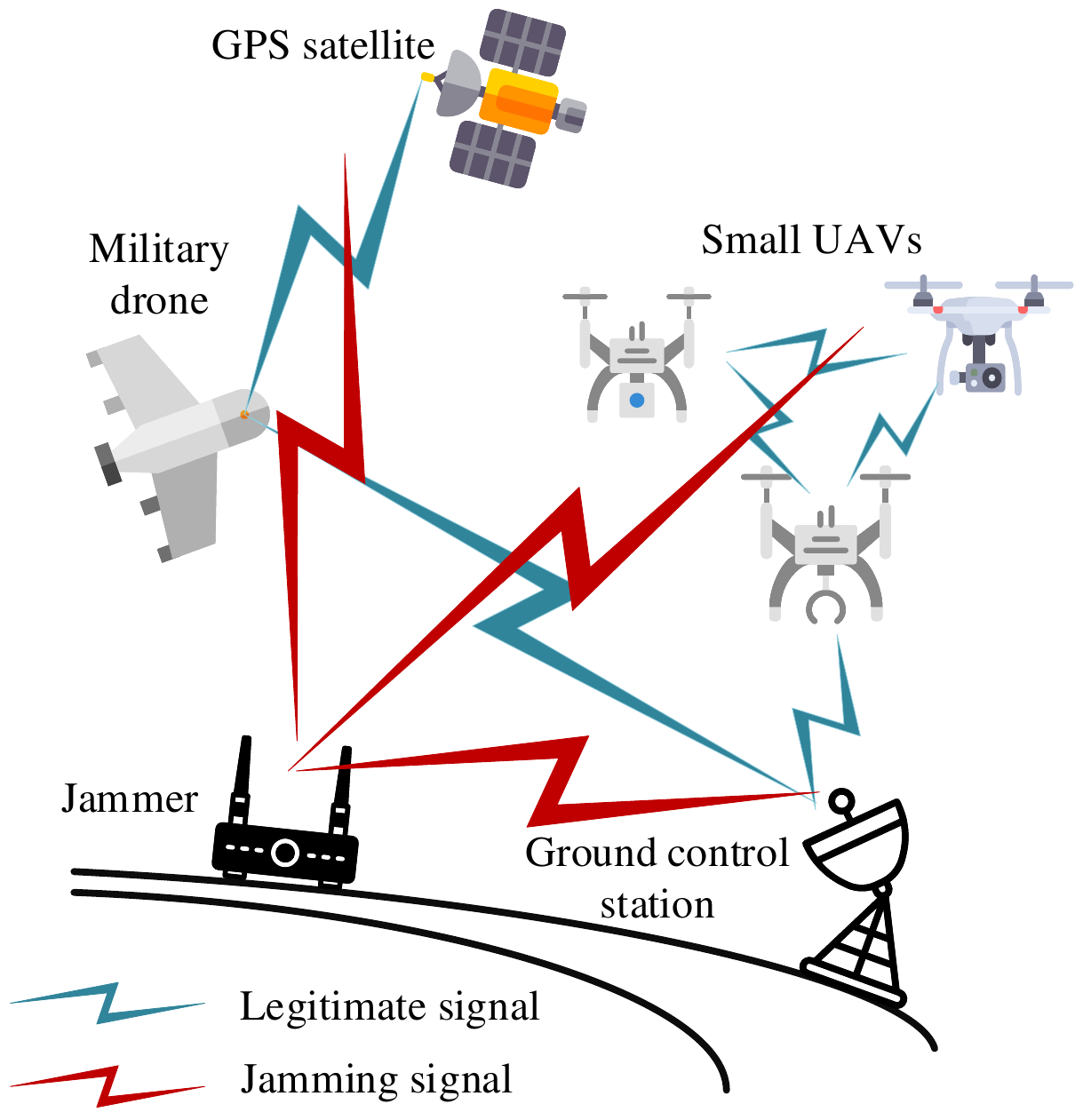}
	\caption{Illustration of jamming attacks in a UAV network.}
	\label{fig:uav_jammer}
\end{figure}

\noindent
\textbf{UAV-Specific Jamming Attacks:}
While UAV networks are vulnerable to the generic jamming attacks in Section~\ref{subsec:wifi_jamming}, we focus on those jamming attacks that were deliberately designed for satellite-based UAV networks. 
Fig.~\ref{fig:uav_jammer} shows an instance of jamming attack on UAV networks,
Since most UAVs rely on GPS signals for positioning and navigation, a jamming attacker may target on their GPS communications to dysfunction UAV networks. 
This class of jamming attacks can pose a severe threat to military UAV networks.

In \cite{hartmann2013vulnerability}, Hartmann et al. investigated the possible theories behind the loss of a military UAV (RQ-170 Sentinel) to Iranian military forces in December 2011.
They argued that a GPS-spoofing attack might have been carried out to spoof the UAV's position estimation, thereby hijacking the UAV's routing decision.

In \cite{rudinskas2009security}, another UAV-specific jamming attack, called control command attack, was studied.
In this attack, a jammer attempts to interrupt the control commands issued by UAVs' ground control station.
Control command attacks can be realized using conventional jamming attacks to block the desired received signal or by sending fake information to lure UAVs following incorrect commands. 
The control command attack can cause the loss of UAVs and mission failure.

\subsection{Anti-Jamming Techniques}

In this subsection, we review the anti-jamming techniques that were dedicatedly designed for VANETs and UAV networks. 
Table~\ref{tab:vehicle} presents a summary of existing anti-jamming mechanisms in vehicular networks.
We elaborate on them in the following.

\noindent
\textbf{VANET-Specific Anti-Jamming Techniques:}
Thus far, very limited work has been done in the design of anti-jamming techniques for VENETs.
Given that most VENETs are using IEEE 802.11p (DSRC) for V2V and V2I communications, a straightforward anti-jamming strategy for VENETs is to switch to another available wireless network (e.g., cellular), provided that such an alternative network is available.
In addition, frequency hopping and data relaying techniques have been studied to secure UAV communications. 
%

%
In \cite{lu2017anti} and \cite{xiao2018uav}, the authors proposed to cope with jamming attacks for VANETs by leveraging UAV devices.
They studied smart jamming attacks in a VANET, where a jammer continuously changes its attack strategy based on the network topology.
To salvage the vehicular communications, a UAV was utilized to relay data for vehicles to the alternative RSUs when the serving RSU is under jamming attacks. 
In this work, a game theory approach was employed to model the interactions between jammer and UAV, where the jammer adaptively selects its transmission power, and the UAV makes decisions for relaying the data. 

In \cite{karagiannis2018jamming}, Karagiannis et al. proposed to use an unsupervised learning algorithm for detecting mobile jammers in vehicular networks.
The key component of the proposed scheme is to use a newly-defined parameter that measures the relative speed of victim vehicles and the jammer vehicle for training a learning algorithm.
They evaluated the performance of the detection algorithm in two jamming attack scenarios:
i) the jammer periodically sends jamming signal and maintains a safe distance from the victim vehicle;
and
ii) the jammer is unaware of the possible detection mechanism and sends constant jamming signal at any distance from the victim vehicle.
Their simulation results show that their proposed algorithm can classify the attacks with the accuracy of $98.9\%$ under constant jamming attack and $44.5\%$ under periodic jamming attack.

In \cite{kumar2019delimitated}, Kumar et al. employed a machine learning-based approach to estimate the location of jammers in a vehicular network.
Particularly, a foster rationalizer was used to detect any undesired frequency changes stemming from jamming attacks on legitimate V2V communications.  
Following the rationalizer, a Morsel filter was applied to remove the noise-like components from jammed signal.
The filtered signal was then injected into a Catboost algorithm to estimate the jammer's vehicle location.
Their simulation results show that the proposed scheme can predict the jamming vehicle's position with an accuracy of $99.9\%$.

\noindent
\textbf{UAV-Specific Anti-Jamming Techniques:}
Despite having attracted more research efforts, UAV networks are still in its infancy, and the research on UAV-specific anti-jamming strategies remains rare.
In \cite{lv2017anti}, the authors performed a theoretical study on anti-jamming power control game for UAV communications in the presence of jamming attacks, where the interactions between a jammer and a UAV are modeled as a Stackelberg game.
An optimal power control strategy for UAV transmissions under jamming attack was obtained using a reinforcement learning algorithm.
In \cite{xu2017anti} and \cite{xu2018one}, Xu et al. studied the anti-jamming problem in a UAV ad-hoc network, in which every transmitter tends to send the data toward its desired receiver while causing interference to other communication links (refers to the co-channel mutual interference).
The problem was formulated as a Bayesian Stackelberg game, in which a jammer is the leader, and UAVs are the followers of the game.
The optimal transmission power of UAVs and jammer were analytically derived using their utility functions optimization.
In \cite{xu2019joint}, a model was developed to take into account the flying process of UAV.
Based on the model, the flying trajectory and transmission power of the UAV were optimized.
In \cite{peng2019anti}, Peng et al. investigated the jamming attacks in a UAV swarm network, where a jammer network consisting of multiple nodes attempts to block the communication between the swarm and its ground control station.
Multiple parameters were taken into account to provide high flexibility for UAV communications to deal with jamming attacks.
In particular, the UAVs exploited the degrees of freedom in the frequency, motion, and antenna-based spatial domains to optimize the link quality in the reception area.
A modified Q-learning algorithm was proposed to optimize this multi-parameter problem.

\section{Jamming and Anti-Jamming Attacks in RIFD Communication Systems}
\label{sec:rfid}

In this section, we first provide a primer of Radio-frequency identification (RFID) communication and then provide a review on RFID-specific jamming and anti-jamming attacks. 

\begin{figure}
	\centering
	\includegraphics[width=2.7in]{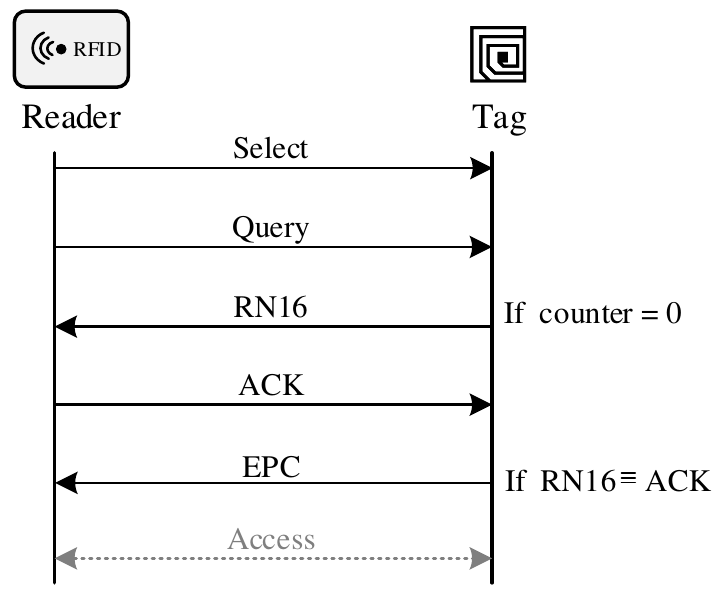}
	\caption{RFID tag inventory and access processes.}
	\label{fig:rfid_mac}
\end{figure}
 
\subsection{A Primer of RFID Communication}

As we enter into the Internet of Everything (IoE) era, RFID emerges as a key technology in many industrial domains with an exponential increase in its applications.  
In RFID communications, an interrogator (reader) reads the information stored in simply-designed tags attached to different objects such as humans, animals, cars, clothes, books, grocery products, jewelry, etc. from short distances using electromagnetic waves.
An RFID tag can be either passive or active.
Passive tags do not have a battery and use energy harvesting for their computation and RF signal transmission.
Active tags, on the other hand, use battery to drive their computation and RF circuits.

Fig.~\ref{fig:rfid_mac} shows the medium access protocol used in GEN2 UHF RFID communications \cite{zhang2020standards}.
Generally speaking, in an RFID system, a reader (interrogator) manages a set of tags following a protocol that comprises the following three phases:

\begin{itemize}
\item
\textit{Select:} The reader selects a tag population by sending \textit{select} commands, by which a group of tags are informed for further inquiry process.

\item
\textit{Inventory:} 
In this phase, the reader tries to identify the selected tags. 
The reader initializes the inventory rounds through issuing \textit{Query} commands. 
The reader and RFID tags communicate using a slotted ALOHA medium access mechanism.
Once a new-selected tag receives a Query command, it sets its counter to a pseudo-random number.
Every time a tag receives the Query command, it decreases its counter by $1$.
When the tag's counter reaches zero, it responds to the reader with a 16-bit pseudo-random sequence (RN16).
The reader acknowledges the RN16 signal reception by regenerating the same RN16 and sending it back to the tag.
When a tag receives the replied RN16 signal, it compares the signal with the original transmitted one. 
If they are matched, the tag will broadcast its EPC, and the reader will identify the tag.

\item
\textit{Access:} When the inventory phase is completed, the reader can access the tag.
In this phase, the reader may write/read to/from the tag's memory, kill, or lock the tag.
From a PHY-layer perspective, an UHF RFID readers use pulse interval encoding (PIE) and DSB-ASK/SSB-ASK/PR-ASK modulation for its data transmission, while a UHF RFID tag uses FM0 or Miller data encoding and backscattering for carrier modulation.
Moreover, the required energy for the tag's backscattering is delivered via the reader's continuous wave (CW) transmissions \cite{epcglobal2013epc}.
\end{itemize}

\begin{figure}
	\centering
	\includegraphics[width=2.8in]{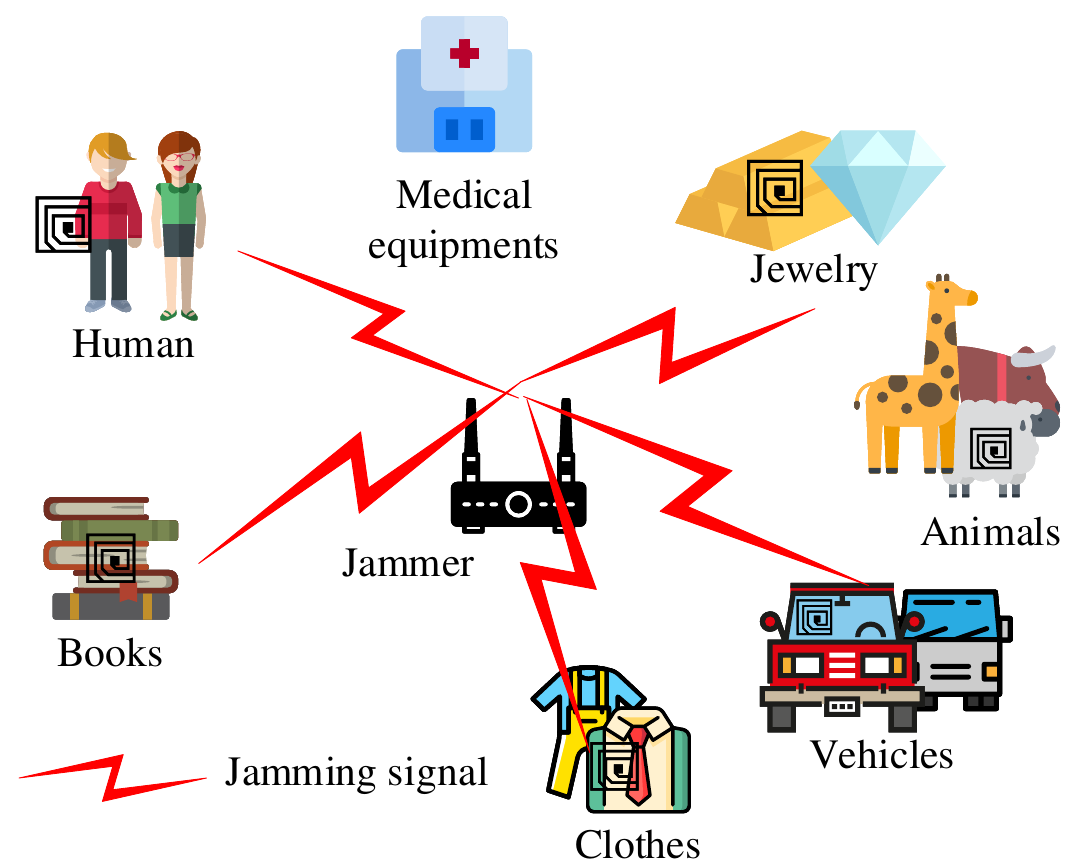}
	\caption{Illustration of jamming attacks in RFID applications.}
	\label{fig:rfid_attack}
\end{figure}

\subsection{Jamming Attacks}

Since RFID communication is a type of wireless communications, it is vulnerable to generic jamming attacks such as constant, reactive, and deceptive jamming threats.
Fig.~\ref{fig:rfid_attack} shows an instance of jamming attack on some applications of RFID technology.
In \cite{fu2010research}, Fu et al. investigated the impact of constant jamming attack on UHF RFID communications and evaluated the system performance through a simulation model.
The results in \cite{fu2010research} showed that when the received RFID reader signal power is $0$~dBm, the received jamming signal power of $-15$~dBm will be sufficient to break down the RFID communications.
Per \cite{mitrokotsa2010classifying}, Zapping attack is one of the well-known security attacks to RFID tags, in which an attacker aims to disable the function of RF front-end circuits in RFID tags.
In a Zapping attack, a malicious attacker produces a strong electromagnetic induction through the tag's circuit by generating a high-power signal in the proximity of the tag's antenna.
The large amount of energy a tag receives may cause permanent damage to its RF circuits.

RFID tags are recently used for electronic voting systems in many countries across the world.
In such voting systems, the votes are written through electronic RFID tags instead of conventional ballot voting papers.
Electronic voting systems were originally designed to improve the accuracy and speed of the vote-counting process.
Given the importance of voting services, research on RFID security receives a large amount of attention from academia and industry.
In \cite{oren2012rfid} and \cite{oren2009attacks}, Oren et al. studied two main security threats on RFID communications: jamming and zapping attacks.
The authors studied these two jamming attacks and evaluated their performances in terms of their maximum jamming ranges.
The author also investigated the impact of a jammer's antenna type on its jamming effectiveness. 
It was shown that a helical antenna yields a larger jamming range compared to a hustler or $39$~cm loop antenna.

In \cite{rieback2006evolution}, Rieback et al. classified the security threats in RFID networks into the following categories.

\begin{itemize}
\item
\textit{Spoofing attack :} 
A spoof attacker may access a tag's memory blocks and alter a tag's information such that it is still meaningful to the reader but carries false data.   

\item
\textit{Denial-of-Service Attack:} 
This attack refers to a security threat where a reader will no longer be able to read data from its corresponding tags.
Per \cite{rieback2006evolution}, DoS attack in RFID communications can be performed either by physically isolating the tags (e.g., wrapping the tags in foils) so the reader can not deliver the required energy for tag's backscattering, or by injecting a large amount of data into medium to overwhelm the reader.

\item
\textit{Replay attack:} 
In \cite{danev2012physical}, Danev et al. introduced a replay attack where the attacker receives the tag's signal, stores it, and re-uses it in different time slots.

\item
\textit{Passive attack :} 
Passive attack is one of the leading security vulnerabilities in RFID communications, as a tag's EPC information can be read by any commercial off-the-shelf or custom-designed interrogators.
In \cite{cho2015consideration}, Cho et al. proposed a brute-force attack in which the attacker uses brute-forcing algorithms to pass the tag's authentication procedure, where it can write to (or read from) the tag.
The recorded data from RFID tags can be further used for object tracking or spoofing attack.
\end{itemize}

\subsection{Anti-Jamming Techniques}

Unlike other types of wireless networks, securing RFID communication against jamming attacks is a particularly challenging task due to the passive nature of RIFD tags or low-energy supply of RFID tags.
Generic anti-jamming techniques (e.g., MIMO-based jamming mitigation, spectrum spreading, and frequency hopping) appear ineffective or unsuitable for RFID systems. 
Most of the existing works focus on authentication issues in RFID networks.
For example, Wang et al. in \cite{wang2020hu} proposed an authentication scheme to cope with RFID replay attack.
Avanco et al. in \cite{avanco2015effective} proposed a low-power jamming detection mechanism in RFID networks.
The proposed mechanism detects malicious activities within the network by exploiting side information such as the received signal power in adjacent channels, the received preamble, and the tag's uplink transmission response.

\section{Jamming and Anti-Jamming Attacks in GPS Systems}
\label{sec:gps}

\begin{figure}
\centering
\includegraphics[width=2.9in]{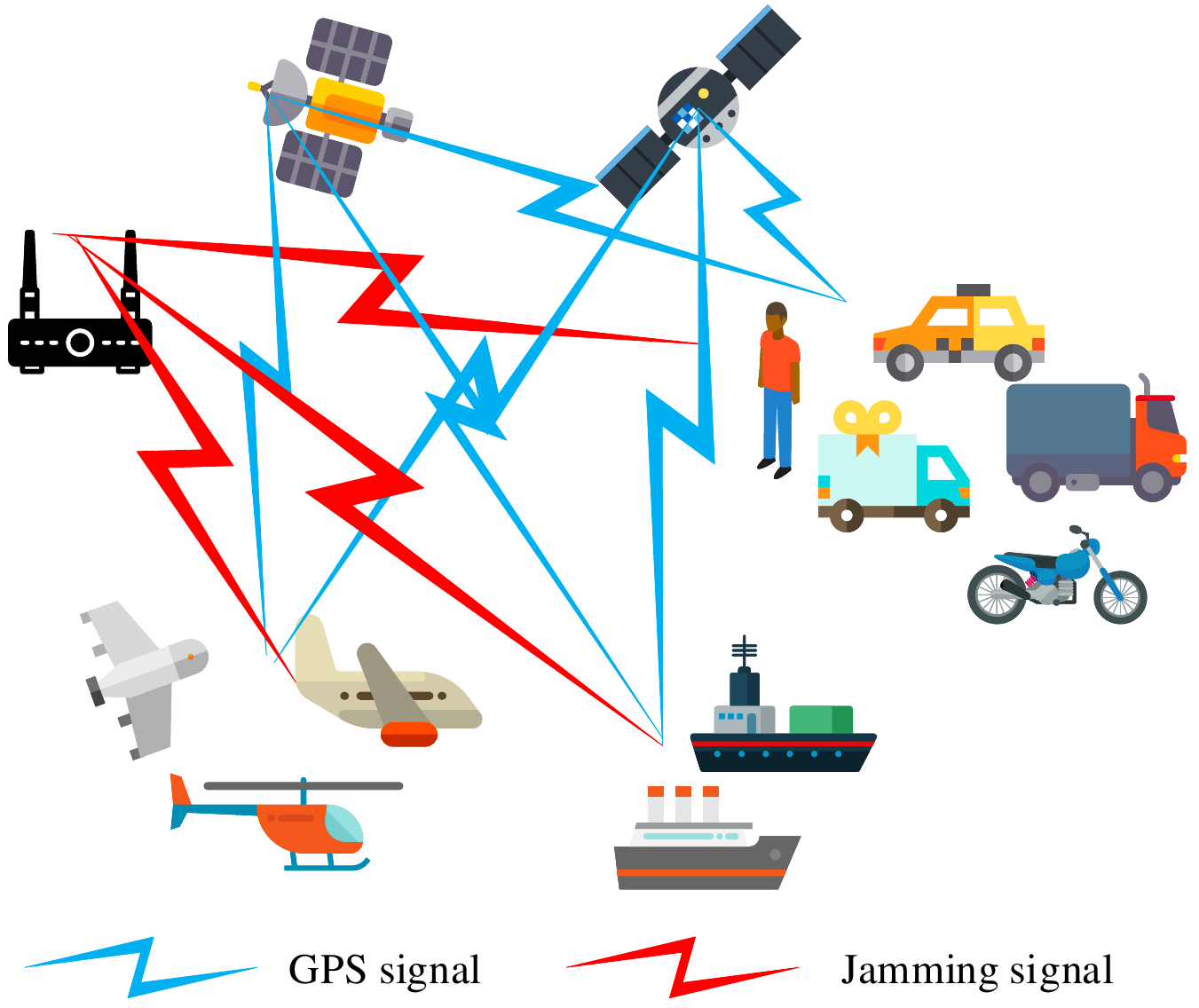}
\caption{Illustration of jamming attacks in GPS system.}
\label{fig:gps_attack}
\end{figure}

In this section, we survey jamming and anti-jamming attacks in the Global Positioning System (GPS).
By the same token, we first offer a primer of GPS communication system and then review existing jamming/anti-jamming attacks deliberately designed for GPS systems.

\begin{figure}[t]
\centering
\includegraphics[width=3.5in]{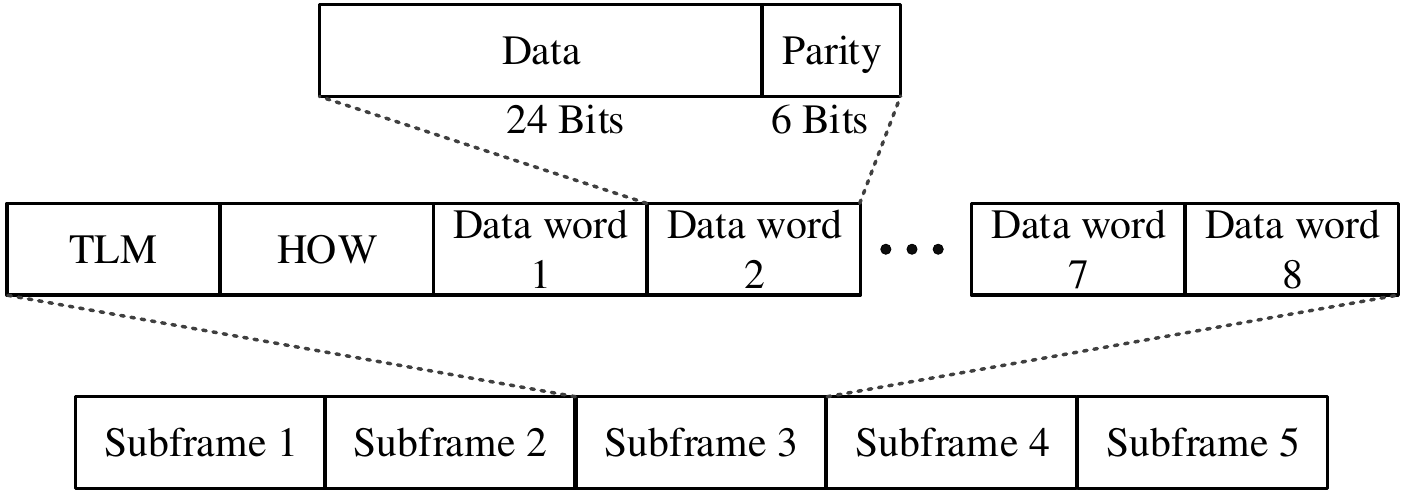}
\caption{GPS navigation message frame structure.}
\label{fig:gps_frame}
\end{figure}

\begin{figure}[t]
\centering
\includegraphics[width=3.5in]{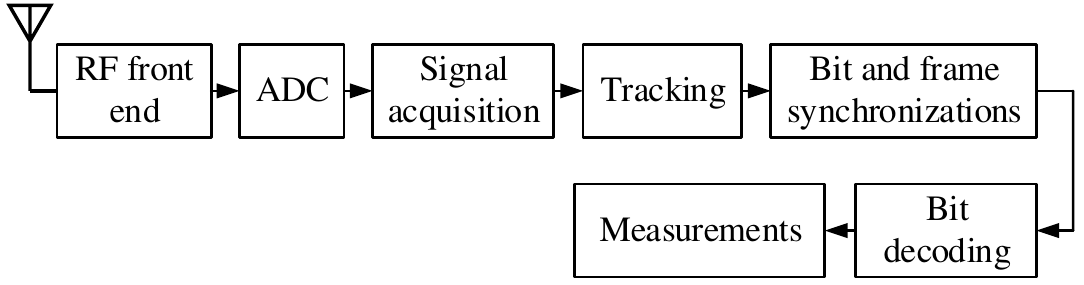}
\caption{The schematic diagram of a GPS receiver \cite{karaim2019ultra}}
\label{fig:gps_receiver}
\end{figure}

\subsection{A Primer of GPS Communication System}

GPS was designed to provide on-earth users with an easy way of obtaining their location and timing information by establishing a set of satellites that continuously orbit around the earth and broadcast their location and timing information.
In GPS, an on-earth user first acquires a satellite's time and location information and then estimate the distance (a.k.a. pseudo-range) from the satellite to itself by calculating the time of signal travel. 
The on-earth user's geographical location can be determined mathematically, considering the locations and estimated pseudo ranges of multiple satellites.
As such, wireless communication in GPS is one-way communication.

Fig.~\ref{fig:gps_frame} shows the structure of a GPS navigation message, which consists of $5$ equally-sized subframes.
Each subframe is composed of $10$ data words, consisting of $24$ raw data bits followed by $6$ parity bits.
The first data word in each subframe is the telemetry (TLM) word, which is a binary preamble for subframe detection.
It also carries administrative status information.
Following the TLM word is Handover (HOW) word and carries GPS time information, which is used to identify the subframe. 
Subframe $1$ carries the information required for clock offset estimation. 
Subframes $2$ and $3$ consist of satellite ephemeris data that can be used by the users to estimate the satellite's location at a particular time accurately. 
Subframes $4$ and $5$ carry almanac data that provides information on all satellites and their orbits on the constellation.

The navigation message is sent at $50$~bps data rate.
All GPS satellites operate in $1575.42$~MHz (L1 channel) and $1227.60$~MHz (L2 channel) and use one of the following signaling protocols:

\begin{itemize}
\item
\textit{Course Acquisition (C/A) Code:} 
C/A signal uses a direct sequence spread spectrum (DSSS) with a 1023-chip pseudo-random spreading sequence, which is replicated $20$ times to achieve higher spreading gain. 
The main lobe C/A signal bandwidth is $2$~MHz and the spreading gain is $10 \times \log_{10}(20 \times 1023) \approx 43$~dB.
The C/A signal is transmitted over the L1 frequency channel and is mainly used to serve civilian users.

\item
\textit{P-Code:}
P-code signal also uses spreading spectrum. 
The length of its spreading code is $10,230$~chips per bit.
P signal is more spread over the frequency and can achieve higher spreading gain ($\approx 53$~dB) than the C/A signal.
The signal bandwidth is $20$~MHz for the P-code signal.
P signal can be transmitted over both L1 and L2 channels and used by the US Department of Defense to serve authorized users.
\end{itemize}

Fig.~\ref{fig:gps_receiver} shows the structure of a GPS receiver.
When a GPS device receives GPS signal from satellites, it uses a signal acquisition module to identify the corresponding satellites.
The signal acquisition will be made using C/A code correlation and exhaustive search over a possible range of Doppler frequencies.
The tracking module modifies the coarse changes in the phase and frequency of the acquired signal.
Following the phase and frequency tracking, the received chips are synchronized and demodulated into bits that are used for measurement.

\subsection{Jamming Attacks}
Given that GPS is a one-way communication system, the potential jamming threat in GPS is mainly posing on on-earth GPS user devices.
Fig.~\ref{fig:gps_attack} shows an instance of jamming attack on some GPS-based navigation applications.
Since GPS satellites are very far from earth, on-earth GPS signals can be easily drowned by jamming signals.
Table~\ref{tab:gps_table} summarizes existing jamming attacks on GPS communications.
We detail them in the following.

In \cite{glomsvoll2014jamming}, the performance of a GPS receiver was studied in the face of jamming attacks. 
A wide-band commercial off-the-shelf radio jammer at L1 carrier frequency has been used to jam an on-board GPS signal.
The jamming signal bandwidth was set to $60$~MHz, and its transmit power was measured to $-35$~dBW.
Carrier-to-Noise Ratio (CNR) was used to measure the received GPS signal strength. 
CNR is a bandwidth-independent metric expressed as the ratio of carrier power to noise power per hertz.
It was shown that, for stationary and semi-stationary GPS receivers, the CNR of $30$~dB/Hz to $-35$~dB/Hz suffices for signal detection.
In a jamming-free scenario, it was measured that the CNR is $44$~dB/Hz at a GPS receiver.  

Signal spoofing is another critical security threat to GPS systems.
GPS spoofing can be used to falsify the navigation estimation and timing synchronization. 
In \cite{warner2002simple,humphreys2008assessing,motella2010performance}, the authors analyzed the performance of civilian GPS receivers under practical GPS spoofing. 
In addition to global location, the GPS system provides time information for GPS receivers.
This time information provides time source in the range from microsecond to seconds for interoperability of many applications, and the system components that are dispersed in different time-zone areas \cite{zeng2012gps}.
In \cite{kuhn2004asymmetric}, Kuhn et al. introduced a potential security threat to GPS communications called ``\textit{selective-delay attack}'', in which the attacker was designed to falsify the timing of a GPS receiver by spoofing a delayed version of the signal.
They analytically showed that the current positioning systems are not resilient to selective-delay attacks.


\subsection{Anti-Jamming Attacks}

Since the GPS communication system employs the spectrum spreading technique for data transmission, it bears 43~dB spreading gain for C/A code (civilian use) and 52~dB spreading gain for P-code (military use). 
The spreading gain enables a GPS receiver to combat jamming attacks to some extent. 
In addition to its inherent spectrum spreading technique, other anti-jamming techniques (e.g., MIMO-based jamming mitigation) can also be used to improve its resilience to jamming attacks.
In what follows, we review existing anti-jamming attacks for GPS in the literature, which are summarized in Table~\ref{tab:gps_table}.

\noindent
\textbf{Jamming Signal Filtering:}
In \cite{zhang2001anti}, Zhang et al. proposed to enhance a GPS receiver's resilience to jamming attack by designing a filtering mask in both time and frequency domains.
Such a filtering mask uses the time and frequency distributions of the jamming signal to suppress its energy while allowing GPS signals to pass.
In \cite{chien2013design}, an adaptive notch filter was designed to detect, estimate, and block single-tone continuous jamming signals.
Their proposed notch filter used a second-order IIR filter designed in the time domain.
In \cite{rezaei2016new}, Rezaei et al. aims to enhance the capability of notch in jamming mitigation filter for a GPS receiver.
The authors proposed to use the short-time Fourier transform (STFT) to increase the time and frequency domain resolution.

\noindent
\textbf{Antenna Array Design:}
Antenna array processing is another technique that is used by GPS receivers to enhance their jamming resilience \cite{li2014robust}.
In \cite{rezazadeh2019compact}, Rezazadeh et al. designed an adaptive antenna system by leveraging the antenna's pattern and polarization diversity to nullify the jamming signal in airborne GPS applications.
The authors built a prototype of the designed antenna and evaluated its performance in the presence of jamming attack.
Their experimental results showed that the implemented antenna is capable of achieving up to $38$~dB jamming suppression gain.
In \cite{zhang2012anti}, Zhang et al. studied the challenges associated with the antenna array design for GPS receiver.
The authors proposed an array-based adaptive anti-jamming algorithm to suppress jamming signal with negligible phase distortion.
In \cite{sun2005self}, Sun et al. harnessed the inherent self-coherence feature of GPS signal to mitigate the jamming signal, as the GPS signal repeats $20$ times within each symbol.
In \cite{agee1990spectral}, a beamforming technique was designed as an anti-jamming solution for GPS communications.
A spectral self-coherence beamforming technique was proposed to design a weight vector to nullify the jamming signal and improve the desired signal detection.

A similar idea was explored in \cite{lu2013global}, where the received signal is first projected onto the jamming signal's orthogonal space, and a so-called CLEAN method \cite{schwarz1978mathematical} was used to extract the received GPS signals.
The CLEAN method is a classical approach that iteratively computes the beamforming vectors to identify the arrival direction of the signals of interest.
It uses the repetitive pattern of the C/A code to construct the beamforming vectors.
The authors evaluated the performance of their proposed anti-jamming technique via both computer simulations and real-world experiments.
They considered four GPS signals arriving at different directions and a continuous wave jamming signal.
The simulation results demonstrated that their proposed algorithm could successfully acquire the $4$ GPS signal streams when jamming signal power to noise power ratio (JNR) is $40$~dB.
Their experimental results show that, using a $4$-antenna array plane, the proposed receiver can correctly determine its position when JNR is $20$~dB.

\noindent
\textbf{Jamming Detection Mechanisms:}
In \cite{gao2017gnss}, Gao et al. proposed to use machine learning for spoofing attack detection in GNSS communications.
They used isometric mapping and Laplacian eigen mapping algorithms for feature extraction, and trained the support vector machine to classify the features and detect the spoofing attack.
\begin{table*}[]
\caption{{A summary of jamming attacks and anti-jamming strategies for GPS communications.}}
\centering
\begin{tabular}{|l|l|l|}
\hline
& \multicolumn{1}{c|}{\textbf{Ref.}}                                     & \multicolumn{1}{c|}{\textbf{Description}}                                                                         \\ \hline
\multirow{4}{*}{\begin{tabular}[c]{@{}l@{}}Jamming \\ attacks\end{tabular}}         & \cite{glomsvoll2014jamming}                                           & Analyzed GPS receiver performance under wide-band commercial jamming attacks                                      \\ \cline{2-3} 
& \cite{warner2002simple,humphreys2008assessing,motella2010performance} & Investigated the GPS signal spoofing.                                                                             \\ \cline{2-3} 
& \cite{zeng2012gps}                                                    & Studied the GPS time information attack.                                                                          \\ \cline{2-3} 
& \cite{kuhn2004asymmetric}                                             & Used spoofing attack to falsify the timing of the military GPS signals.                                           \\ \hline
\multirow{10}{*}{\begin{tabular}[c]{@{}l@{}}Anti-jamming\\ techniques\end{tabular}} & \cite{zhang2001anti}                                                  & Designed a time-frequency mask using the time and frequency distributions of the jamming signal.                  \\ \cline{2-3} 
& \cite{chien2013design}                                                & Used adaptive notch filter design to detect, estimate, and block single-tone continuous jamming signals.          \\ \cline{2-3} 
& \cite{rezaei2016new}                                                  & Used short-time Fourier transform (STFT) to enhance resolution in notch filter design.                            \\ \cline{2-3} 
& \cite{li2014robust}                                                   & Designed an adaptive antenna array.                                                                               \\ \cline{2-3} 
& \cite{rezazadeh2019compact}                                           & Designed an adaptive antenna system based on pattern and polarization diversity.                                  \\ \cline{2-3} 
& \cite{zhang2012anti}                                                  & Proposed an adaptive array-based algorithm with negligible phase distortion.                                      \\ \cline{2-3} 
& \cite{sun2005self}                                                    & Used a beamforming technique considering inherent self-coherence feature of the GPS signals.                      \\ \cline{2-3} 
& \cite{agee1990spectral}                                               & Used a self-coherence beamforming technique to nullify the jamming signal and enhance the desired signal quality. \\ \cline{2-3} 
& \cite{lu2013global}                                                   & Used received signal projection onto the orthogonal space of jamming signal.                                      \\ \cline{2-3} 
& \cite{gao2017gnss}                                                    & Proposed a learning-based spoofing attack detection.                                                              \\ \hline
\end{tabular}
\label{tab:gps_table}
\end{table*}

\section{Open Problems and Research Directions}
\label{sec:open_problems}

In this section, we first list some open problems in designing efficient anti-jamming techniques and then point out some promising research directions toward securing wireless communication networks against jamming attacks.

\subsection{Open Problems}

Despite the significant advancement of wireless communication and networking technologies in the past decades, real-world wireless communication systems (e.g., Wi-Fi, cellular, Bluetooth, ZigBee, and GPS) are still vulnerable to malicious jamming attacks. 
As wireless services become increasingly important in our society, the jamming vulnerability of wireless Internet services poses serious security threats to existing and future cyber-physical systems.
This vulnerability can be attributed to the lack of practical, effective, and efficient anti-jamming techniques that can be deployed in real-world wireless systems to secure wireless communication against jamming attacks. 
One may argue that Bluetooth is equipped with the frequency hopping technique, and ZigBee/GPS is equipped with spectrum spreading technique, and (therefore) these networks can survive in the presence of jamming attacks.
This argument, however, is not valid.
Bluetooth can only work in the face of narrow-band jamming attack, and ZigBee/GPS can only work under a low-power jamming attack. 
These wireless systems as well as the most prevailing Wi-Fi and cellular networks, can be easily paralyzed by a jamming attack using commercial off-the-shelf SDR devices.

In what follows, we describe some open problems in the design of anti-jamming techniques, with the aim of spurring more research efforts on advancing the design of jamming-resistant wireless communication systems. 

\subsubsection{Effectiveness of Anti-Jamming Techniques}
One open research problem is to design effective anti-jamming techniques for wireless networks.
Existing anti-jamming techniques (e.g., frequency hopping, spectrum spreading, retransmission, and MIMO-based jamming mitigation) have a limited ability to tackle jamming attacks. 
For example, neither frequency hopping nor spectrum spreading technique is able to salvage wireless communication services when jamming signal is covering the full spectrum and stronger than useful signal. 
The state-of-the-art MIMO-based technique can offer at most 30~dB jamming mitigation capability for two-antenna wireless receivers.
This indicates that if jamming signal is 22~dB stronger than the useful signal, the receivers in a wireless network will not be capable of decoding their packets under jamming attacks.
Therefore, a natural question to ask is how to design effective anti-jamming techniques for wireless networks so that those wireless networks can be immune to jamming attacks, regardless of jamming signal power, bandwidth, sources, and other configuration parameters.

\subsubsection{Efficiency of Anti-Jamming Techniques}
Another open problem is to improve the efficiency of anti-jamming techniques. 
For example, frequency hopping can cope with narrow-band jamming attacks, but it significantly reduces spectral efficiency. 
Bluetooth uses frequency hopping to be immune to unknown interference and jamming attack, at the cost of using only one of 79 channels at one time. 
Spectrum spreading technique has been used in ZigBee, GPS, and 3G cellular networks. 
It allows these networks to be immune to low-power jamming attacks. 
However, their jamming immunity does not come free. 
It significantly reduces the spectral efficiency by expanding the signal bandwidth using a spreading code. 
Retransmission may be salvage wireless communication, but it also reduces the communication efficiency in the time domain.
MIMO-based jamming mitigation also lowers the spatial degrees of freedom that can be used for useful signal transmission. 
Therefore, a question to ask is how to improve the communication efficiency of wireless networks when they employ anti-jamming techniques to secure their communications.
As expected, the jamming resilience will not come for free.
A more reasonable question is how to achieve the tradeoff between communication efficiency and jamming resilience of a wireless network.

\subsubsection{Practicality of Anti-Jamming Techniques}
An important problem that remains open is to bridge the gap between theoretical study (or model-based analysis) and practical implementation. 
In the past decades, many research works focus on the theoretical investigation of anti-jamming techniques using approaches such as game theory and cross-layer optimization. 
Despite offering insights to advance our understanding of anti-jamming design, such theoretical results cannot be deployed in real-world wireless network systems due to their unrealistic assumptions (e.g., availability of global channel information, prior knowledge of jamming actions) and prohibitively high computational complexity.
Securing real-world wireless networks (e.g., Wi-Fi, cellular, ZigBee, Bluetooth, and GPS) calls for the intellectual design of anti-jamming strategies that can be implemented in realistic wireless environments where computational power and network-wide cooperation are limited. 
Particularly, PHY-layer anti-jamming techniques have a stringent requirement on their computational complexity.
This is because PHY-layer anti-jamming techniques should have an ASIC or FPGA implementation in modern wireless chips, which have a strict delay constraint for decoding each packet.

\subsubsection{Securing Wireless Communication System by Design}
A conventional anti-jamming approach for wireless communication systems is composed of the following three steps:
i) wireless devices detect the presence of jamming attacks,
ii) wireless devices temporarily stop their communications and invoke an anti-jamming mechanism, 
and
iii) wireless communication resumes under the protection of its anti-jamming mechanism.
This approach, however, is not capable of maintaining constant wireless connection under jamming attack due to the separation of jamming detection and countermeasure invocation, and the disconnection of wireless service may not be intolerable in many applications such as surveillance and drone control on the battlefield. 
Realizing this limitation, securing wireless communication \emph{by design} has emerged as an appealing anti-jamming paradigm and attracted a lot of research attention in recent years. 
The basic idea behind this paradigm is to take into account the anti-jamming requirement in the original design of wireless systems. 
By doing so, a wireless communication system may be capable of offering constant wireless services without disconnection when suffers from jamming attacks. 
For this paradigm, many problems remain open and need to be investigated, such as the way of designing anti-jamming mechanisms and the way of striking a balance between communication efficiency and jamming mitigation capability.

\subsection{Research Directions}

Jamming attack is arguably the most critical security threat for wireless networking services as it is easy to launch but hard to defend. 
The limited progress in the design of jamming-resilient wireless systems underscores the grand challenges in the innovation of anti-jamming techniques and the critical need for securing wireless networks against jamming attacks.
In what follows, we point out some promising research directions.

\subsubsection{MIMO-based Jamming Mitigation}
Given the potential of MIMO technology that has demonstrated in Wi-Fi and 4/5G cellular networks, the exploration of practical yet efficient MIMO-based jamming mitigation techniques is a promising research direction towards securing wireless networks and deserves more research efforts.
The past decade has witnessed the explosion of MIMO research and applications in wireless communication systems. 
With the rapid advances in signal processing and antenna technology, MIMO has become a norm for wireless devices.
Most commercial Wi-Fi and cellular devices such as smartphones and laptops are now equipped with multiple antennas for MIMO communication.
Recent results in  \cite{zeng2017enabling} show that, compared to frequency hopping and spectrum sharing, MIMO-based jamming mitigation is not effective in jamming mitigation but also efficient in spectrum utilization.
In addition, the existing results from the research on MIMO-based interference management (e.g., interference cancellation, interference neutralization, interference alignment, etc.) can be leveraged for the design of MIMO-based anti-jamming techniques. 
In turn, the findings and results from the design of MIMO-based anti-jamming techniques can also be applied to managing of unknown interference (e.g., blind interference cancellation) in Wi-Fi, cellular, and vehicular networks.

\subsubsection{Cross-Domain Anti-Jamming Design}
Most existing anti-jamming techniques exploit the degree of freedom in a single (time, frequency, space, code, etc.) domain to decode in-the-air radio packets in the presence of interfering signals from malicious jammers. 
For instance, channel hopping, which is used in Bluetooth, manipulates radio signals in the frequency domain to avoid jamming attack; spectrum spreading employs a secret sequence in the code domain to whiten the energy of narrow-band jamming signal to enhance a wireless receiver's resilience to jamming attacks; MIMO-based jamming mitigation aims to project signals in the spatial domain so as to make useful signal perpendicular to jamming signals. 
However, these single-domain anti-jamming techniques appear to have a limited ability of handling jamming signals due to a number of factors, such as the available spectrum bandwidth, the computational complexity, the number of antennas, the resolution of ADC, the nonlinearity of radio circuit, and the packet delay constraint. 
One research direction toward enhancing a wireless network's resilience to jamming attacks is by jointly exploiting multiple domains for PHY-layer signal processing and MAC-layer protocol manipulation. 
This direction deserves more research efforts to explore practical and efficient anti-jamming designs.

\subsubsection{Cross-Layer Anti-Jamming Design}
For constant jamming attacks, most existing countermeasures rely on PHY-layer techniques to avoid jamming signal or mitigate jamming signal for signal detection.
With the growth of smart jamming attacks that target on specific network protocols (e.g., preamble/pilot signals in Wi-Fi network and PSS/SSS in cellular network), cross-layer design for anti-jamming strategies becomes necessary to thwart the increasingly sophisticated jamming attacks.
It calls for joint design of PHY-layer signal processing, MAC-layer protocol design, and network resource allocation as well as cross-layer optimization to enable efficient wireless communications in the presence of various jamming attacks.

\subsubsection{Machine Learning for Anti-Jamming Design}
Machine learning has become a powerful technique and has been applied to many real-world applications such as image recognition, speech recognition, traffic prediction, product recommendations, self-driving cars, email spam, and malware filtering.
It is particularly useful for solving complex engineering problems whose underlying mathematical model is unknown.
In recent years, machine learning techniques have been used to secure wireless communications against jamming attacks (e.g., \cite{xiao2018uav,xiao2017reinforcement}) and produced some pioneering yet exciting results. 
Therefore, the design of learning-based anti-jamming techniques is a promising research direction that deserves more research efforts for an in-depth investigation.

\section{Conclusion}
\label{sec:conclusion}
This survey article provides a comprehensive review of jamming attacks and anti-jamming techniques for Wi-Fi, cellular, cognitive radio, ZigBee, Bluetooth, vehicular, RFID, and GPS wireless networks. 
For each network, we first offered a primer of its PHY and MAC layers and then elaborated on its vulnerability under jamming attacks, followed by an in-depth review on existing jamming strategies and defense schemes.
Particularly, we offered informative tables to summarize existing jamming attacks and anti-jamming techniques for each network, which will help the audience to grasp the fundamentals of jamming and anti-jamming strategies. 
We also listed some important open problems and pointed out the promising research directions toward securing wireless networks against jamming attacks.
We hope such a survey article will help the audience digest the holistic knowledge of existing jamming/anti-jamming research results and facilitate the future design of jamming-resilient wireless communication systems.


\bibliographystyle{ieeetr}

\end{document}